\documentclass[structabstract]{aa}
\usepackage{txfonts}
\usepackage{aas_macros}
\usepackage{natbib}
\usepackage{color}
\usepackage[normalem]{ulem}
\bibpunct{(}{)}{;}{a}{}{,} 
\usepackage{psfrag}
\usepackage{version}

\begin{document}

\title{Variations on a theme -- the evolution of hydrocarbon solids:}
\subtitle{III. Size-dependent properties -- the optEC$_{\rm (s)}(a)$ model \\[0.2cm] 
{\small Combined version of the published papers: A\&A, 542, A98 (2012) -- DOI: 10.1051/0004-6361/201118483 and \\ \hspace*{5.3cm} A\&A, 545, C3 (2012) DOI: 10.1051/0004-6361/201118483e\\[0.2cm] 
Data files are available at the CDS via \\[-0.1cm] 
http://cdsarc.u-strasbg.fr/viz-bin/qcat?J/A+A/545/C2 and http://cdsarc.u-strasbg.fr/viz-bin/qcat?J/A+A/545/C3}} 

\author{A.P. Jones \inst{1,2}}
           
    \institute{Institut d'Astrophysique Spatiale, CNRS/Universit\'e Paris Sud, UMR\,8617, Universit\'e Paris-Saclay, Orsay F-91405, France\\
               \email{ Anthony.Jones@ias.u-psud.fr}              }

    \date{Received 18 November 2011 / Accepted 1 April 2012}

   \abstract
{The properties of hydrogenated amorphous carbon (a-C:H) dust evolve in response to the local radiation field in the interstellar medium and the evolution of these properties is particularly dependent upon the particle size.}
{A model for finite-sized, low-temperature amorphous hydrocarbon particles, based on the microphysical properties of random and defected networks of carbon and hydrogen atoms, with surfaces passivated by hydrogen atoms, has been developed.}
{The eRCN/DG and the optEC$_{\rm (s)}$ models have been  combined, adapted and extended into a new optEC$_{\rm (s)}$(a) model that is used to calculate the optical properties of hydrocarbon grain materials down into the sub-nanometre size regime, where the particles contain only a few tens of carbon atoms.}
{The optEC$_{\rm (s)}$(a) model predicts a continuity in properties  from large to small (sub-nm) carbonaceous grains. Tabulated data of the size-dependent optical constants (from EUV to cm wavelengths) for a-C:H (nano-)particles as a function of the bulk material band gap [$E_{\rm g}$(bulk)], or equivalently the hydrogen content, are provided. The effective band gap [$E_{\rm g}$(eff.)] is found to be significantly larger than $E_{\rm g}$(bulk) for hydrogen-poor a-C(:H) nano-particles and their predicted long-wavelength ($\lambda > 30\,\mu$m) optical properties differ from those derived for interstellar polycyclic aromatic hydrocarbons (PAHs).}
{The optEC$_{\rm (s)}$(a) model is used to investigate the size-dependent structural and spectral evolution of a-C(:H) materials under ISM conditions, including: the IR-FUV extinction, the 217\,nm bump and the infrared emission bands. 
The model makes several predictions that can be tested against observations. }


\maketitle

\section{Introduction}

The evolution of  hydrocarbon solids is a key issue in astrophysical studies and these materials have received considerable interest as a model for the properties of the solid carbonaceous matter in the interstellar medium \citep[ISM, {\it e.g.},][]{1990QJRAS..31..567J,1995ApJ...445..240D,1997ApJ...482..866D,2004A&A...423..549D,2004A&A...423L..33D,2005A&A...432..895D,2008A&A...490..665P,2008A&A...492..127S,2009ASPC..414..473J,2010A&A...519A..39G,2011A&A...529A.146G,2011A&A...525A.103C}, in circumstellar media \citep[{\it e.g.},][]{2003ApJ...589..419G,2007ApJ...664.1144S} and in the Solar System \citep[{\it e.g.},][]{2011A&A...533A..98D}. 

Hydrocarbon solids darken with ultraviolet (UV) irradiation and thermal annealing \citep[{\it e.g.},][]{1985OpEffinAS.120..258I,1984JAP....55..764S} due to the evolution of the band gap or optical gap energy, $E_{\rm g}$, of the material in response to the local conditions. The result is a loss of hydrogen atoms from the structure and an evolution towards more aromatic materials. The evolution of the band gap is therefore at the heart of the inherent variability of a-C(:H) properties that are of interest in their utility in unravelling  the nature of hydrocarbon grains in the ISM.

As elucidated in detail in \cite{jones2011A,jones2011B}, hereafter called papers~I and II, this work was conceived with the application of the Random Covalent Network (RCN) theory and defective graphte (DG) model for the structure of amorphous hydrocarbon materials \citep{1979PhRvL..42.1151P,1980JNS...42...87D,1983JNCS...57..355T,1988JVST....6.1778A,1990JAP....67.1007T} to the astrophysical context \citep[][and papers~I and II]{1990MNRAS.247..305J}.
As in papers~I and II, an amorphous hydrocarbon particle is taken to be a finite-sized,  macroscopically-structured ({\it i.e.}, a contiguous network of atoms), solid-state material consisting of only carbon and hydrogen atoms. The designation a-C(:H) is used to imply the suite of hydrogen-poor (a-C) to hydrogen-rich (a-C:H) solid carbonaceous materials whose optical and electrical properties were extensively reviewed by  \citep{1986AdPhy..35..317R}.
In paper~I it was shown that the eRCN and DG structural models can be used to predict the structural and spectral properties of the suite of a-C(:H) solids. In paper~II this was used as the basis for the optEC$_{\rm (s)}$ model, which determines the optical properties (complex refractive index) for bulk a-C(:H) materials and predicts important observable consequences for their evolution in the ISM. This paper presents the optEC$_{\rm (s)}$(a) model (optical property prediction tool for the Evolution of Carbonaceous (s)olids as a function of r(a)dius) that allows a determination of the size-dependent optical properties of a-C(:H) materials down to particles containing perhaps only a few tens of carbon atoms.

This paper considers the nature and evolution of hydrogenated amorphous carbons, a-C(:H) or HAC, under extraterrestrial conditions and is organised as follows: 
Sect.~\ref{sect_size_structure_effects} introduces size- and surface-dependent structural properties, 
then in Sect.~\ref{size-dep_props} the size-dependent optical properties of a-C(:H) particles are derived,
in Sect.~\ref{sect_optECsa_data} the complex indices of refraction, $m(a,E_{\rm g},\lambda)=n+ik$, the optEC$_{\rm (s)}$(a)  data, are presented and discussed,  
Sect.~\ref{sect_ast_impl} presents the astrophysical implications of these data, 
Sect.~\ref{sect_predictions} summarises the principal predictions of the model, 
Sect.~\ref{sect_limitations} emphasises the limitations of the optEC$_{\rm (s)}$(a) data and
Sect.~\ref{sect_conclusions} concludes and summarises the work.

The Appendices~\ref{appendix_surface_H_effects} to \ref{appendix_heat_cap} present the supporting and more detailed technical aspects of the adopted solid-state physics approach, which are not essential to an understanding of the astrophysical aspects of the work.

\section{Structural and size-dependent properties}
\label{sect_size_structure_effects}

Papers~I and II considered the structure, composition, spectra and optical properties of bulk a-C(:H) solids, which can be applied to particles in the ISM with radii as small as 100\,nm. However, the properties of smaller particles will be increasingly dependent upon their size \citep[{\it e.g.},][]{2001ASPC..231..171J}. This section considers in detail the effects of size upon the surface and ``bulk'' properties of a-C(:H) finite-sized particles.

\subsection{The specific density of a-C(:H) materials}
\label{sect_aCH_density}

The optical and physical properties of amorphous hydrocarbons were extensively studied by \cite{1984JAP....55..764S} who showed that their density varies as a function of the annealing temperature, which is accompanied by a reduction in the hydrogen atom content, $X_{\rm H}$, and a concomitant reduction in the band gap ($E_g \simeq 4.3 \, X_{\rm H}$, see  Sect.~\ref{size-dep_props}). The following relationship between the material density and the band gap can be derived from the \cite{1984JAP....55..764S} data, 
\begin{equation}
\rho(E_{\rm g}) \approx 1.3  + 0.4 \, {\rm exp}[ -( E_{\rm g} + 0.2 ) ] \ \ \ \ \ {\rm g \, cm^{-3}}, 
\label{rhoXH}
\end{equation}
which yields a  value of $\simeq 1.3$\,g\,cm$^{-3}$ at large $E_{\rm g}$ and $\simeq 1.6$\,g\,cm$^{-3}$ for $E_{\rm g} = 0$.  Note that these densities are somewhat lower than the values generally used for interstellar carbon dust and that this could  ease the carbon abundance constraints on  dust models.

\subsection{Surface hydrogenation effects}
\label{sect_surf_hydrogenation_effects}

In the case of macroscopic, eRCN/DG-described materials all of the associated hydrogen and carbon atoms are ``internal'' to the structure, {\it i.e.}, the surface effects are not considered. However, for particles a significant fraction of the atoms may be at the surface ({\it e.g.}, see Appendix~\ref{appendix_surface_H_effects}). For carbon atoms this leads to incomplete coordination and therefore to the presence of ``dangling'' bonds, C(---)$_n$ where $1 \leqslant n \leqslant3$. These ``dangling'' bonds must be passivated and in a hydrocarbon material, in the absence of surface re-structuring (see below), this can only be by the addition of extra C(---H)$_n$ bonds at the surface. The extra hydrogen atoms can be accounted for by including a surface component to $X_{\rm H}$, which is designated as $X_{\rm H}^s(a)$, where $a$ is the particle radius. This component adds H atoms but is not allowed to alter the bulk structure, which would lead to a change in the band gap, $E_{\rm g}$, inconsistent with, and in violation of, the model. A new, surface-dependent hydrogen atom fraction, can now be defined,  
\begin{equation}
X_{\rm H}^\prime = [\ X_{\rm H} + X_{\rm H}^s(a)\ ], 
\label{eq_surf_H}
\end{equation}
which is not re-normalised, because the surface H atoms are counted as ``extra''. Eq.~(\ref{eq_surf_H}) trivially reduces to $X_{\rm H}$ for the bulk network for large particles or when no surface effects are considered. Appendix~\ref{appendix_surface_H_effects} gives a detailed analysis of size and surface hydrogenation effects in the eRCN/DG model; there it is concluded that, for particles with radii $\gtrsim 30$\,nm, the effects of surface hydrogenation can be ignored and the bulk optical properties used to describe them. 

It is possible for hydrocarbon particle surfaces to reconstruct without the addition of extra H atoms \cite[{\it e.g.},][]{2001ASPC..231..171J}. For example, a surface $sp^3$ aliphatic C$-$C pair, with a dangling bond on each C atom, could reconstruct to form a $sp^2$ olefinic C$=$C pair without the need for additional H atoms. However, this type of re-construction  is inconsistent with the model because it would effectively alter the bulk properties of sub-nm particles (see Appendix \ref{appendix_surface_H_effects}). 

\subsection{The aliphatic CH$_2$/CH$_3$ ratio in finite-sized particles}

\begin{figure}
 \resizebox{\hsize}{!}{\includegraphics{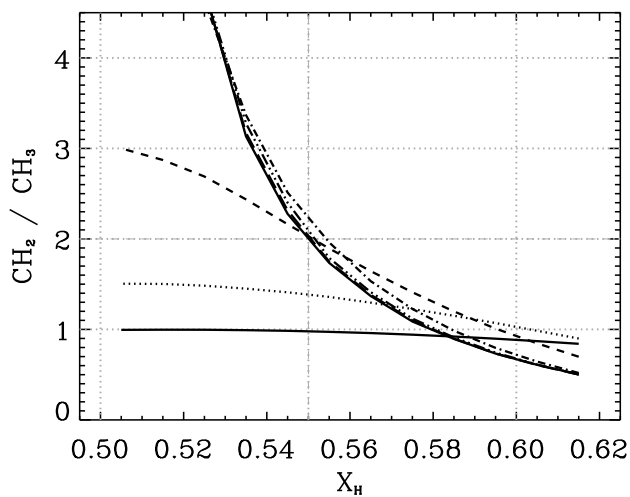}}
 \caption{ The CH$_2$/CH$_3$ abundance ratio as a function of $X_{\rm H}$ and radius: 100\,nm (curved solid line), 30\,nm (long-dashed), 10\,nm (triple-dot-dashed), 3\,nm (dot-dashed), 1\,nm (short-dashed), 0.5\,nm (dotted) and 0.33\,nm (flat solid line),  for the aliphatic carbon component  in the eRCN model (see paper~I). }
 \label{fig_decomp_4_III}
\end{figure}

In the ISM it appears that the observed $3.3-3.6\,\mu$m absorption spectrum towards the Galactic Centre and the aliphatic CH$_2$/CH$_3$ ratio are consistent with amorphous hydrocarbon grains \cite[{\it e.g.},][see also paper~I]{2002ApJS..138...75P,2005A&A...432..895D}. Fig.~\ref{fig_decomp_4_III} shows the CH$_2$/CH$_3$ abundance ratio as a function of $X_{\rm H}$ and radius ($a = 100$, 30, 10, 3, 1, 0.5 and 0.33\,nm) for the eRCN a-C:H model aliphatic carbon component (see paper~I). What Fig.~\ref{fig_decomp_4_III} shows is that, except for particles with radii $\leqslant 1$\,nm ({\it i.e.}, the short-dashed, dotted and lower [$\sim$ horizontal] solid lines that deviate from the general trend in the figure),  the CH$_2$/CH$_3$ ratio depends only on $X_{\rm H}$ and is practically independent of size. Thus, H-rich, a-C:H particles significantly smaller than 100\,nm, but larger than 1\,nm, could be consistent with the observed CH$_2$/CH$_3$ abundance ratio. However, it should be recalled that the smallest carbonaceous grains in the ISM ($a \lesssim 20$\,nm) are probably hydrogen-poor and highly aromatic ({\it e.g.}, see Sect.~\ref{sect_proc_timescales}) and, even though their CH$_2$/CH$_3$ ratio may be significant, the atomic fraction of aliphatic CH$_2$ and CH$_3$ groups is small for $X_{\rm H} < 0.2$  and their CH$_2$/CH$_3$ ratio likely to be very different from that of larger grains. Hence, the conclusion in paper~I, that the interstellar absorption observations are consistent with large a-C:H particles having an atomic hydrogen content $>57$\%, with a CH$_2$/CH$_3$ abundance ratio $\approx 0.5-1.3$, appears to hold true, provided that the particles in question are larger than $\simeq 40$\,nm. 

\subsection{The structural properties of very small grains}

Clearly the effects of size on the structural properties of a-C(:H) particles will be most pronounced for the smallest particles and, following from the above discussions (see also Appendix~\ref{appendix_surface_H_effects}), it is of interest to concentrate on the properties of particles smaller than 10\,nm. 
The model takes into account the material density dependence on $E_{\rm g}$ ($\equiv X_{\rm H}$, Eq.~\ref{rhoXH}) and also includes an effective radius term ($a_{\rm eff} = a^{0.8}$) to simulate lower effective densities for particles smaller than 1\,nm, which could arise from structural relaxation effects (see below), but that do not engender any re-structuring of the material. 

It is found that surface hydrogenation (passivation) effects can be safely ignored for particles with $a \gtrsim 30$\,nm, {\it i.e.}, the fraction of H atoms on the surfaces of large interstellar grains is negligible ($\lesssim$ a few percent). However, for particles as small as 0.5\,nm, there can be as many H atoms on the surface as in the interior ({\it e.g.}, see Appendix~\ref{appendix_surface_H_effects}, Fig.~\ref{fig_Fs_vs_a}). Given that the ion- and UV-irradiation of a-C(:H) leads to its aromatisation, and that experiments find an $\approx5$\% ($X_{\rm H} = 0.05$) lower limit to the H atom content \citep{1989JAP....66.3248A,1996MCP...46...198M,2011A&A...528A..56G,2011A&A...529A.146G}, H-poor a-C will consist primarily of $sp^2$ C atoms. For such materials ($X_{\rm H} \lesssim 0.2$) the DG model, rather than the eRCN model, must be applied and for $X_{\rm H} = 0.05$ this model predicts a carbon atom hybridisation ratio $R = X_{sp3}/X_{sp2} = 0.04$, equivalent to an $sp^2$ [$sp^3$] carbon atom fraction of 0.91 [0.04] (see the lower dashed line, for the DG model,  in Fig.~2 of paper~I). For a 0.5\,nm radius, H-poor particle there will be $\approx 25$ times as many H atoms on the surface as in the bulk (see Fig.~\ref{fig_XHprime_vs_a}), which might at first glance seem rather unreasonable. Stoichiometrically, the {\em bulk} structure of a particle with $X_{\rm H} \sim 0.05$, could {\em schematically} be represented by a coronene (C$_{24}$H$_{12}$) and two pyrene molecules (2\,C$_{16}$H$_{10}$) that have lost 30 of their 32 peripheral H atoms, {\it i.e.}, C$_{24}$H + C$_{16}$H + C$_{16}$ (C$_{56}$H$_2$), $\equiv X_{\rm H} \approx 0.04$. If the bulk material were to be reduced to a particle of radius $\sim$\,0.5\,nm, the three molecules ``piled up'' to create a pyrene-coronene-pyrene sandwich and the 30 H atoms replaced, a surface-to-bulk H atom ratio of $30/2 = 15$, rather similar to that predicted by the model is recovered (Appendix~\ref{appendix_surface_H_effects}). 

What the above discussion indicates, perhaps rather surprisingly, is that the adopted surface H atom passivation model is, apparently, applicable down to particles, or large molecules, with as few as $50-60$ carbon atoms. However, the optEC$_{(s)}$ model optical property determination (paper~II) is strictly no longer valid because of the transition from bulk-like band structure towards discrete molecular bands. This is the interesting interface between solid-state and molecular physics where a bulk material optical property description is probably no longer valid (see later discussions). Nevertheless, this bulk-material approach may yield a useful ``stop-gap'' until there is a better understanding of the optical properties of particles or clusters in the solid-to-molecule transition domain.

\subsection{Experimental results on clustering in a-C(:H)}

In their laser ablation-produced carbonaceous particle UV-irradiation experiments \cite{2011A&A...528A..56G} typically find, in all of their materials (with pre-irradiation values of $E_{\rm g}$ in the range $1.1-1.65$\,eV), the presence of strongly distorted graphene layers and small fullerene-like carbon structures.\footnote{Suggesting an intriguing link between a-C(:H) particle evolution, via UV photolysis, and the formation of fullerenes.} Post UV-irradiation they find $E_{\rm g} = 1.54 \pm 0.17$\,eV but note that this may be due to a scattering effect and that the true values of $E_{\rm g}$ probably lie in the range $0.25-0.45$\,eV, corresponding to $X_{\rm H} = 0.05-0.1$. In their HRTEM images they are able to discern aromatic clusters containing $3-8$ rings (with diameters $\simeq 0.4-0.7$\,nm). 

As \cite{2004PhilTransRSocLondA..362.2477F}, \cite{2007JAP...102g4311H} and \cite{2007JChPh.126o4705H}  point out, $sp^2$ chain and cage-like clusters and $sp^1$ (cumulenic) chains may be rather common structures in some a-C:H materials. 

\cite{2003ApJ...594..869P} have theoretically studied an interesting suite of structures that they call locally aromatic polycyclic hydrocarbons (LAPHs), which lie in the compositional range C$_{19}$H$_{22}$ to C$_{36}$H$_{32}$. These LAPHs contain aromatic domains with only a few rings per domain that are linked by bridging aliphatic ring systems. They have shown that these molecular species should have spectral properties rather similar to HAC and a-C(:H) materials. It is possible that within the astrophysical context that the aromatic components of LAPH-like species could also be bridged olefinic carbon-containing rings and chains. 

Rather interestingly, \cite{1984JAP....55..764S} and \cite{1996ApJ...472L.123S} have shown that a-C:H/HAC decomposition leads to low density materials ($\rho \lesssim 1.5$\,g~cm$^{-3}$). \cite{1996ApJ...472L.123S}  show, in particular, that the solid decomposition product appears to be an aerogel-like material consisting of weakly-connected aromatic ``proto-graphitic'' or polycyclic aromatic hydrocarbon(PAH)-like clusters in a ``friable network''.  Thus, the simplistic, radius-defined number of carbon atoms in the particle (see Eq.~\ref{eq_Natom_estimator}) is probably not going to be valid for particles containing less that a few hundred atoms and it would be better to think more in terms of an effective radius approach, as above, or to just consider their properties as a function of the number of carbon atoms as is done for interstellar ``PAH'' molecules. The effective radius approach, {\it i.e.}, $a_{\rm eff} = a^{-0.8}$, leads to about a factor of two increase in the volume and therefore in the carbon atom content for particles of radius $a = 0.4$\,nm (see Fig.~\ref{fig_NCatom_vs_a}).

\section{Size-dependent optical properties}
\label{size-dep_props}

As discussed in paper~II, laboratory studies of amorphous hydrocarbons indicate a linear relationship between the band gap, $E_{\rm g}$, of these materials and their atomic hydrogen fraction, $X_{\rm H}$, \citep{1990JAP....67.1007T}, {\it i.e.}, 
\begin{equation}
X_{\rm H} \simeq  \frac{E_{\rm g}[{\rm eV}]}{4.3},
\label{Eg4XH}
\end{equation}
and also with the number of aromatic rings, $N_{\rm R}$, in an aromatic domain or cluster \citep{1987PhRvB..35.2946R}, 
\begin{equation}
N_{\rm R} \simeq \left( \frac{5.8}{E_{\rm g}[{\rm eV}]} \right)^2. 
\label{eq_NR_Eg}
\end{equation}
The presence of hydrogen leads to broad resonances in the a-C:H optical spectra that lie well away from the gap. Therefore, it is not the presence of hydrogen atoms that directly contributes to the gap. However, their presence completely determines the $sp^3$ to $sp^2$ carbon atom ratio and hence the nature of the carbon atom clustering into aliphatic, olefinic and aromatic domains \citep[see Fig.~1 in paper~I and also][]{1986AdPhy..35..317R,2004PhilTransRSocLondA..362.2477F}. 

\begin{figure}
 \resizebox{\hsize}{!}{\includegraphics{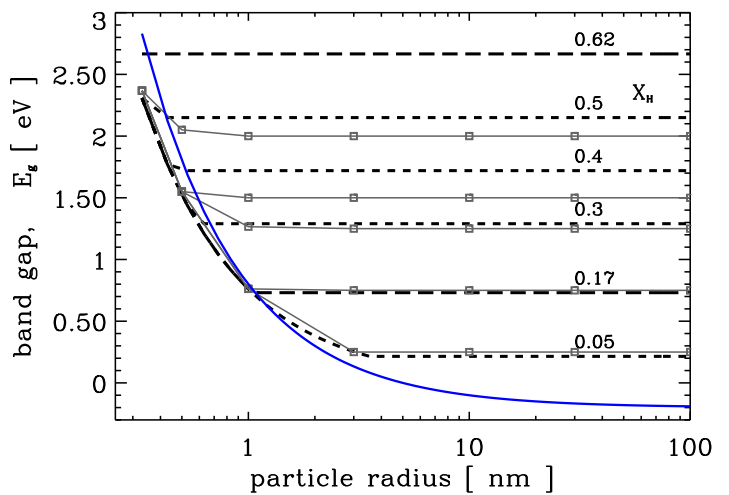}}
 \caption{Size effects on the band gap of eRCN/DG particles as a function of $X_{\rm H}$ (black dashed lines); the long-dashed lines indicate the upper and lower validity limits for the eRCN model, $X_{\rm H} = 0.62$ and 0.17, respectively. The grey lines with square data points indicate the size-dependent band gap derived using $\pi-\pi^\ast$ band constraints (for $E_{\rm g}{\rm (bulk)} = 0.25$, 0.75, 1.25, 1.50, 2.00\,eV, $\equiv X_{\rm H} = 0.06$, 0.17, 0.29, 0.35 and 0.47, respectively, from bottom to top) and the solid (blue) line shows the approximate empirical limit, $E_{\rm g}{\rm [eV]} = \{(a{\rm [nm]})^{-1}-0.2\}$, above which no aromatic cluster size effects need to be taken into account (see Sect.~\ref{appendix_pi_pi}).}
\label{Eg_size_effects}
\end{figure}

Paper~I showed that the number of carbon atoms $n_{\rm C}$ per cluster and the radius of the most compact aromatic domain, $a_{\rm R}$, can be expressed as  
\begin{equation}
n_{\rm C} = 2 N_{\rm R} + 3.5 \surd N_{\rm R} + 0.5  \ \ \ {\rm and}
\label{eq_nC_NR}
\end{equation}
\begin{equation}
a_{\rm R} = 0.09 [2 N_{\rm R} + \surd N_{\rm R} + 0.5]^{0.5} \ \ {\rm nm}.  
\label{eq_aR}
\end{equation}
As noted in paper~II, the particle size imposes strict limits on the minimum possible band gap, which is determined by the largest aromatic cluster, {\it i.e.}, aromatic clusters cannot be bigger than the particle. Fig.~\ref{Eg_size_effects} shows the limiting values of $E_{\rm g}$ as a function of the particle size for different values of $X_{\rm H}$ ($\equiv E_{\rm g}$). In conclusion, aromatic cluster limiting-size effects can safely be ignored for particles with radii greater than a few nm. For large particles $E_{\rm g}$ is constrained by the H atom fraction, which in this work is determined by the eRCN and DG models. 

A band gap size-dependence is not unique to a-C(:H) materials and a size-dependent band gap has been experimentally-determined for (nano-)diamonds down to radii of 2\,nm \citep[see][and references therein, for a discussion of the size-dependent properties of nano-diamonds]{2001ASPC..231..171J}.

For the smaller particles $E_{\rm g}$ can be determined, as a function of size, using Eqs.~(\ref{Eg4XH}) to (\ref{eq_aR}).  Fig.~\ref{Eg_size_effects} clearly illustrates that hydrogenated amorphous carbon nano-particles will always have a significant band gap, one that is larger than for bulk matter of the same composition. This effect is simply due to the maximum allowed aromatic cluster size. The maximum band gap is here completely determined by the bulk hydrogen content, $X_{\rm H}$ (Eq.~\ref{Eg4XH}), interior to the surface of the particle. The ``extra'' hydrogen atoms required to passivate the ``dangling'' C(---)$_n$ bonds at the particle surface do not affect the bulk structure. 

In conclusion, the surface hydrogenation of a-C(:H) particles: {\em does not} affect their band gap, which is determined by the ``bulk'' H atom content, but {\em does} affect their spectra because of the addition of ``extra''  H atoms in surface CH$_n$ groups.  However, the {\em size} of the particles {\em does} affect the band gap because it limits the maximum allowed size of the band gap-determining aromatic clusters.

The following sections (and the Appendices) consider the effects of size on the annealing behaviours of of a-C(:H) particles and, in particular, its effect on the visible-UV optical properties and IR spectral properties.

\subsection{Size-dependent visible-UV electronic properties}
\label{sect_visUV_el_props}

The wide-band optical properties of a-C(:H) materials at energies greater than $\sim0.5$\,eV show two peaks: a $\pi-\pi^\star$ peak at $\sim 4$\, eV and $\sigma-\sigma^\star$ at $\sim 13$\, eV, and an additional peak at $\sim 6.5$\, eV attributed to C$_6$, ``benzene-like'' aromatic clusters in the structure \citep[see following Sect.~\ref{sect_pi_pi_C6_sig_sig}, paper~II and][for more details]{1986AdPhy..35..317R}. The optEC$_{(\rm s)}$ model presented in paper~II uses these bands as the basis for the de-construction of the visible-UV electronic properties of a-C(:H) materials. In their optical property derivation \cite{2007DiamondaRM...16.1813K} also fix their $\pi-\pi^\star$ and $\sigma-\sigma^\star$ band energies. Recently, \cite{2011A&A...528A..56G} decomposed the HAC visible-UV spectra measured in their experiments into $\pi-\pi^\star$ ($2.5-4.5$\,eV), $\sigma-\sigma^\star$ ($> 6$\,eV) and $n-\sigma^\star$ ($5.0-5.5$\,eV) components. However, and unlike our approach, they allow their component bands to vary, as indicated by the energy ranges shown in brackets. 

The thermal- or photo-annealing of a-C(:H) materials leads to a strengthening of the $\pi-\pi^\ast$ with respect the $\sigma-\sigma^\ast$ and $\sim 6.5$\, eV (C$_6$) peaks. These trends are the result of increasing aromaticity with increased annealing (see paper II). The following sections discuss how the observed behaviour of the $\pi-\pi^\ast$, $\sigma-\sigma^\ast$ and $\sim 6.5$\, eV (C$_6$) bands are affected by particle size.

\subsubsection{The $\pi$--$\pi^\star$, C$_6$ and $\sigma$--$\sigma^\star$ bands}
\label{sect_pi_pi_C6_sig_sig}

In this work a phenomenological model is adopted for the $\pi$--$\pi^\star$ band, the band arising from the inherent size distribution of aromatic clusters within the given a-C(:H) material \citep[{\it e.g.},][]{1986AdPhy..35..317R}. 
The bulk $\pi$--$\pi^\star$ bands are predominantly due to $sp^2$ clusters with an even number of carbon atoms because they are more stable and do not have transitions deep within the band gap. The form of the $\pi$--$\pi^\star$ band is dependent upon the band gap, $E_{\rm g}$, and must also be size-dependent because it is the size of the largest aromatic clusters, in major part, that determine the band gap of the material. The aromatic clusters are limited by the particle size, and critically so for nano-particles, where the largest aromatic clusters obviously cannot be bigger than the particle size. Fig.~\ref{Eg_size_effects} shows how the particle band gap is directly affected by this limitation on the size of the largest aromatic clusters.

A band attributable to six-fold ({\it i.e.}, ``benzene'' ring-like) aromatic clusters in a-C(:H), here designated the  C$_6$  band, also contributes to the wide-gap optical properties of a-C(:H) \citep[{\it e.g.},][]{1986AdPhy..35..317R}. However, {\em this band is not due to benzene} because the aromatic clusters are an intrinsic component of the structural network and are not fully hydrogenated as per benzene (C$_6$H$_6$). Note that the adopted C$_6$ band could include important contributions from clusters other than six-fold aromatic rings (see Appendix~\ref{appendix_visUV_el_props}, Sect.~\ref{appendix_all_bands} and Sect.~\ref{appendix_C6}) and also from the normally weak $\pi-\sigma^{\star}$ transitions involving the promotion of electrons from bonding $\pi$ orbitals to anti-bonding $\sigma^{\star}$ orbitals, which would indeed lead to the larger than predicted band width. The $\sigma$--$\sigma^\star$ band could also exhibit some size-dependence effects but no such  effect is considered here because of the current lack of suitable laboratory constraints. 

Appendix~\ref{appendix_visUV_el_props}, following on from the approach presented in paper~II, gives the technical details of how particle size limits the aromatic cluster size distribution and how this in turn affects the optical properties. Principally, this results in significantly reduced absorption at wavelengths longwards of the $\pi$--$\pi^\star$ band ({\it i.e.}, energies $< 0.5$\,eV) as the particle radius decreases. 

\begin{figure} 
 \resizebox{\hsize}{!}{\includegraphics{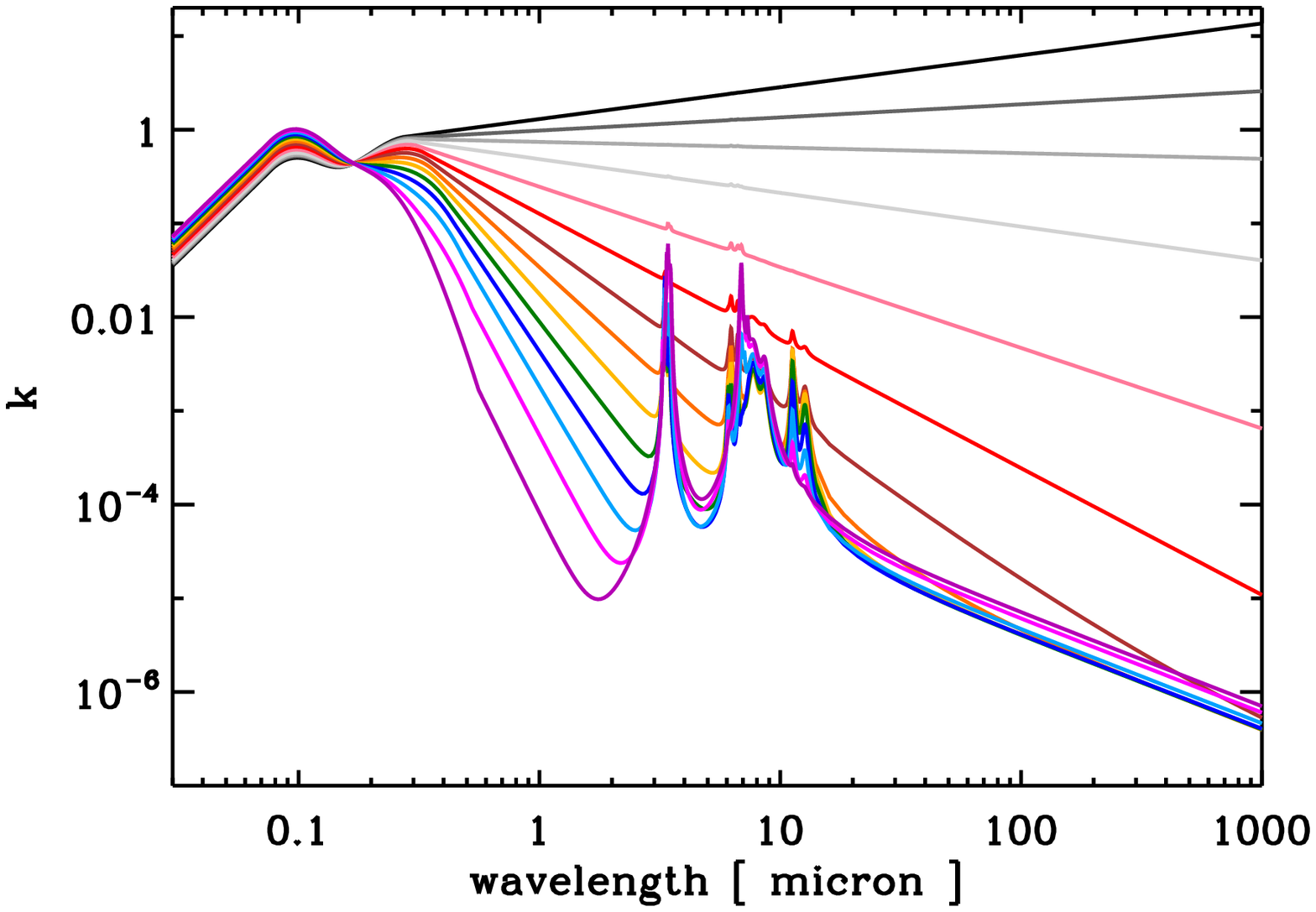}}
 \resizebox{\hsize}{!}{\includegraphics{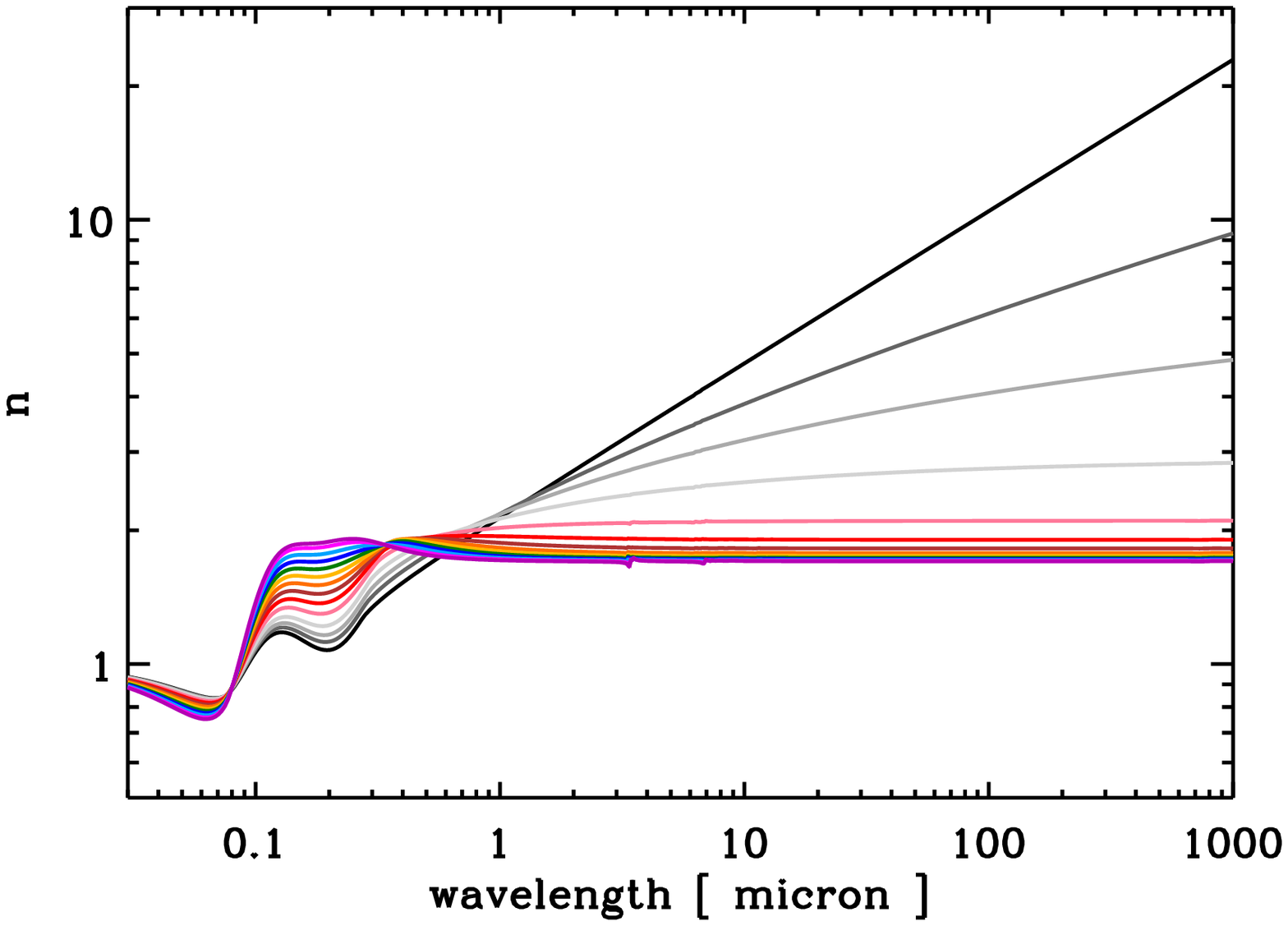}}
 \caption{The imaginary ($k$, upper plot) and real ($n$, lower plot) parts of the refractive index for 100\,nm radius particles as a function of $E_{\rm g}$ for large gap (2.67\,eV, lower purple lines  at $\lambda = 2\,\mu$m) to small gap ($-0.1$\,eV, upper black lines at $\lambda = 2\,\mu$m) a-C(:H) materials (see Table~1 
 for the full colour-coding scheme). }
 \label{fig_nk_100nm_eg}
\end{figure}
\begin{figure} 
 \resizebox{\hsize}{!}{\includegraphics{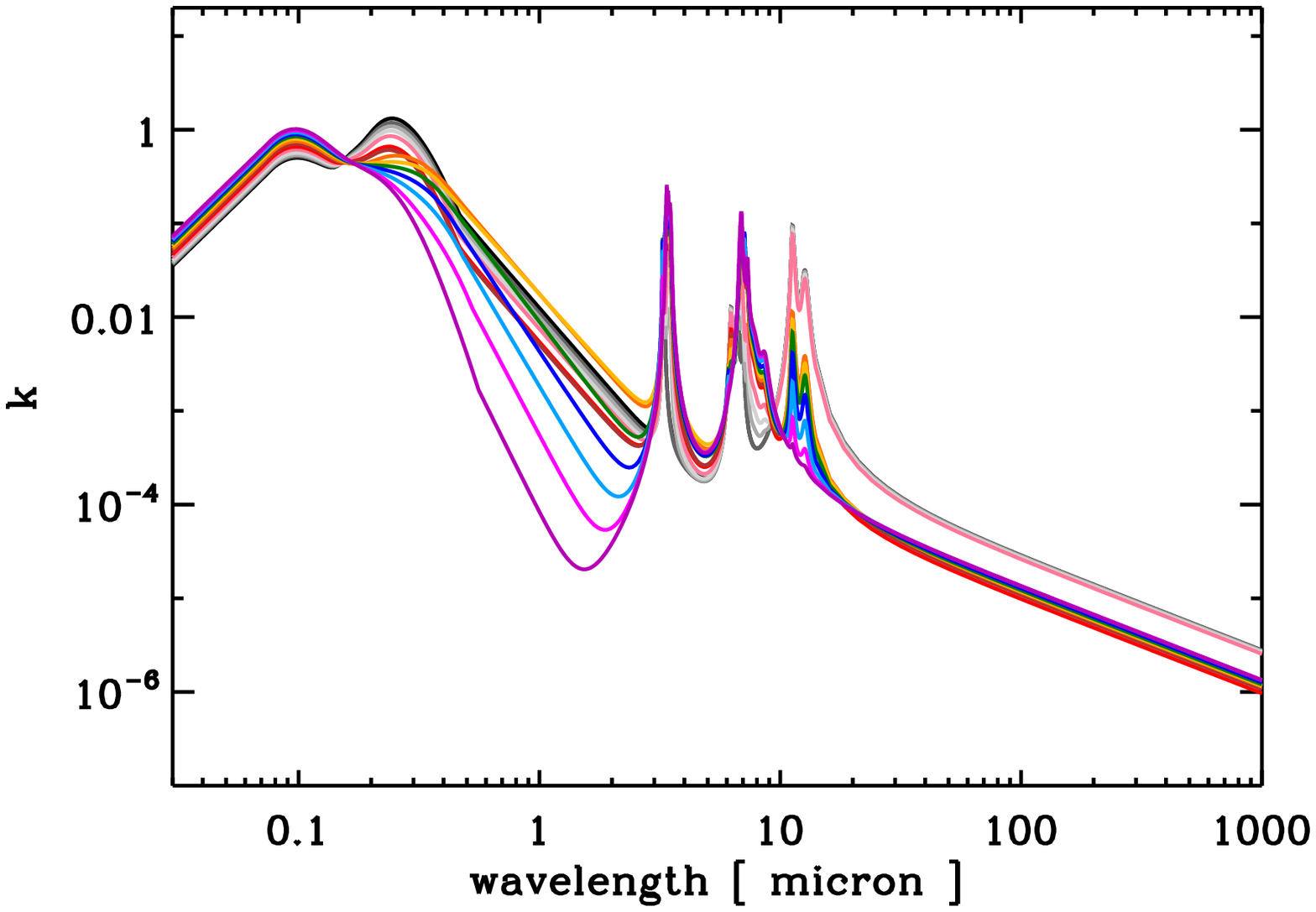}}
 \resizebox{\hsize}{!}{\includegraphics{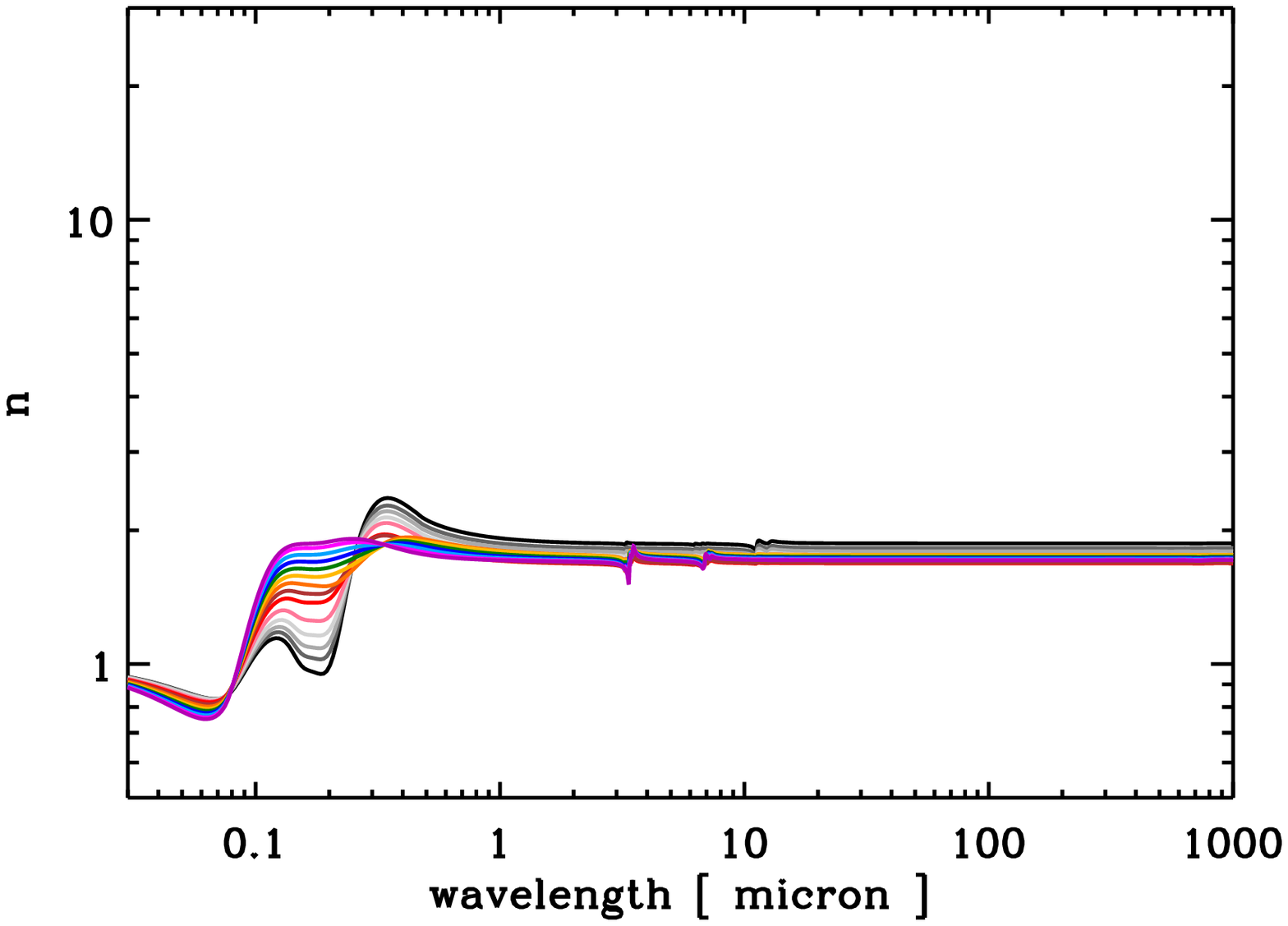}}
 \caption{Same as for Fig.~\ref{fig_nk_100nm_eg} but for 0.5\,nm radius particles.  Note that, at $\lambda = 2\,\mu$m, $n$ and $k$ change little with size for a-C:H materials with $E_{\rm g} \geq 1.5$\,eV (lower purple to yellow lines in the upper $k$ plot) but bunch-up for $E_{\rm g} < 1.5$\,eV. Note also that, for $\lambda > 20\,\mu$m, the $n$ and $k$ data appear to be almost independent of the band gap for $E_{\rm g} \geq 0.5$\,eV.  }
 \label{fig_nk_0.5nm_eg}
\end{figure}

\begin{table}
\caption{The optEC$_{\rm (s)}$(a) material band gap, $E_{\rm g}$, hydrogen atom fraction, $X_{\rm H}$, and colour coding scheme.}
\begin{center}
\begin{tabular}{lcll}
                                                               &                              &         &    \\[-0.35cm]
\hline
\hline
                                                              &                              &           &    \\[-0.35cm]
 $E_{\rm g}$  [ eV ]                               &    $X_{\rm H}$  &   colour  &               \\[0.05cm]
\hline
                                                             &                            &              \\[-0.35cm]
   \hspace*{-0.3cm} $-0.1$  [$E_{g-}$]  &       0.00            &   black              &                             \\
    0.0                                                    &        0.00            &   dark grey        &                   \\
    0.1                                                    &        0.02             &   mid grey        &                    \\
    0.25                                                  &        0.05             &   lightgrey        &   $|$             \\
    0.5                                                    &        0.11             &   pink               &   $|$  a-C     \\
    0.75                                                  &        0.17             &   red                &   $|$              \\
    1.0                                                    &        0.23             &   brown            &                      \\
    1.25                                                  &        0.29             &   orange          &   $|$              \\
    1.5                                                    &        0.35             &   yellow            &   $|$              \\
    1.75                                                  &        0.41             &   green             &   $|$             \\
    2.0                                                    &        0.47             &   blue               &    $|$  a-C:H  \\
    2.25                                                  &        0.52             &   cobalt            &    $|$             \\
    2.5                                                    &        0.58             &   violet             &    $|$             \\
    2.67 [$E_{g+}$]                                &        0.62             &   purple           &     $|$            \\\hline
\hline
                                                             &                            &        &     \\[-0.25cm]
\end{tabular}
\end{center}
\label{table_colour_code}
\end{table}

\section{The optEC$_{\rm (s)}(a)$ model refractive index data}
\label{sect_optECsa_data}

This paper extends the earlier work of papers~I and II to include the effects of both band gap and particle size on the derived complex refractive index, $m(n,k)$. Figs.~\ref{fig_nk_100nm_eg} and \ref{fig_nk_0.5nm_eg} show the derived $n$ and $k$ data for 100\,nm and 0.5\,nm radius particles and for $E_{\rm g}$(bulk) from $-0.1$ to $2.67$\,eV. 
The full colour-coding scheme is given in Table~\ref{table_colour_code}.  The size-dependent refractive index data for the other particle radii ($a = 30, 10, 3, 1$ and 0.33\,nm) are given in Appendix~\ref{appendix_n_and_k}. The $n$ and $k$ data for all of the considered particle radii ($a = 100, 30, 10, 3, 1, 0.5$ and 0.33\,nm) are  tabulated in the accompanying pairs of ASCII files for $n$ and $k$. These data, presented as a function of the band gap ($E_{\rm g} \equiv 4.3 X_{\rm H}$) and particle radius from EUV to cm wavelengths, were derived using the formalism described in paper~II, which has here been extended to include size-effects (see Sect.~\ref{sect_visUV_el_props} and Appendix~\ref{appendix_visUV_el_props}). As in the earlier work the wavelength- and composition-dependent absorption coefficient, $k(\lambda,E_{\rm g},a)$, is calculated and then the Kramers-Kronig relations used to calculate the accompanying values of $n(\lambda,E_{\rm g},a)$. 

The first thing to be seen in Fig.~\ref{fig_nk_100nm_eg} is, not unexpectedly given the analysis of Sect.~\ref{sect_size_structure_effects} (and Appendix~\ref{appendix_surface_H_effects}), that the optical properties of 100\,nm radius particles are identical to those of the bulk materials derived in paper~II. 
However, as can be seen by comparing Figs.~\ref{fig_nk_100nm_eg} and \ref{fig_nk_0.5nm_eg}\footnote{Note that in Figs.~\ref{fig_nk_100nm_eg} and \ref{fig_nk_0.5nm_eg} and Figs.~\ref{fig_nk_30nm} to \ref{fig_nk_0.33nm} the same $x$- and $y$-axis ranges are used in order to allow a direct comparison between the different data sets.}
for 100\,nm and 0.5\,nm radius particles, the principal changes in the $n$ and $k$ data, manifest as an increased contrast between the bands and the continuum for the smaller particles, {\it i.e.}, the bands are stronger but  the underlying continuum considerably weaker for low band gap materials. This effect arises from two independent effects, namely: 
\begin{enumerate}
  \item increased surface hydrogenation as size decreases (see Sect.~\ref{sect_surf_hydrogenation_effects} and Appendix~\ref{appendix_surface_H_effects}) and 
  \item a pronounced weakening of the underlying continuum for $E_{\rm g} \gtrsim 1.5$\,eV and wavelengths $\gtrsim 0.5\,\mu$m as the maximum-allowable, particle-radius-determined aromatic domain size decreases (see Sect.~\ref{size-dep_props} and Appendix~\ref{appendix_pi_pi}).
\end{enumerate}
In essence, what this resembles is a transition in behaviour from bulk solid to molecule-like. For the smallest considered particles, containing a few tens of carbon atoms, only the optical-UV bands are seen at short wavelengths and the chemical structural bands at IR wavelengths ($\lambda > 3\,\mu$m). 

Note, in particular, that for particles with radii $\gtrsim 10$\,nm ({\it i.e.}, Figs.~\ref{fig_nk_100nm_eg} and \ref{fig_nk_30nm} to \ref{fig_nk_10nm}) that $n$ and $k$ are rather invariant but that they vary significantly for smaller particles, where the most obvious effect is clearly a dramatic decrease in the continuum at wavelengths longwards of $\approx 0.5\, \mu$m.

\subsection{Comparison of nm-sized a-C(:H) particles with PAHs}
\label{sect_cf_aCH_PAH}

Given that a-C(:H) nano-particles and PAHs are viable and complementary alternatives for the carriers of some of the interstellar dust observables it is useful to compare their derived optical properties.  


\begin{figure} 
 \resizebox{\hsize}{!}{\includegraphics{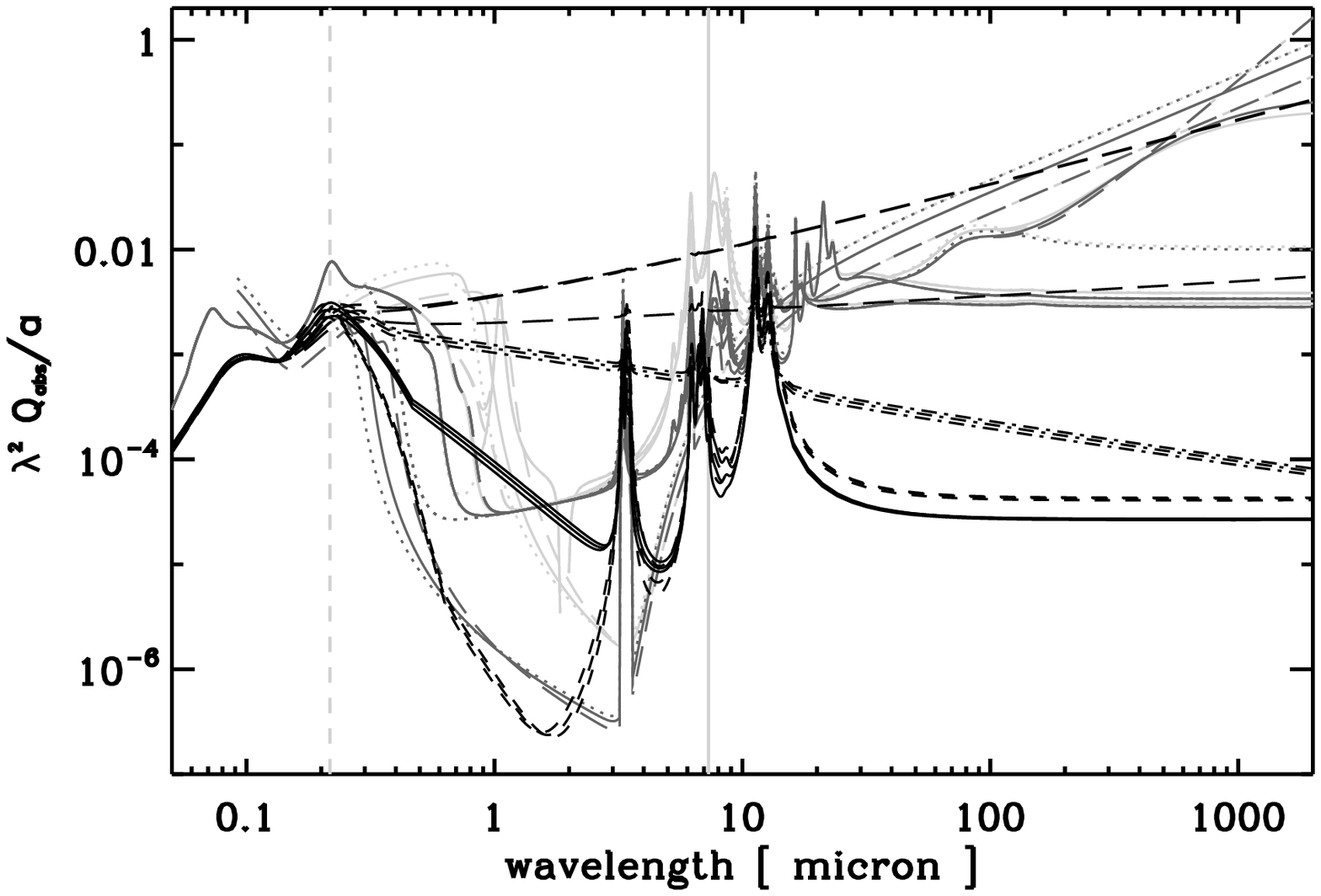}}
 \caption{ $\lambda^2$Q$_{\rm abs}$/a for 0.33\,nm (short-dashed), 0.5\,nm (solid), 1\,nm (dashed-dotted) and 3\,nm (long-dashed) radius particles: for a-C(:H)  with $E_{\rm g}{\rm (bulk)} = 0.5$, 0.25 and 0.1\,eV (black lines, from lower to upper at $100\,\mu$m) and neutral and cation PAHs  with the same number of C atoms \citep[dark and light grey solid lines, respectively,][]{1990A&A...237..215D,2001ApJ...551..807D,2007ApJ...657..810D,2011A&A...525A.103C}. 
The dashed, vertical, grey line shows the peak position of the UV bump and the solid, vertical, grey line shows the upper wavelength limit for the well-constrained a-C(:H) IR features.  }
 \label{fig_cf_aCH_PAH}
\end{figure}

Fig.~\ref{fig_cf_aCH_PAH} shows the $\lambda^2$Q$_{\rm abs}$/a values for hydrogen-poor, ``end point''  evolution, a-C(:H) nano-particles with radii of 0.33, 0.5, 1 and 3\,nm ({\it i.e.}, $X_{\rm H} = 0.02-0.11 \equiv E_{\rm g}{\rm (bulk)} = 0.5$, 0.25 and 0.1\,eV) and compares these data with with that for the equivalent PAHs, {\it i.e.}, for the same number of carbon atoms per particle, for four different PAH model datasets \citep{1990A&A...237..215D,2001ApJ...551..807D,2007ApJ...657..810D,2011A&A...525A.103C}. In this plot the $\lambda^2$ long-wavelength wings of the IR band Drude profiles appear flat, as per a-C(:H) and the  \cite{2001ApJ...551..807D,2007ApJ...657..810D} models, whereas the \cite{1990A&A...237..215D} and \cite{2011A&A...525A.103C} PAH data absorption coefficients increase in strength with increasing wavelength in this plot. The \cite{1990A&A...237..215D} data are those that most closely match the 0.33\,nm radius a-C(:H) particle data (short-dashed lines) in the $1-3\,\mu$m MIR minimum region. In general, the comparison of the data in Fig.~\ref{fig_cf_aCH_PAH} indicates:
\begin{itemize}
  \item a generic agreement for $\lambda \lesssim 10\,\mu$m, especially for the 0.5\,nm radius particles, 
  \item that a-C(:H) materials exhibit about a factor of $\sim 2$ weaker UV-EUV absorption than the \cite{2011A&A...525A.103C} PAHs 
  but comparable to that of the other PAH models,  
  \item similar  IR band intensities in the $3\,\mu$m region, 
  \item that large [small] a-C(:H) particles generally seem to have stronger [weaker] $1-3\,\mu$m MIR absorption than the PAH model data,  
  \item that nm and sub-nm sized a-C(:H) particles have weaker FIR-mm absorption than that predicted for the interstellar PAHs, 
  \item that $1-3$\,nm a-C(:H) particles will absorb more energy than any of the modelled PAHs at MIR wavelengths ($\simeq 1-6\,\mu$m), and 
  \item that the predicted long-wavelength ($\lambda \gtrsim 30\,\mu$m) behaviour for all of the PAH models is rather well bracketed by that of the 3\,nm a-C(:H) particles. 
\end{itemize} 
As shown in Fig.~14 in paper~II, the predicted UV cross-section per C atom for a-C(:H) materials agree well with that for the {\em bulk materials} graphite, diamond and other hydrocarbon solids. However, the $\pi-\pi^\ast$ and $\sigma-\sigma^\ast$ UV features in a-C(:H) materials are broader but only about half as intense as those of graphite and diamond, as expected for disordered/amorphous materials. In the UV region, the \cite{2011A&A...525A.103C} PAH absorption is also stronger, by about a factor of two, than that for a-C(:H) particles with the same number of C atoms (Fig.~\ref{fig_cf_aCH_PAH}). The determination of the FIR-cm optical properties of PAH and a-C(:H) {\em nano-particles} is not without difficulties and should probably still be considered as something of an open issue. Hence, it is not surprising that, at these long wavelengths, there are  differences in the predicted properties of a-C(:H) nano-particles and PAHs and also between the different PAH models. 

In general, the predicted optical properties of sub-nm, interstellar PAHs at FIR-cm wavelengths are implicitly influenced by those of {\em bulk} graphite \citep[{\it e.g.},][]{1990A&A...237..215D,2001ApJ...551..807D,2007ApJ...657..810D,2011A&A...525A.103C}. As shown in paper~II (Sect.~4.1.2, Eq. 22), the optical properties of a-C(:H) materials are completely determined by the band gap, $E_{\rm g}$, with the long-wavelength absorptivity increasing as $E_{\rm g}$ decreases. Further, and as shown here, the FIR-cm optical properties of H-poor/aromatic-rich nano-particles also depend on the effective, or size-dependent, band gap, $E_{\rm g}$(eff) (see Sect.~\ref{size-dep_props} and Appendix~\ref{appendix_visUV_el_props}, Fig~\ref{fig_Eg_eff_vs_Eg}). The long-wavelength absorption of a-C(:H) particles is strongly size-limited, for $a \lesssim 10$\,nm, because $E_{\rm g}$(eff) is determined by the largest aromatic clusters, which cannot be larger than the particle. In Fig.~\ref{fig_cf_aCH_PAH} it is evident that the predicted FIR-cm behaviour of the interstellar PAH models compares well with that of the 3\,nm a-C(:H) particles with $E_{\rm g}{\rm (bulk)} = 0.1-0.5$\,eV (long-dashed lines\footnote{Note that the data for the $E_{\rm g}{\rm (bulk)} = 0.1$ and 0.25\,eV particles almost completely overlap (upper long-dashed lines).}) containing $\simeq 10^4$ C atoms (see the last itemised point above). The optEC$_{\rm (s)}(a)$ model predicts that smaller aromatic-rich particles  have weaker FIR-cm absorption because of particle-size-limitations on the aromatic domain dimensions (see above).  Thus, and based on these assumptions, it is not clear how interstellar PAHs can be such strong absorbers, at FIR-cm wavelengths, unless they contain $\gtrsim 10^4$ C atoms and have $a_{\rm R} \gtrsim 0.09\surd n_{\rm C}$\,nm, {\it i.e.}, radii $\gtrsim 9$\,nm.
The extrapolation of bulk graphite data, in the derivation of the interstellar PAH properties at FIR-cm wavelengths, therefore seemingly requires molecules that are larger than those assumed in current interstellar PAH models. However, current PAH models do include long wavelength bands, such as those that have been investigated experimentally \citep[{\it e.g.},][]{2006PCCP....8.3707P} and theoretically \cite[{\it e.g.},][]{2006A&A...460...93M,2006PCCP....8.3707P} in order to determine their FIR spectral properties and to explain the anomalous microwave emission \citep{2010A&A...509A..12Y}. However, the  band widths for $2-4$ aromatic ring PAHs are rather narrow, {\it i.e.}, FWHM $\lesssim 30$\,cm$^{-1}$, at FIR wavelengths \citep[in the $10-700 $\,cm$^{-1}$ region, as experimentally-measured by][]{2006PCCP....8.3707P}. Such bands have not been incorporated into the  optEC$_{\rm (s)}(a)$ model because they do not appear to be required to explain the currently-available {\em bulk material} laboratory data but that is not to say that they may not be observable when appropriate data on a-C(:H) nano-particles  become available. In any event, it is likely that the  analogous (pseudo-phonon) bands in a-C(:H) particles would be broader due to the inherent disorder in the structure.  

The optEC$_{\rm (s)}(a)$ model would appear to provide a means of calculating the optical properties of nm- and sub-nm-sized hydrocarbon particles until such time as appropriate laboratory data become available. However, the predicted FIR-cm optical properties of nm and sub-nm a-C(:H) particles are inconsistent with those predicted by current ``interstellar PAH'' models.

\subsection{The optEC$_{\rm (s)}(a)$ model data validity range and limitations}

For the considered particles the equivalent number of carbon atoms per particle (of radius $a$), for $X_{\rm H} = 0.5$, is $\approx 10^8$ ($100$\,nm), $\approx 10^7$ ($30$\,nm), $\approx 10^5$ ($10$\,nm), $\approx 10^4$ ($3$\,nm), $\approx 300$ ($1$\,nm), $\approx 60$ ($0.5$\,nm) and $\approx 30$ ($0.33$\,nm). 
It seems reasonable to assume that the complex refractive index data derived here are valid down to the nano-particle size domain ({\it i.e.}, for $a \geqslant 1$\,nm), where the particles contain at least several hundreds of carbon atoms. The calculated optical properties, {\it e.g.}, $n$ and $k$, for particle sizes in the ``molecular domain'' ($a \lesssim 1$\,nm) are probably less certain because the material response will start to resolve into more distinct, ``molecular structure''-dependent bands, particularly at short wavelengths. Nevertheless, and given the inherent experimental difficulties in isolating and studying such small particles, the data for small particles ($a = 0.5$ and 0.33\,nm) are provided here as something of a ``stop-gap'' measure. Given the results of the a-C(:H) {\it vs.} PAH comparison (Sect.~\ref{sect_cf_aCH_PAH}), the data for sub-nm a-C(:H) particles, nevertheless, do not appear to be too unreasonable. 

As pointed out in paper~II, the adopted linear behaviour for the long-wavelength optical properties is empirical and no ``real'' dust analogue material will show such a characteristic and perfectly linear behaviour. 
Unfortunately, current data do not allow better constraints to be imposed on the modelling of low-temperature-formed, hydrogenated carbon particles. 
In particular, no contribution from long-wavelength bands ($\lambda \gtrsim 13\,\mu$m) is currently included in the model, {\it i.e.}. as seen in the DDOP a-C(:H) data\footnote{The Jena group's Databases of Dust Optical Properties: 
\\ http://www.astro.uni-jena.de/Laboratory/Database/databases.html.} 
and as pointed out in paper~II.

\subsection{Size-dependent optical properties and the UV bump}
\label{sect_optUV_spect}

Figs.~\ref{fig_Qs_vs_size_1_eg} and \ref{fig_Qs_vs_size_5_eg} show the efficiency factors for extinction, Q$_{\rm ext}$ (thick solid), scattering, Q$_{\rm sca}$ (dotted), and absorption, Q$_{\rm abs}$ (dashed), for 100\,nm and 1\,nm radius particles, as a function of inverse wavelength.\footnote{Appendix~\ref{appendix_vis_UV} shows the data for the 30, 10, 3, 0.5 and 0.33\,nm radii particles.}  Each coloured line represents one of the 14 different band gap materials ($E_{\rm g} = -0.1$ to 2.67\,eV: see Table~\ref{table_colour_code} for the full colour code).  These figures also show the ratio Q$_{\rm sca}$/Q$_{\rm ext}$ (thin solid lines), which indicates the fractional contribution of scattering to the extinction, {\it i.e.}, the albedo. Note, that for large ($a \gtrsim 100$\,nm), hydrogen-rich ($X_{\rm H} \geq 0.35$), a-C:H particles the extinction will be dominated by scattering in the $\sim 0.3-3\, \mu$m ($\sim 0.3-3\, \mu$m$^{-1}$) region (Fig.~\ref{fig_Qs_vs_size_1_eg}). For somewhat smaller particles ($a \simeq 30$\,nm) this also remains true but only for the hydrogen-richest materials ($X_{\rm H} \simeq 0.6$, see Fig.~\ref{fig_Qs_vs_size_2}) and over a narrower wavelength range $\sim 0.5-2\, \mu$m ($\sim 0.5-2\, \mu$m$^{-1}$). As expected, and for all a-C:H materials, the scattering contribution to the extinction, the albedo, decreases rapidly with decreasing particle size. 

For particles with radii smaller than 10\,nm (Figs.~\ref{fig_Qs_vs_size_4} to \ref{fig_Qs_vs_size_7}), and as the hydrogen content and the band gap decrease, a clear ``bump'' appears in the spectra in the region of the observed UV extinction bump at 217\,nm. The observed variation in the central position of the interstellar UV extinction bump at 217\,nm is indicated by the thick, vertical,  grey line and the range of observed full width at half maximum, FWHM, by the vertical, lighter grey bands on either side \citep[{\it e.g.},][]{2004ASPC..309...33F,2007ApJ...663..320F}. In particular, it should be noted that poorly-hydrogenated ({\it i.e.}, aromatic-rich) particles with radii $\simeq 1$\,nm (Fig.~\ref{fig_Qs_vs_size_5_eg}) would appear to give an excellent match to the UV bump position. However, it is clear that the predicted peak position of the bump does depend on particle size, occurring close to 4\,$\mu$m$^{-1}$ (5\,$\mu$m$^{-1}$) in 10\,nm (0.33\,nm)-radius particles and, in all cases, the a-C(:H) UV bump width is larger than the width of the observed feature in these Q$_{\rm ext}$ plots.

\subsubsection{Experimental constraints on the UV bump}
 
Experiments show that a-C(:H) materials in general produce broad UV absorption peaks, with widths and positions that depend on the nature of the material and its post-formation processing. It is clear that this broad feature has its origin in the aromatic component of these solids. This section outlines some important experimental constraints from studies of laboratory analogues of interstellar carbons that provide critical information on the nature of the UV bump-carrying particles in the ISM. 

\cite{1996ApJ...464L.191M}  have shown that UV and ion irradiation of HAC solids leads to an increase in the strength of the measured UV bump in these materials as a result of the increased aromatic carbon content in the structure. They attributed the bump at 215\,nm in their experiments to $\pi-\pi^\star$ electronic transitions in the $sp^2$ carbon atom clusters in the structure. However, it is also apparent that aromatic molecules can also produce a UV bump \cite[{\it e.g.},][]{1992ApJ...393L..79J}.  \cite{1998ApJ...507..874D,1999ApJ...522L.129D} proposed that the UV bump has its origin in dehydrogenated coronene and carbon nano-particles containing up to several hundred carbon atoms and 
\cite{2004ApJ...612L..33D} have been able to experimentally reproduce a UV bump feature in their experiments.  
In their thorough experimental investigation \cite{2005A&A...432..895D} find a UV bump in their a-C:H materials at 4.2\,$\mu$m$^{-1}$.

In a recent and comprehensive experimental study of UV-irradiated HAC materials \cite{2011A&A...528A..56G}, based on their earlier work \citep{2007Carbon.45.1542L}, confirm the suggestion by \cite{1996ApJ...464L.191M} that the UV extinction bump is consistent with UV-irradiated hydrocarbon nano-particles in the ISM. \cite{2011A&A...528A..56G} show that the ``distinguishable UV band'' at 217\,nm results from structural modifications, arising from UV irradiation, which alter the electronic density of states and, in particular, the $\pi-\pi^\star$ transitions, which they state  {\em `` \ldots probably results from $\pi-\pi^\star$ electronic transitions due to the existence of the aromatic structures motivated by the UV radiation within the HAC material''.} However, and as pointed out by \cite{2011A&A...528A..56G}, there does still appear to be the remaining problem that the required abundance of carbon in the carriers is too high and that the bump in the laboratory data is perhaps still rather too large. 

The above-cited experimental works serve to rather convincingly show that it is indeed the detailed nature of the aromatic clusters in hydrocarbon solids that is at the origin of observed UV extinction bump in the ISM. This appears to be entirely consistent with the adopted description of the inherent aromatic cluster distribution and the derived a-C(:H) optical properties resulting from the model (presented in full in Appendix~\ref{appendix_visUV_el_props}). 
However, it is also clear that current models and experiments on interstellar carbon dust analogues are not yet been able to reproduce the observed constancy of the peak position and the rather narrow range of widths of the interstellar UV bump . It seems that the current understanding of the microphysics of likely UV bump-carrying particles is therefore currently incomplete.

\subsubsection{Observational constraints on the UV bump}

In a detailed study of the IR-UV dust extinction morphology \cite{2004ASPC..309...33F} and \cite{2007ApJ...663..320F}  show that the position of the observed UV bump, at 217.85$\pm0.91$\,nm ($\simeq4.5903\pm0.0085\,\mu$m$^{-1}$, $1\sigma$ range), and its width are uncorrelated. Further, \cite{2007ApJ...663..320F} find that the UV bump FWHM varies significantly between sight-lines ($\simeq1.00\pm0.15\,\mu$m$^{-1}$) and seems to exhibit a bi-modal distribution.  \cite{2004ASPC..309...33F} and \cite{2007ApJ...663..320F} also show that the bump is strongest for curves with intermediate levels of FUV extinction and weaker for lower or higher values than this ``mean'' value of the FUV extinction. Sect.~\ref{sect_ext_abs_em} gives a more detailed comparison with other observational aspects, including the UV bump. 

%
\begin{figure}
 \resizebox{\hsize}{!}{\includegraphics{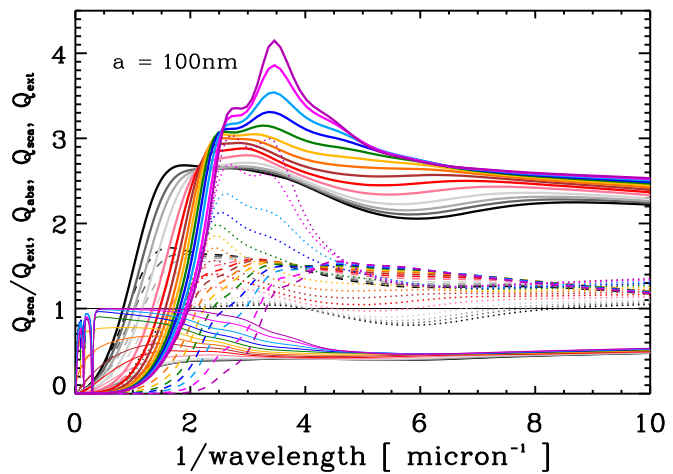}}
 \caption{Q$_{\rm ext}$ (thick solid lines), Q$_{\rm sca}$ (dotted lines), Q$_{\rm abs}$ (dashed lines) and Q$_{\rm sca}$/Q$_{\rm ext}$  ({\it i.e.}, the albedo,  thin solid lines),  with $E_{\rm g}$ decreasing from top to bottom at $\lambda^{-1} = 6\,\mu$m$^{-1}$,   as a function of  inverse wavelength for all the considered band gap materials (see Table~\ref{table_colour_code}) for 100\,nm radius particles. The thin horizontal black line shows the limit where Q$_{\rm sca}$/Q$_{\rm ext} = 1$.}
\label{fig_Qs_vs_size_1_eg}
\end{figure}
%
\begin{figure} 
 \resizebox{\hsize}{!}{\includegraphics{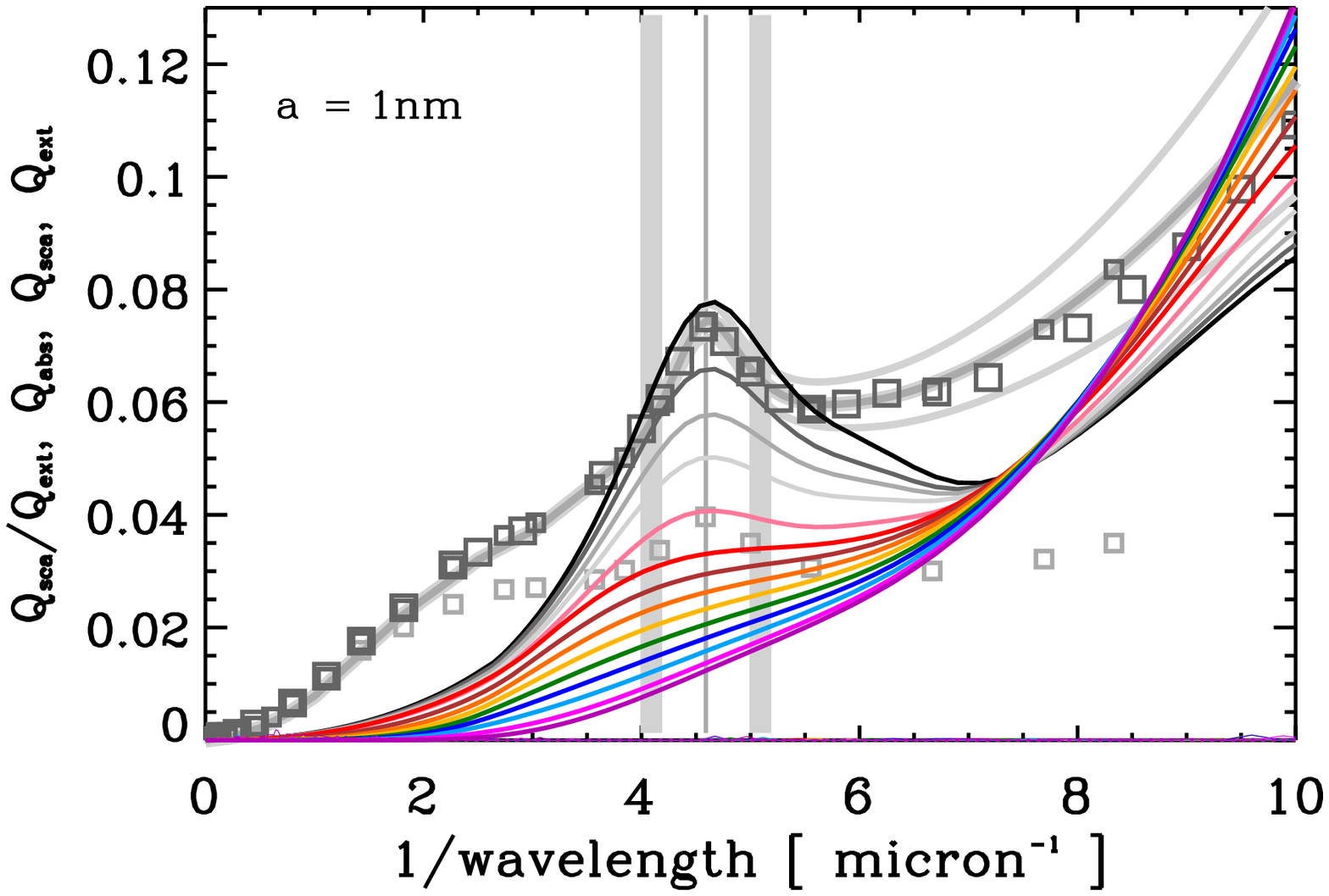}}
 \caption{Same as for Fig.~\ref{fig_Qs_vs_size_1_eg} but for 1\,nm radius particles,  with $E_{\rm g}$ increasing from top to bottom at the peak of the UV bump (at $4.6\,\mu$m$^{-1}$).   The vertical grey line shows the central position of the UV bump at 217\,nm and the vertical, lighter grey bands on either side indicate the full width of the observed UV bump \citep{2004ASPC..309...33F,2007ApJ...663..320F}. The grey squares show the diffuse ISM extinction curve for $R_{\rm V} = 3.1$ (dark) and 5.1 (light) from \cite[][large squares]{1979ARA&A..17...73S} and \citep[][small squares]{1990ARA&A..28...37M}. The grey curves indicate the average galactic extinction, and its variation (upper and lower grey curves), as derived by \cite{2007ApJ...663..320F}. }
\label{fig_Qs_vs_size_5_eg}
\end{figure}

\subsection{The optEC$_{\rm (s)}$(a) particle $2-14\,\mu$m spectra}
\label{sect_2to14_IR_spect}

%
\begin{figure} 
 \resizebox{\hsize}{!}{\includegraphics{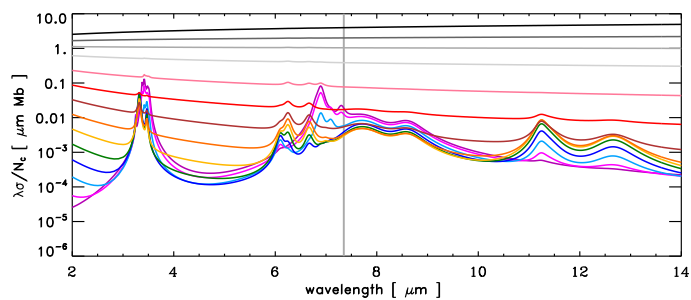}}
 \caption{The predicted spectra of optEC$_{\rm(s)}$(a)-modelled particles of radius 100\,nm, presented as the wavelength times absorption coefficient per carbon atom, $\lambda \alpha / N_{\rm C}$, in the $2-14\,\mu$m region (1\,Mb$= 10^{-18}$\,cm$^{2}$).  Note how the spectra evolve from aliphatic-rich with prominent IR bands to aromatic-rich with no evident IR bands but a much stronger continuum.  N.B. The bands with central positions long-ward of the vertical grey line ($\lambda(\nu_0) > 7.3\,\mu$m)  are not yet well-determined by laboratory measurements.}
 \label{fig_alpha_spectrum_100nm_eg}
\end{figure}
%
\begin{figure} 
 \resizebox{\hsize}{!}{\includegraphics{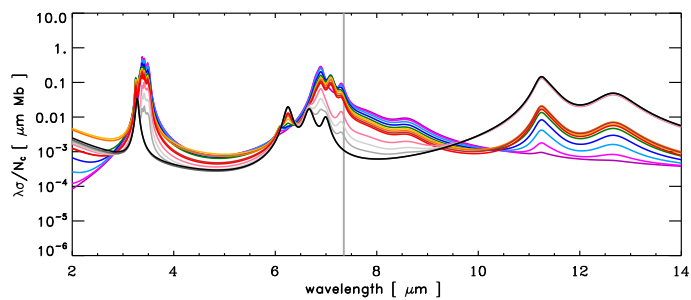}}
 \caption{As per Fig.~\ref{fig_alpha_spectrum_100nm_eg} but for particles of radius 0.5\,nm. Note how the spectra bunch up and how the IR band positions evolve from aliphatic-rich (with maxima in the $\approx 3.4$ and $\approx 7\,\mu$m regions) to aromatic-rich (with maxima at 3.3, 6.3, 6.7, 7.0, 11.3 and $12.7\,\mu$m).  }
 \label{fig_alpha_spectrum_0.5nm_eg}
\end{figure}

The  eRCN-DG model-derived IR, optEC$_{\rm (s)}$(a) spectra of 100 and 0.5\,nm radius particles in the $2-14\,\mu$m region are shown in Figs.~\ref{fig_alpha_spectrum_100nm_eg} and \ref{fig_alpha_spectrum_0.5nm_eg}.\footnote{Appendix~\ref{appendix_size_dep_spectra} shows the complementary spectra  for 30, 10, 3, 1 and 0.33\,nm radius particles.} As expected, the derived spectra are practically particle size-independent for radii  $\gtrsim 10$\,nm and identical to the bulk materials spectra presented in paper~I. The only major difference is an increase in the aromatic CH stretch at $3.28\,\mu$m, which is due to the addition of surface hydrogenation to the DG-modelled particles, where no aromatic CH component was present in the bulk material model (see paper~I, Sect.~2.3).  

For particles smaller than 10\,nm, and with decreasing size, the following systematic changes in the spectra can be seen:
\begin{itemize} 
  \item An increase in the aromatic CH stretch at $3.28\,\mu$m, due to surface hydrogenation in H-poor particles. 
  \item A weakening of the olefinic CH mode at 3.32\,$\mu$m due to  surface hydrogenation to CH$_2$ with weaker IR modes.
  \item A change in the aliphatic CH$_n$ bands   due to hydrogenation to CH$_{(n+1)}$, where $(n+1) \leqslant3$. 
  \item 3\,$\mu$m region spectra,   for small H-poor particles (grey lines), that resemble the AIB spectra, {\it viz.}, a strong $3.28\,\mu$m band with an adjacent $3.45-3.55\,\mu$m plateau with superimposed peaks (see Sect.~\ref{sect_3mic_spectra} for a more detailed discussion). 
\end{itemize}

There is some experimental evidence to support this general scenario in that \cite{2006ApJ...653L.157H} have shown that carbon nano-particles, containing $10^2 - 10^3$ carbon atoms (radii $\simeq 0.8-1.5$\,nm, {\it e.g.}, see Fig.~\ref{fig_NCatom_vs_a}) can produce IR bands at 6.2, 7.7, 7.8 and 8.6\,$\mu$m. In a follow-up study they also found that the 3.29\,$\mu$m emission band is consistent with polycyclic aromatics with peripheral aliphatic side-chains \citep{2008ApJ...677L.153H}, generically consistent with the last item above. 

Note that the hydrocarbon grain deconstruction and spectral prediction tool presented here, and in papers~I and II, could be used to fit observational data, in a physically meaningful way, {\em but only if all of the positions,  strengths and widths of the CC and CH modes in a-C:H and, perhaps most importantly, the correct band assignments can be made.}
Alternatively, the observational data could be de-constructed into component bands and those then fitted to a `sequential' and evolutionary range of spectra to reveal the nature and sizes of the responsible carriers and their evolution from region to region. However, a note of caution is needed here as not all the observed bands necessarily originate in the same carriers, and thus careful band assignments are essential in this case if data over-interpretation is to be avoided. 

\begin{figure} 
 \resizebox{\hsize}{!}{\includegraphics{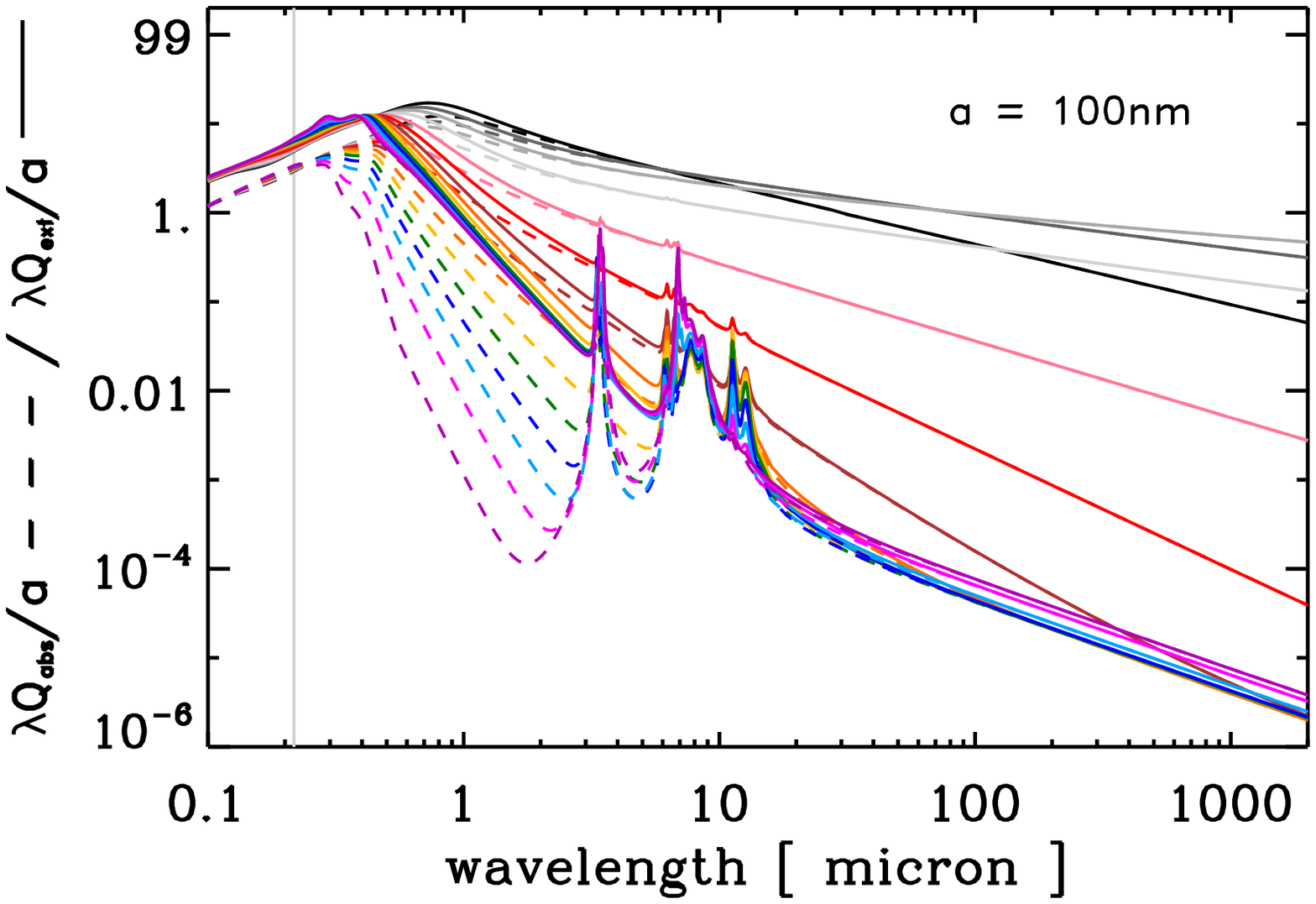}}
 \resizebox{\hsize}{!}{\includegraphics{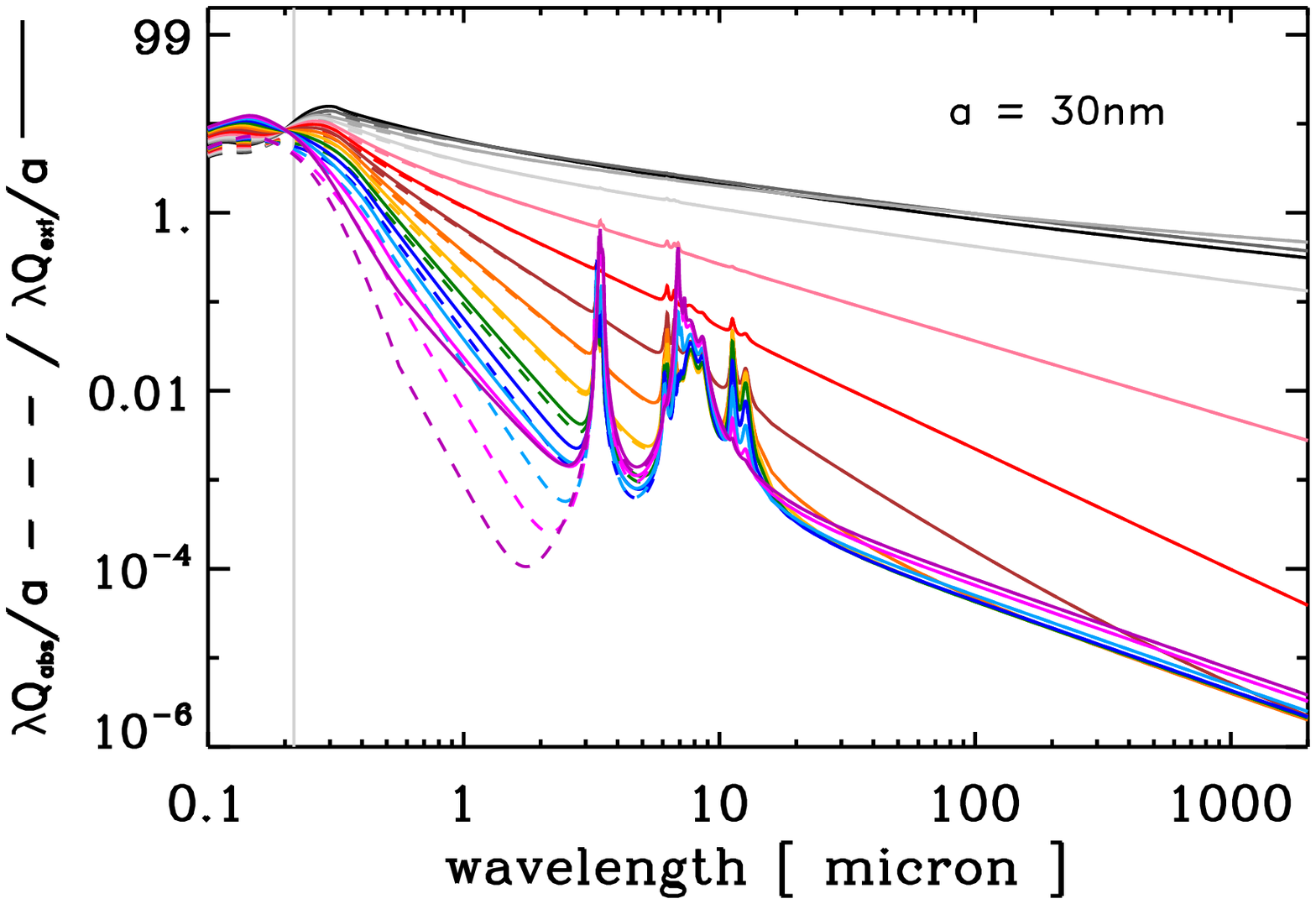}}
 \resizebox{\hsize}{!}{\includegraphics{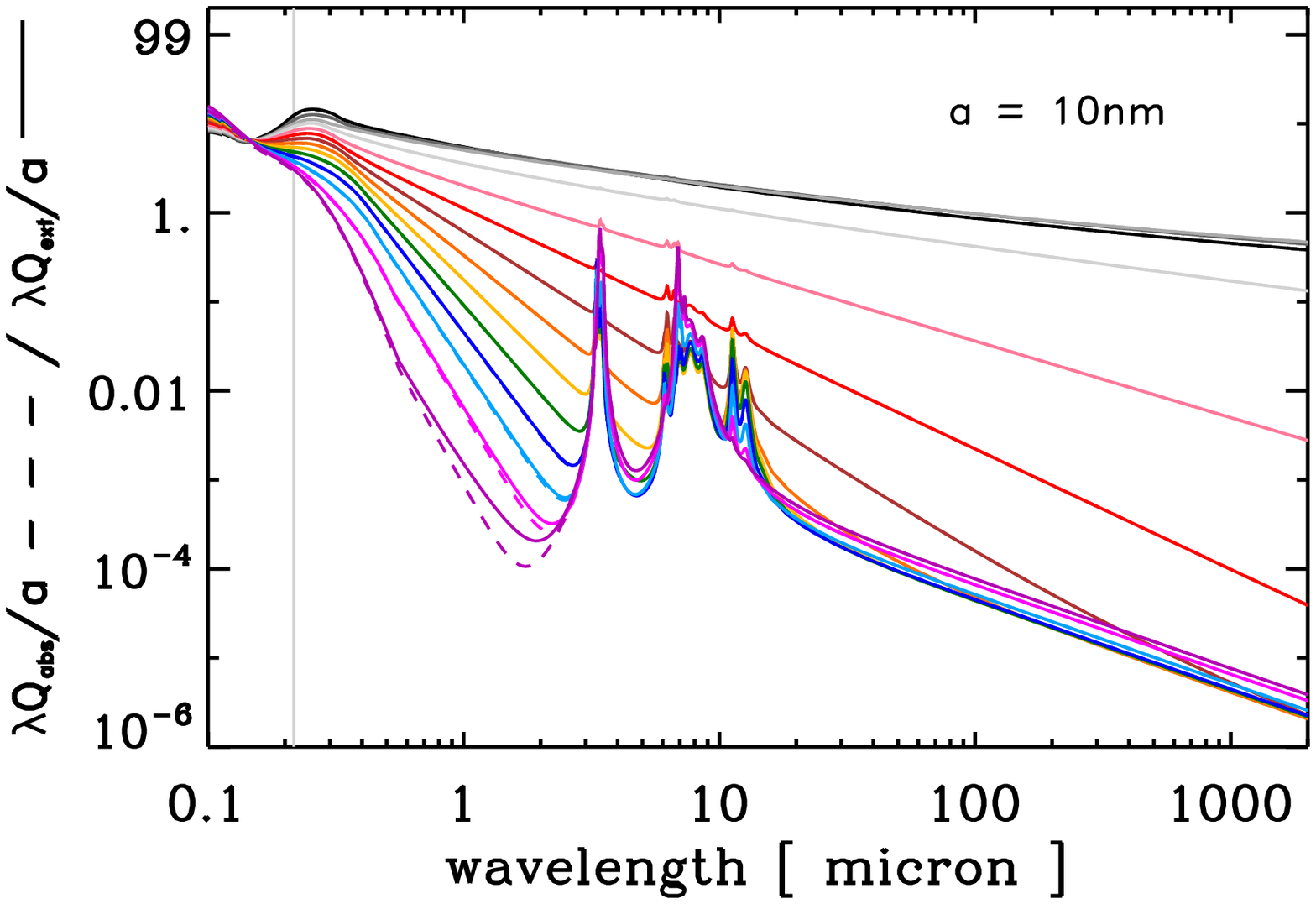}}
 \caption{$\lambda$Q$_{\rm ext}/a$ (solid) and $\lambda$Q$_{\rm abs}/a$ (dashed),  for  100, 30 and 10\,nm radius particles, as a function of wavelength, for all of the considered band gap materials (see Table~1). 
 At $\lambda = 2\,\mu$m,  $E_{\rm g}$ decreases from 2.67\,eV (lower purple line) to $-0.1$ to $0.25$\,eV (the four upper grey and black lines).  The vertical grey line marks the position of the UV extinction bump at 217\,nm.}
\label{fig_app_Qs_vs_size_3}
\end{figure}
%
\begin{figure} 
 \resizebox{\hsize}{!}{\includegraphics{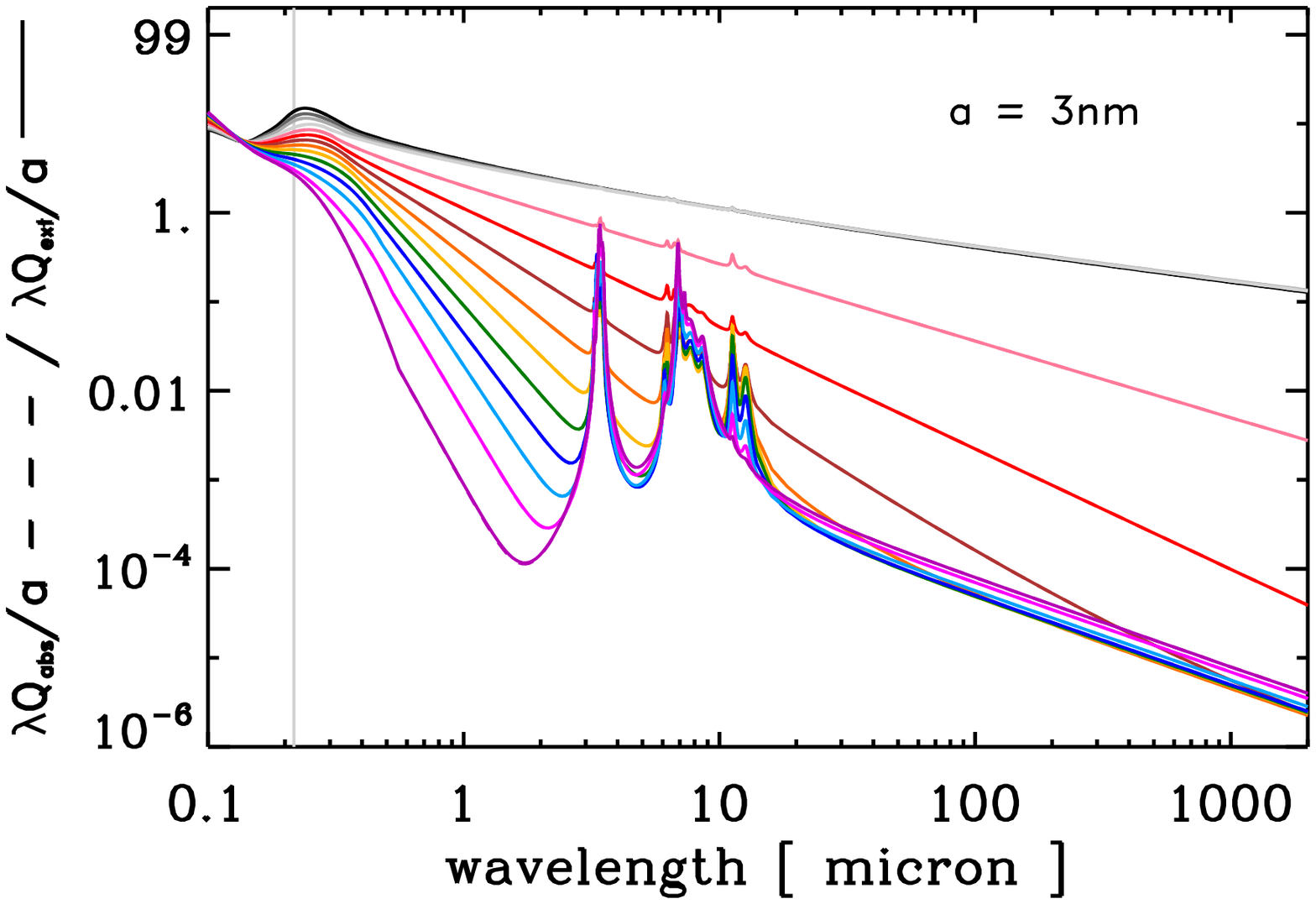}}
 \caption{Same as for Fig.~\ref{fig_app_Qs_vs_size_3} but for 3\,nm radius particles. Note that the data for $E_{\rm g} \leq 0.25$\,eV overlap for $\lambda > 1\,\mu$m. }
\label{fig_app_Qs_vs_size_4}
\end{figure}
%
\begin{figure} 
 \resizebox{\hsize}{!}{\includegraphics{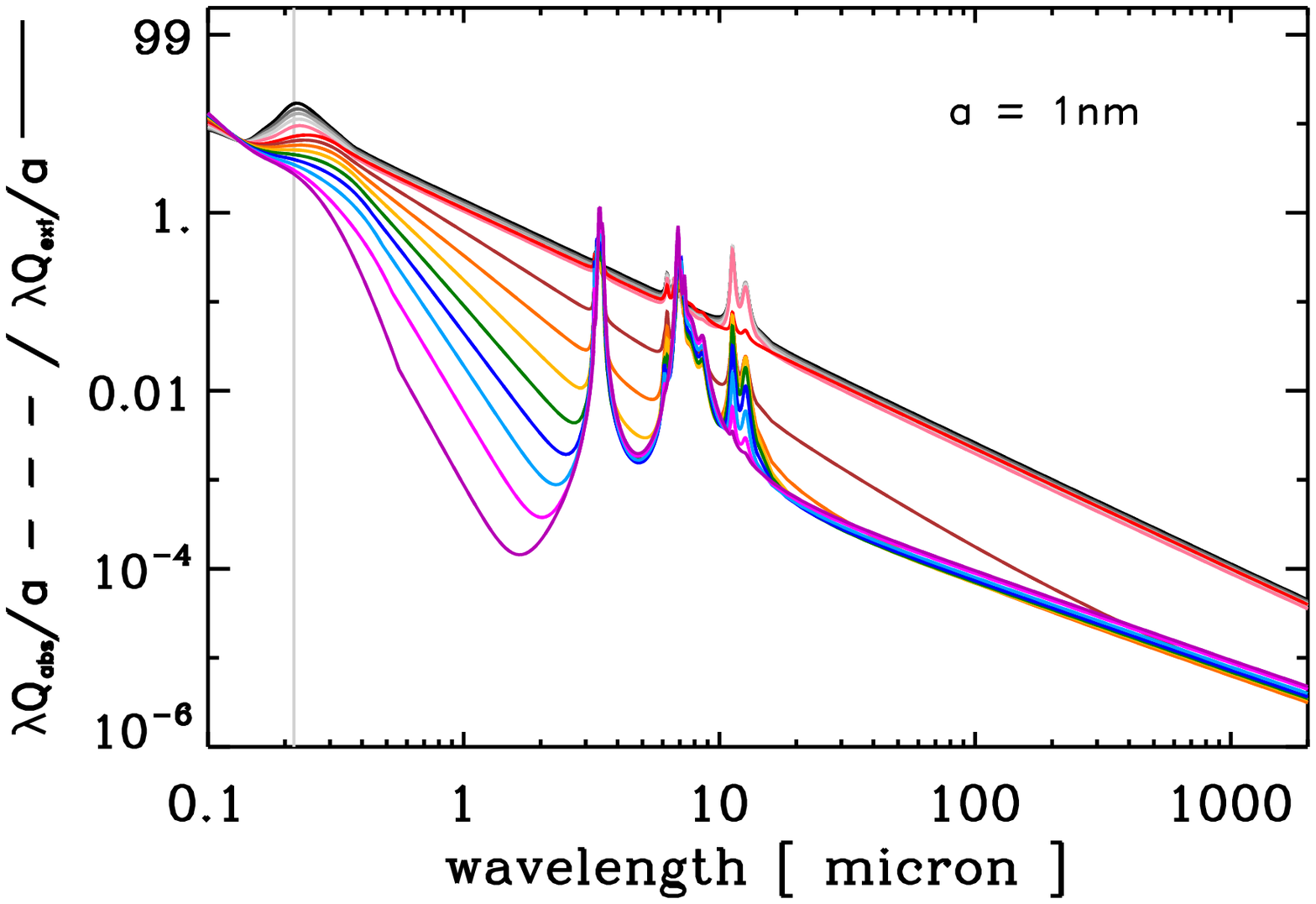}}
 \caption{Same as for Fig.~\ref{fig_app_Qs_vs_size_3} but for 1\,nm radius particles. Note that the data for $E_{\rm g} \leq 0.75$\,eV almost completely overlap for $\lambda > 1\,\mu$m. }
\label{fig_app_Qs_vs_size_5}
\end{figure}
%
\begin{figure} 
 \resizebox{\hsize}{!}{\includegraphics{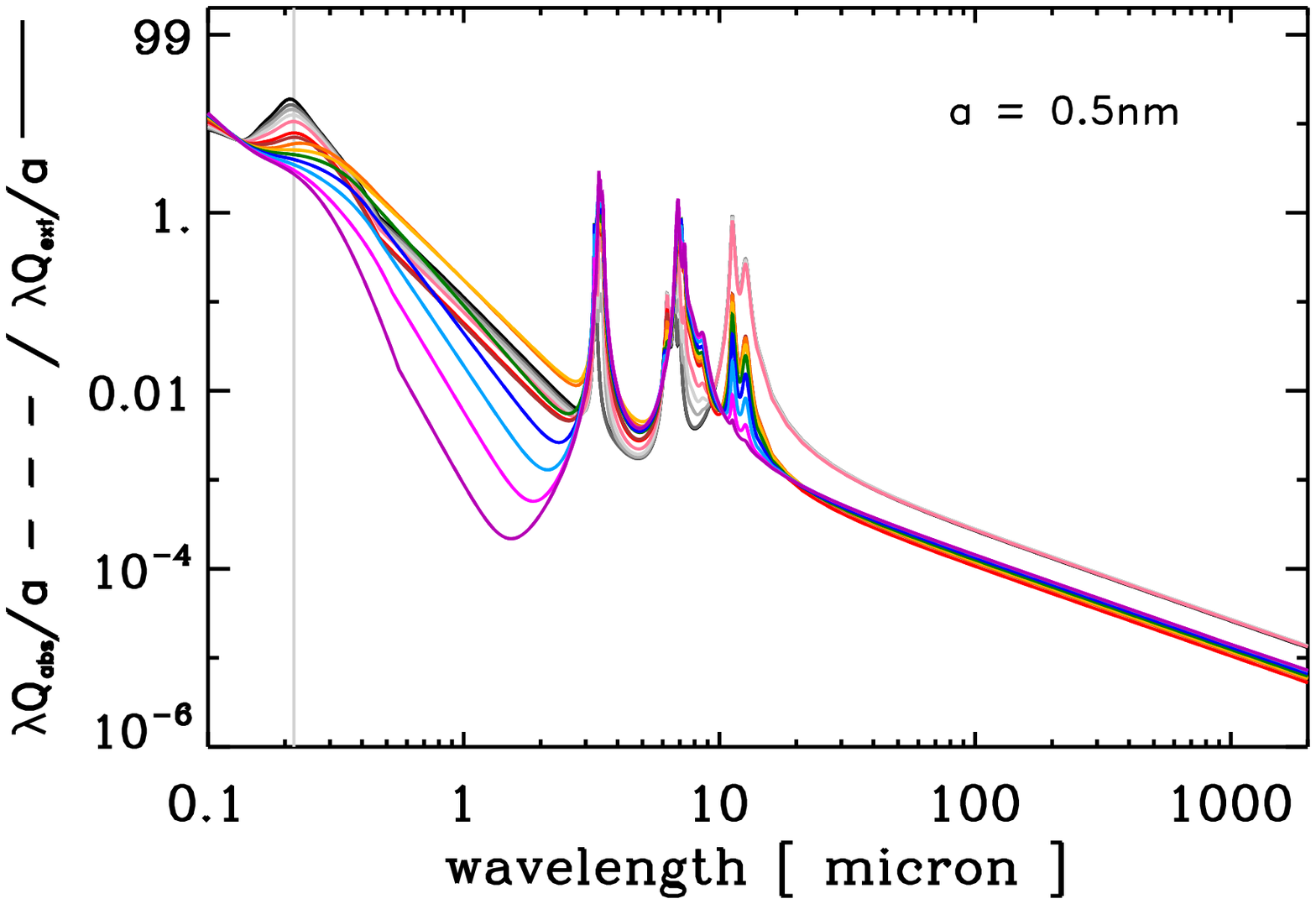}}
  \caption{Same as for Fig.~\ref{fig_app_Qs_vs_size_3} but for 0.5\,nm radius particles. Note the bunching up of all of the data, except for the H-rich materials ($E_{\rm g} > 2$\,eV) in the $0.5-3\,\mu$m region and at long wavelengths. }
\label{fig_app_Qs_vs_size_6}
\end{figure}
%

\subsection{Size-dependent spectra from EUV to mm wavelengths}


The previous sub-sections (Sect.~\ref{sect_optUV_spect} and Sect.~\ref{sect_2to14_IR_spect})  presented the optical and UV extinction properties for the optEC$_{(s)}$(a) data. Here the same data is shown over a larger wavelength range in order to explore the IR to mm wavelength behaviour. Figs.~\ref{fig_app_Qs_vs_size_3} to \ref{fig_app_Qs_vs_size_6} show the efficiency factors for extinction (solid) and absorption (dashed) for particles with radii of 100, 30, 10, 3, 1 and 0.5\,nm, respectively, as a function of wavelength and plotted as $\lambda Q_i/a$ for clarity ($i=$ ext or abs). Each coloured line represents one of the 14 different band gap materials ($E_{\rm g} = -0.1$ to 2.67\,eV, see Table~\ref{table_colour_code} for the full colour code). The absorption properties ($Q_{\rm abs}$) of particles with radii $\geqslant 10$\,nm are essentially identical for $\lambda \gtrsim 0.5 \,\mu$m, as are their extinction properties ($Q_{\rm ext}$) at long wavelengths ($\lambda \gtrsim 50 \,\mu$m), except for the 100\,nm particles with the smallest band gap ($E_{\rm g} = -0.1$). However, at shorter wavelengths ($\lambda \simeq 0.5-3 \,\mu$m), and as expected, scattering becomes more important as size increases. For 100\,nm radius particles some scattering is still apparent at wavelengths as long as $50\,\mu$m, particularly for  large band gap materials ($E_{\rm g} \gtrsim 2$\,eV). The expected size effects on the optical properties in the visible-UV region are also apparent, {\it i.e.}, the peak in $Q_{ext}$ shifts to shorter wavelengths with decreasing particle radius. 

At first glance the IR-mm spectra show a systematic evolutionary sequence from high to low $E_{\rm g}$ and from large to small sizes, with IR bands superimposed on continua that strengthen as the band gap decreases.  In detail, the principal effects seen in these data are: weakly-varying properties for particles with $a \geqslant 10$\,nm and a decreasing continuum contribution with decreasing particle size for $a < 10$\,nm. The latter effect is akin to a transformation of optical properties from bulk-like (bands and important continua) to molecule-like (strong bands and weak continua). In the latter case, using a bulk-matter approach to derive the optical properties of species that contain of the order of only tens of carbon atoms, would appear to be less than ideal for such small sizes ({\it e.g.}, see the detailed comparison in Sect.~\ref{sect_cf_aCH_PAH}).

\section{Astrophysical implications}
\label{sect_ast_impl}

In addition to the astrophysical implications already discussed in papers~I and II, the effects of size on the optical and spectral properties have important and further consequences for the nature and evolution of the a-C(:H) dust observables in the ISM. This section re-visits these implications within the context of the size-dependent properties of a-C(:H) materials.

\subsection{Structural variations and spectral properties}
\label{sect_3mic_spectra}

\begin{figure}
 \resizebox{\hsize}{!}{\includegraphics{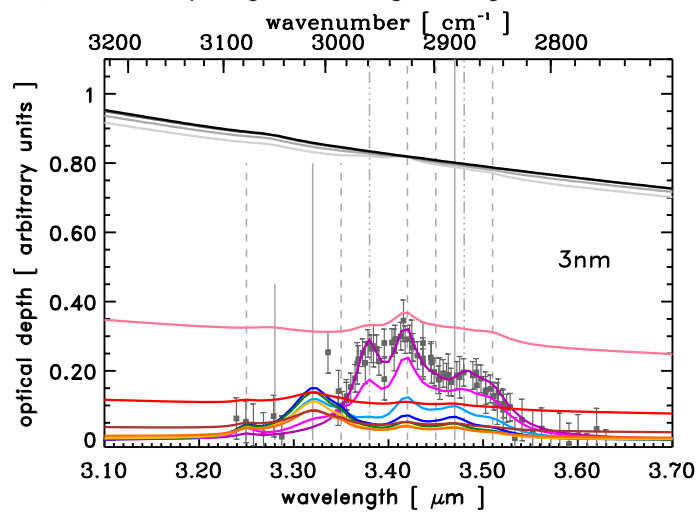}}
 \caption{The 3\,$\mu$m region spectrum, presented as optical depth, for $a = 3$\,nm a-C(:H) particles for increasing band gap,  $E_{\rm g} = -0.1$ to 0.25\,eV (upper, grey-black), $E_{\rm g} = 0.5$\,eV (middle, pink), $E_{\rm g} = 0.75$\,eV (lower, red, y-axis intercept at $\approx 0.12$) and $E_{\rm g} = 1$ to 2.67\,eV (lower, yellow-purple, bottom to top at $\approx 3.45\,\mu$m).  The thin, grey, vertical lines indicate the band origins (see paper I): aromatic (short), olefinic (medium) and aliphatic (long), CH (solid), CH$_2$ (dashed) and CH$_3$ (dashed-triple dotted).  The data with error bars (in grey) are for the diffuse ISM line of sight towards the Galactic Centre source \object{IRS6E} and \object{Cyg OB2 No. 12} \citep[taken from][]{2002ApJS..138...75P}. For comparison the observational data are scaled to the $E_{\rm g} = 2.67$\,eV data.}
 \label{fig_spectra_3nm}
\end{figure}

\begin{figure}
 \resizebox{\hsize}{!}{\includegraphics{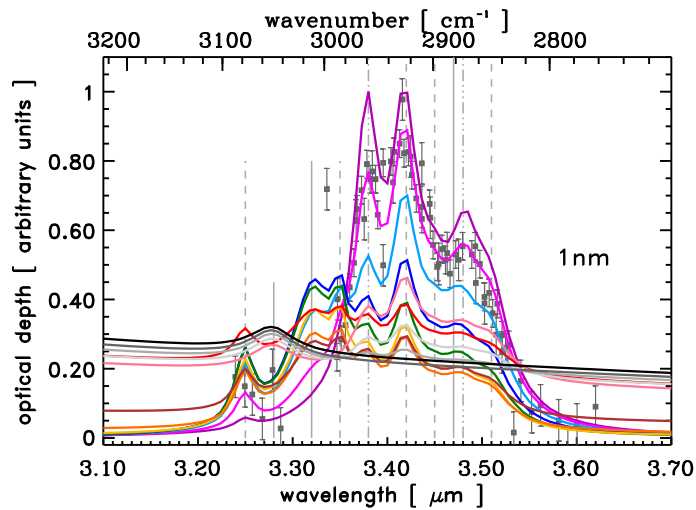}}
 \caption{Same as Fig.~\ref{fig_spectra_3nm} but for $a = 1$\,nm and  for increasing band gap, $E_{\rm g}$ ($-0.1$ to 2.67\,eV), from bottom to top at $\approx 3.45\,\mu$m. For comparison the observational data are scaled to the $E_{\rm g} = 2.5$\,eV data. }
 \label{fig_spectra_1nm}
\end{figure}

\begin{figure}
 \resizebox{\hsize}{!}{\includegraphics{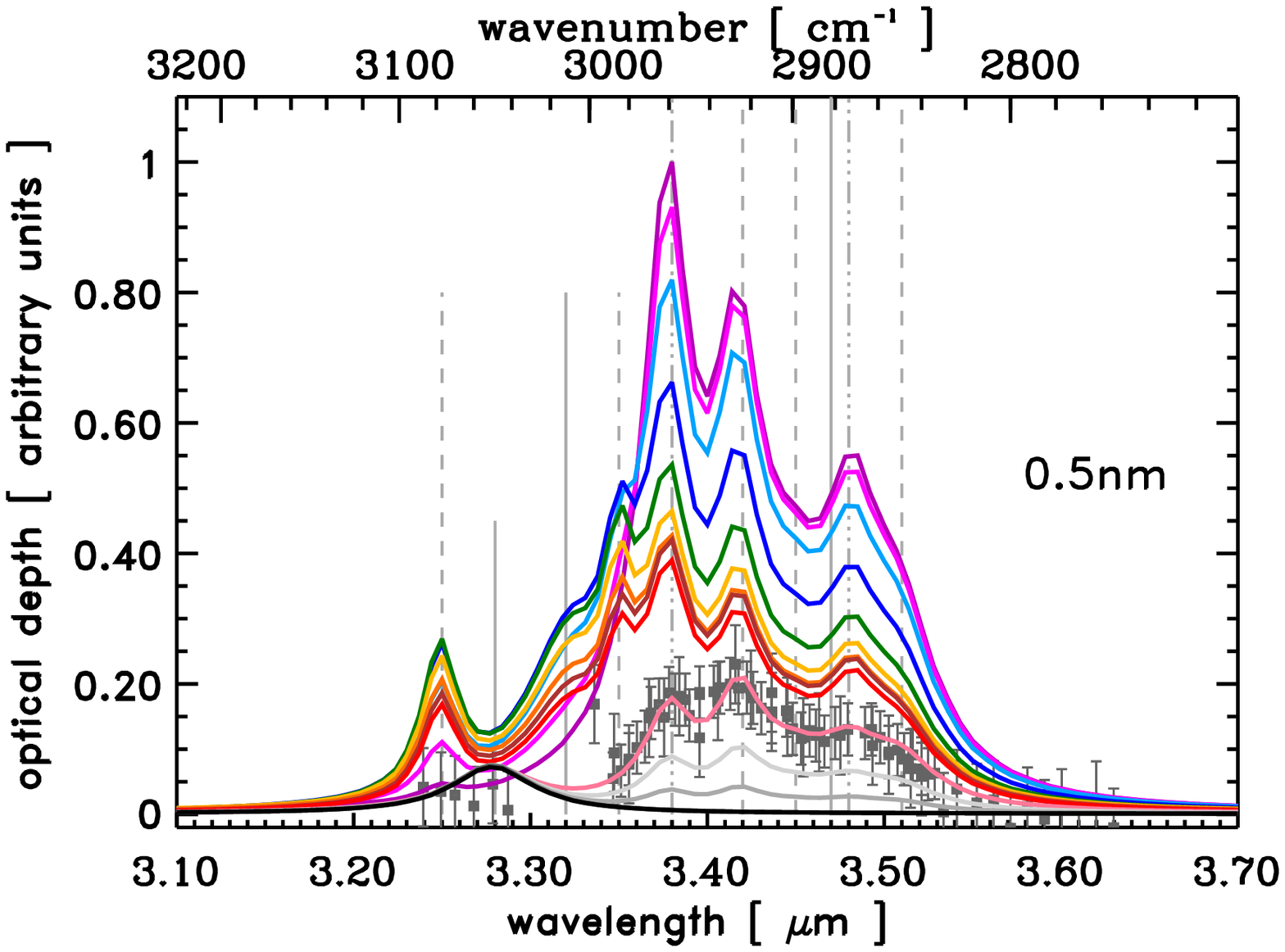}}
 \caption{Same as Fig.~\ref{fig_spectra_3nm} but for $a = 0.5$\,nm.  For comparison the observational data are scaled to the $E_{\rm g} = 0.5$\,eV data. }
 \label{fig_spectra_0.5nm}
\end{figure}

\begin{figure}
 \resizebox{\hsize}{!}{\includegraphics{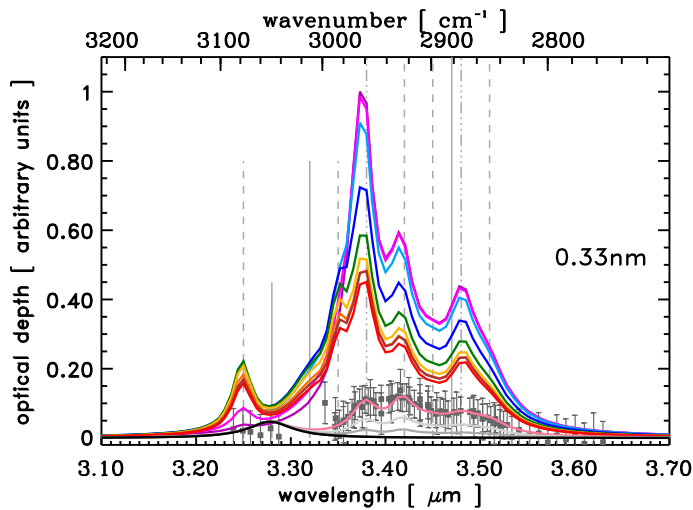}}
 \caption{Same as Fig.~\ref{fig_spectra_3nm} but for $a = 0.33$\,nm.  For comparison the observational data are scaled to the $E_{\rm g} = 0.5$\,eV data. }
 \label{fig_spectra_0.33nm}
\end{figure}

Given the important, but not exclusive, diagnostic importance of the $3\,\mu$m region in determining the nature of  hydrocarbon dust in the ISM \citep[{\it e.g.},][]{2002ApJS..138...75P} Figs.~\ref{fig_spectra_3nm} to \ref{fig_spectra_0.33nm} show the spectra of small a-C(:H) particles in the $3\,\mu$m wavelength region, presented as  optical depth.\footnote{Note that the 3.28\,$\mu$m (2960\,cm$^{-1}$) band has here been reduced in strength by 20\% to $4\times10^{-18}$\,cm$^2$ compared to the value assumed in paper~I.} These data illustrate the intrinsic weakness of aromatic band CH at $3.28\,\mu$m compared to the olefinic and aliphatic CH$_n$ modes. Given that there seems to be a natural ``block'' on the evolution to materials with band gaps smaller than $\approx 0.2$\,eV ($\equiv X_{\rm H} \simeq 0.05$) during ion or UV irradiation \citep[{\it e.g.},][]{1989JAP....66.3248A,1996MCP...46...198M,2011A&A...528A..56G,2011A&A...529A.146G}, there ought then to be a natural block on their spectral evolution. For these limiting compositions ({\it i.e.} $E_{\rm g} \simeq 0.1 - 0.25$\,eV) the optEC$_{\rm (s)}(a)$ model predicts spectra (the light grey lines in Figs.~\ref{fig_spectra_1nm} to \ref{fig_spectra_0.33nm}), for small particles ($a \lesssim 1$\,nm), that exhibit an aromatic CH band with satellite aliphatic CH$_n$ bands but little evidence for olefinic CH$_n$ bands. {\em The optEC$_{\rm (s)}(a)$ model therefore predicts that the aromatic CH band should always be associated with aliphatic (but not olefinic) {\rm CH$_n$} side bands and/or a plateau in the $\approx 3.35-3.55\,\mu$m region.} The lack of olefinic CH bands is because, in the eRCN/DG model, the $sp^2$ component in low $X_{\rm H}$, sub-nm particles is predominantly in small aromatic domains that are linked by aliphatic, rather than olefinic, bridging species. Thus, the optEC$_{\rm (s)}(a)$ model then also predicts that no IR emission band spectrum, arising from hydro-carbonaceous dust, and exhibiting only a $3.28\,\mu$m aromatic CH band, in the $3\,\mu$m region will be observable in the ISM. Some evidence that seems to support this prediction, and therefore favour an a-C(:H) origin for the IR emission bands, comes from AKARI observations of the IR emission features in M\,82, which show no pure aromatic CH band but only {\em aromatic and aliphatic} CH {\em bands that always appear together}, even in the very energetic superwind region \citep{2012arXiv1203.2794Y}. In contrast, the PAH and (surface-hydrogenated) graphitic interstellar dust models both, implicitly, allow for the possibility of a ``pure'' $3.28\,\mu$m aromatic CH emission band in the $3\,\mu$m region, which is apparently never observed. 


Figs.~\ref{fig_spectra_1nm} to \ref{fig_spectra_0.33nm} predcit that a-C(:H) nano-particles ($a \leqslant 1$\,nm), with band gaps in the range $0.75-1.75$\,eV ($\equiv X_{\rm H} \simeq 0.2-0.4$), will show an olefinic CH$_2$ band at $3.35\,\mu$m, which is not apparent in the observational data. This band is not evident in the predicted spectra for a-C(:H) particles with radii $\geqslant 3$\,nm, nor for bulk a-C(:H) materials (see paper~I), which would therefore be consistent with the observed  $3-4\,\mu$m interstellar absorption features. However, in the ISM the a-C(:H) nano-particles that would show the $3.35\,\mu$m band ({\it i.e.}, those with $a \lesssim 1$\,nm)  will be rapidly UV-EUV photolysed (in $\lesssim 10^6$\,yr, see Sect.~\ref{sect_proc_timescales}) to aromatic-rich particles, with and associated aliphatic component, that do not exhibit a $3.35\,\mu$m olefinic CH$_2$ band (see the spectra for nano-particles with $E_{\rm g}$(bulk) $= 0.1-0.5$\,eV, the pink, light-grey and mid-grey lines in Figs.~\ref{fig_spectra_1nm} to \ref{fig_spectra_0.33nm}). 

As already pointed out, 3\,nm radius a-C(:H) particles with $E_{\rm g}$(bulk) $\leqslant 0.5$\,eV will be relatively strong continuum absorbers (see Sect.~\ref{sect_cf_aCH_PAH}). In Figs.~\ref{fig_spectra_1nm} to \ref{fig_spectra_0.33nm} we also note that, for these same particles, the $3.28\,\mu$m aromatic CH band is intrinsically rather weak in absorption compared to the  aliphatic CH$_{n=1-3}$ bands in the adjacent $3.35-3.55\,\mu$m region. Recall that within a-C(:H) particles, it is the small band gap, aromatic clusters that will be the UV-visible, photon-absorbing species and that they appear to be bridged, principally, by short, ``isolating'', aliphatic chains. It is therefore likely that the absorbed UV-visible photon energy could be localised in an aromatic cluster before ``vibrational-leakage'' to the rest of the particle through the aliphatic linkages. Thus, the $3.28\,\mu$m aromatic CH band could be stronger in emission than in absorption because it arises from ``locally-hot'' aromatic clusters. Such a process could perhaps be considered as a sort of {\em Locally-Stochastic Emission Process} (L-StEP) arising from localised-absorption within interstellar a-C(:H) particles. 

The above-discussed small particle, or carbon cluster, mixed aromatic-aliphatic structures ought to be rather stable in the ISM. However, should they be further-fragmented by UV-EUV photons, or by ion or electron collisions, then they will preferentially lose their more fragile aliphatic components, thus releasing ``free-flying'', aromatic, PAH-like species into the gas. \cite{1994ApJ...420..307J} showed that small PAHs ($N_{\rm C} \lesssim 30-40$) will dissociate, before they can relax by IR emission, in H{\footnotesize I} regions and will therefore be rapidly destroyed in the low-density, diffuse ISM. Thus, the fragmentation products of sub-nm, a-C(:H) particles, {\it i.e.}, aromatic clusters with few aromatic rings ($N_{\rm R} = 2-3$) and only $\simeq 10-20$ $sp^2$ carbon atoms, are likely to be rather short-lived in the ISM. The smallest a-C(:H) particles, that could be at the origin of the observed IR emission bands, may therefore represent something of an ``end of the road'' evolution for carbon clusters because their disintegration products, {\it viz.}, small free-flying PAH-like species,will be quickly destroyed in the diffuse ISM \citep[{\it e.g.},][]{1994ApJ...420..307J,2010A&A...510A..36M,2010A&A...510A..37M}

In conclusion, it is likely that pure, perfect, planar PAHs are not likely to be a major component of dust in the ISM but rather that the observable characteristics generally attributed to them ({\it i.e.}, principally the IR emission bands) can be better explained by a distribution of nm and sub-nm a-C(:H) particles. Consequently, searches for particular PAH molecules will probably prove fruitless. Further, graphite (containing little or no hydrogen) is probably not a viable material for interstellar carbonaceous dust and {\em small} graphite particles are therefore unlikely to be present in the ISM \citep[{\it e.g.},][]{2008A&A...492..127S}.

\subsection{a-C(:H) processing timescales}
\label{sect_proc_timescales}

\begin{figure}
 \resizebox{\hsize}{!}{\includegraphics{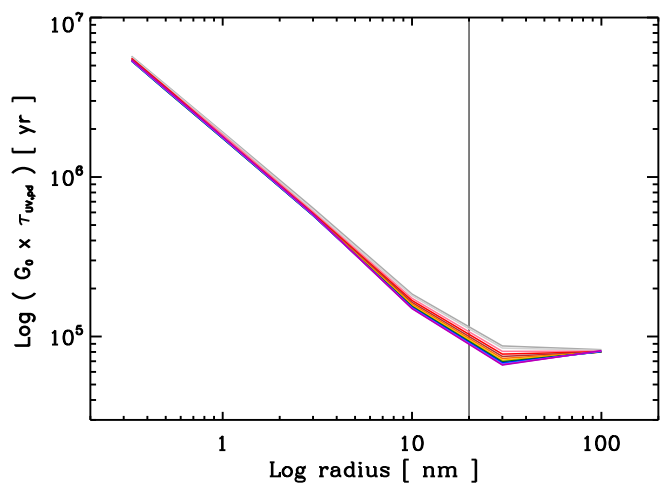}}
 \caption{a-C(:H) particle photo-processing time-scales, $\tau_{\rm UV,pd}$,  multiplied by the ISRF factor $G_0$,  as a function of particle radius for increasing band gap from top ($-0.1$\,eV, grey) to bottom (2.67\,eV, purple). A photo-darkening efficiency, $\epsilon = 0.1$, was assumed here.}
 \label{fig_tauUVpd}
\end{figure}
The size-dependent a-C(:H) particle photo-processing time-scales, $\tau_{\rm UV,pd}$, can be derived as per Eq.~(31) of paper~II by inserting appropriate values for the particle absorption efficiency, $Q_{\rm abs}(a,E)$, and the photo-darkening efficiency, $\epsilon$ (here we assume $\epsilon = 0.1$ but the exact value is uncertain). 
Fig.~\ref{fig_tauUVpd} shows $\tau_{\rm UV,pd}$, multiplied by the ISRF factor $G_0$,  as a function of particle radius for a-C(:H) materials ($E_{\rm g} = -0.1$ to 2.67\,eV), where surface hydrogenation has been included. Presenting the a-C(:H) evolution in this way, Fig.~\ref{fig_tauUVpd} indicates that in the diffuse ISM ($G_0 = 1$) the photo-processing time-scales will likely be $\approx 10^5$\,yr  for $10-100$\,nm radius particles and $\gtrsim 10^6$\,yr for $a \lesssim 1$\,nm.  These data should be scaled accordingly for photo-darkening efficiencies, $\epsilon$, other than the value of 0.1 that has been adopted here. The longer time-scales for smaller particles simply reflects their lower values of $Q_{\rm abs}(a,E)$ at UV wavelengths. Thus, it appears that all a-C(:H) particles are photo-aromatised on rather short time-scales ($\lesssim 10^6$\,yr) in the ISM. 
However, this estimation does not include the effects of thermal processing during the stochastic heating of small a-C(:H) particles, which could be important but has not yet been evaluated. 
In extreme radiation field regions, {\it i.e.}, photo-dissociation regions (PDRs) with $G_0 \simeq 10^3$ ($10^4$),  the carbonaceous nano-particle photo-processing time-scales will be $\approx 10^3$ ($\approx 10^2$)\,yr, and it is in such environments that the loss of aromatic emission bands is observed \citep{1998ASPC..132...15B}, thus indicating extreme carbonaceous nano-particle processing.  

The effects of photo-processing and photo-darkening, the $sp^3$ to $sp^2$ transformation, resulting from exposure to ISRF UV-EUV photons, could possibly be counterbalanced by hydrogen atom addition to the structure as a result of (energetic) collisions in the interstellar medium but this possibility is not discussed here, other than to refer the reader to the discussion presented in Sect.~5.2 of paper~II. 

The EUV-photolysis of a-C(:H) particles is key to their evolution in the ISM and needs to be incorporated into dust models. Carbonaceous dust evolution must therefore be treated dynamically, not only in terms of the evolution of the dust size distribution \citep[{\it e.g.},][]{1996ApJ...469..740J,2008A&A...492..127S} but also time-dependently in terms of the evolution of its chemical composition and structure \citep[{\it e.g.},][papers I and II]{2009ASPC..414..473J}. 
To this end, in the following section, provides a brief and preliminary guide to the usage of the optEC$_{\rm (s)}(a)$ data.

\subsubsection{Which optEC$_{\rm (s)}(a)$ data?}
\label{sect_which_data}

The evolution of a-C:H dust in the ISM is tied to systematic variations its band gap, $E_{\rm g}$, which are driven by the dehydrogenating effects of extreme UV (EUV) $10-13.6$\,eV photon absorption (EUV-photolysis) and possibly by thermally-driven dehydrogenation during the large temperature fluctuations induced by stochastic heating following UV photon absorption. 

A full investigation of the likely a-C(:H) optical properties in the ISM must await an in-depth modelling of the observed properties of carbonaceous dust in the ISM using the optEC$_{\rm (s)}(a)$ data. However, the user is currently recommended to use the minimum band gap material for photo-processed a-C(:H) grains as indicated by the likely lower limit to the H atom fraction ({\it viz}, $X_{\rm H} \simeq 0.05$) as derived by  experiment \citep{1989JAP....66.3248A,1996MCP...46...198M,2011A&A...528A..56G,2011A&A...529A.146G}. For the cores of larger particles, that cannot be photo-processed ({\it i.e.}, at depths $\gtrsim 20$\,nm), the material band gap should reflect the properties of the material prior to photon-irradiation in the ISM and may be an historical vestige of the grains at their time of formation. 
 
The required optEC$_{\rm (s)}(a)$ model optical property data, $m(a,\lambda)$, should be chosen from those tabulated for the given radius and the required {\em bulk} material band gap for that radius. Note that  it may be necessary to interpolate if data for the required radius or band gap, other than those tabulated, are needed.  However, in this case, care will need to be exercised to ensure that the interpolated $n$ and $k$ data are consistent with the Kramers-Kronig relationship.

\subsection{a-C(:H) extinction, absorption and emission}
\label{sect_ext_abs_em}

The use of the optEC$_{\rm (s)}(a)$ data in modelling the likely properties of small interstellar carbon dust particles engenders certain relationships between the observable behaviours, {\it e.g.}, principally, the FUV extinction rise, the UV bump at 217\,nm and the aromatic infrared bands (AIBs).   
The derived data can be used to investigate the characterising properties of the a-C(:H) materials that could be responsible for these observables.
The following very general criteria are adopted for each: 
\begin{itemize}
  \item {\em FUV extinction} -- particles with an extinction rise in the FUV but no discernable bump at 217\,nm,  
  \item {\em 217\,nm UV bump} -- particles that produce a pronounced UV bump, with a peak close to 217\,nm, and  
  \item {\em AIB carriers} -- particles that show a clear band at 3.28\,$\mu$m and also weaker features and a plateau in the $3.35-3.55\,\mu$m region.
\end{itemize}

\begin{figure}
 \resizebox{\hsize}{!}{\includegraphics{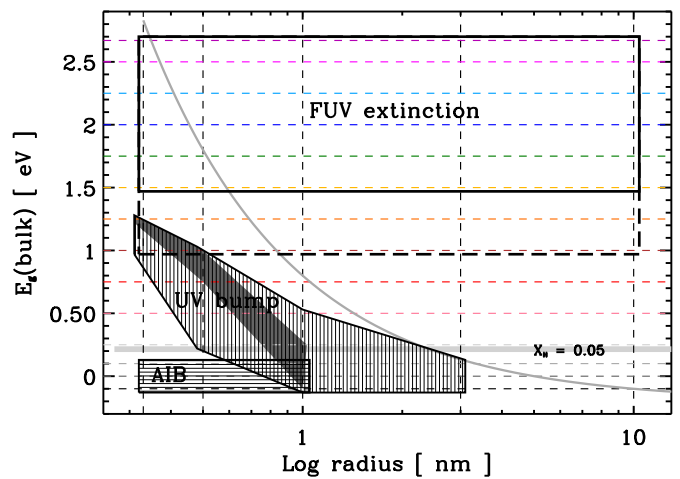}}
 \caption{Schematic diagram of the primary contributions of a-C(:H) particles to 
 the FUV extinction (upper solid and long-dashed boxes), 
 the UV extinction bump (vertical-lined box and dark shaded area) and 
 the AIBs (horizontal-lined box) as function of the equivalent bulk material band gap $E_{\rm g}({\rm bulk})$ and the particle radius. The horizontal $X_{\rm H} \simeq 0.05$ thick grey line indicates the possible minimum band gap attainable for a-C(:H) materials.}
 \label{fig_Eg_correlations}
\end{figure}

Fig.~\ref{fig_Eg_correlations} schematically shows the $E_{\rm g}-a$ parameter spaces occupied by the particles that could be responsible for the FUV extinction rise, the 217\,nm UV bump and the AIBs. The grey curve indicates the minimum possible band gap, given by $E_{\rm g}({\rm min}) = \{ (a/[1\,{\rm nm}])^{-1}-0.2$\}\,eV. However, and for clarity, the different component data are plotted as a function of the bulk material band gap, $E_{\rm g}$(bulk), because plotting as a function of the size-dependent band gap would place all the zones below the grey curve on the curve. 
In Fig.~\ref{fig_Eg_correlations} the upper solid and long-dashed boxes indicate the particles that can explain the FUV extinction but that do not produce a significant bump at 217\,nm. The vertical-lined box indicates particles that show a bump in the UV in the $4.4-4.8\,\mu$m$^{-1}$ region and the dark-shaded inset area shows those particles that give a much closer fit to the observed UV bump peak position. The horizontal-lined box indicates particles that show spectra qualitatively consistent with the observed IR emission bands.  

With the selection criteria adopted here, Fig.~\ref{fig_Eg_correlations} shows that the FUV extinction, UV bump and AIB carrying particles appear to be rather de-coupled. Note that, for a-C(:H) materials to be responsible for the AIBs, the particles would apparently have to be small ($a \lesssim 1$\,nm), and aromatised ($E_{\rm g} \simeq 0$\,eV) and bulk-dehydrogenated below the 5\% H atom fraction limit. The optEC$_{\rm (s)}(a)$ data indicate that the FUV extinction particles ought to be larger and have larger band gaps  than the UV bump carriers. The UV bump carriers probably have properties intermediate between those of the FUV extinction and AIB carriers. There is, nevertheless, likely to be overlap between the different populations, which probably do not occupy well-defined zones in the $E_{\rm g}-a$ parameter space as indicated in Fig.~\ref{fig_Eg_correlations}.

\subsubsection{A brief comparison with experiment}
  
\cite{1996ApJ...472L.123S} experimentally investigated HAC particles at the point at which they are undergoing decomposition. They found that the IR spectra under these limiting  conditions show the major spectral features seen in the emission bands. In this case the decomposition transition species are found to contain ``proto-graphitic'' islands ({\it i.e.}, PAH-like or PAH radical species) with tens to hundreds of aromatic rings (dimensions 1-5\,nm) and which appear to be a good candidate for the carriers of the IR emission bands. The work presented here is then in agreement with this HAC particle decompositional processing scheme and indicates that small, highly-aromatic a-C(:H)/HAC particles are indeed the ``end of the road'' for carbon dust evolution in the ISM.  

Given that the smallest a-C(:H) particles will have small heat capacities, be stochastically-heated by ISRF photons and undergo large temperature excursions upon single photon absorption, the particle heat capacities in this solid-molecule interfacial domain will need to be carefully calculated. Appendix~\ref{appendix_heat_cap} gives a possible approach to this heat capacity issue.

\subsubsection{A brief comparison with observations}

The data, as presented here, are qualitatively consistent with many observable dust properties, namely: 
the form of the FUV extinction, the lack of correlation between 217\,nm UV bump peak position and width, broader UV bumps where the FUV curvature is greater and stronger FUV extinction as the bump weakens \citep[{\it e.g.},][]{2004ASPC..309...33F,2007ApJ...663..320F}.  In particular, the optEC$_{\rm (s)}$(a) data predict a general, but not completely de-coupled, anti-correlation between the intensity of the bump at 217\,nm and the FUV extinction rise ({\it e.g.}, see Fig.~\ref{fig_Qs_vs_size_5_eg}). It is also clear that the derived UV bump at $\approx 4.6\,\mu$m$^{-1}$ is too large, with respect to the observed interstellar feature, and that its position is size-dependent. The variation in the bump width appears to be rather small, which is consistent with observations.\footnote{A ``perfect''  match to the observed UV bump width could, rather straight-forwardly, be obtained in the optEC$_{\rm (s)}$(a) data by introducing an empirical, size-dependence into the calculation of the $\pi-\pi^\ast$ band (Appendix~\ref{appendix_pi_pi}); this temptation has been resisted here.} Clearly, a better understanding of the nano-physics of the likely bump carriers is required. Many  theoretical and laboratory data-based models for the UV bump, which require specific sizes  or use PAH molecules \citep[{\it e.g.},][]{1984ApJ...285...89D,1990MNRAS.243..570S,1992A&A...266..513V,1992ApJ...393L..79J,1995ApJS..100..149M,1996ApJ...464L.191M,1999ApJ...522L.129D,2011A&A...528A..56G,2011A&A...525A.103C} are also not without some difficulties. 

A fuller investigation of these effects is clearly required but is beyond the scope of this paper. A detailed modelling of the interstellar extinction and emission, using the DustEM model  \citep{2011A&A...525A.103C}, will be undertaken in follow-up papers.

\subsection{Re-formation and accretion issues}

The indication of enhanced scattering in the outskirts of molecular clouds implies that the formation and accretion of H-rich a-C:H (mantles) in the ISM, or just carbonaceous dust re-formation in general, as apparently required by observations \cite[{\it e.g.},][]{2011A&A...530A..44J}, could indeed occur in and on the outskirts of molecular clouds, a scenario that was proposed long ago \cite[{\it e.g.}, see ][and references therein]{1990QJRAS..31..567J}. 
This is perhaps supported by the analysis of observational data by \cite{1994A&A...284..956B}, which shows that when matter cycles in and out of the denser molecular regions of the ISM the AIB carriers {\em ``are formed from or transformed into particles sharing the same absorption features.''} This suggests that the variations in the mid-IR emission result from changes in the small particle size distribution, which conserve the dust mass within the Rayleigh limit for UV radiation ($a \lesssim 20-30$\,nm) and implies that the smallest particles grow by accretion in dense molecular clouds but do not coagulate onto large grains.  

Under quiescent conditions the accretion of (carbonaceous) mantles onto dust is an important process, and one that depends upon the total surface area available and leads to the same mantle thickness on all grains, independent of radius.  Then, given that, for a power-law-type dust size distribution typical of the ISM \citep[{\it e.g.},][]{1977ApJ...217..425M,1984ApJ...285...89D,1990ARA&A..28...37M,2011A&A...525A.103C} most of the available surface will be in the small grains, it is on the smaller grain surfaces that a significant fraction of the mass will accrete. This will have a consequent and major impact on the modification of the dust size distribution.  As \cite{2007A&A...476..263G} show, for ice mantle accretion, the resulting dust mass distribution can end up with a significant fraction of the accreted mass on the smallest particles in the size distribution (resulting in a shift to significantly larger sizes) and yield a somewhat bi-modal dust mass distribution.  Thus, the accretion of H-rich carbonaceous mantles onto (predominantly) the smallest pre-existing grains in the ISM (in denser regions) is entirely consistent with the observational effects noted by \cite{1994A&A...284..956B} whereby the smallest grains are most affected but still remain within the Rayleigh limit for UV radiation. Given that the accretion timescale is $\approx 10^9/n_{\rm H}$\,yr, mantle accretion in less than one million years requires gas densities $\gtrsim 10^3$\,cm$^{-3}$, which are not unreasonable.

\section{Predictions of the optEC$_{(s)}$(a) model}
\label{sect_predictions}

The optEC$_{(s)}$ model made several predictions deriving from the band gap evolution of a-C(:H) bulk materials (see papers I and II). Additional predictions, arising from the optEC$_{(s)}$(a) model, and specifically-relating to the likely size-dependent properties of hydrocarbon grains in the ISM, include:  
\begin{enumerate}
  \item  Large a-C(:H) grains ($a \gtrsim 20$\,nm) cannot be completely UV-photolysed 
  and should retain their formation composition in their cores.  
  \item Small grains ($a \lesssim 20$\,nm) will ultimately be  UV-photolysed 
  to aromatic-rich materials with $E_{\rm g} \sim 0.2-0.25$\,eV.
  \item As a-C(:H) grains are progressively dehydrogenated and aromatised by UV photolysis there will be a 
  concomitant:
  \begin{itemize}
    \item decrease in the UV extinction, 
    \item increasing UV bump peaking in the $208-227$\,nm region,  
    \item increase in the FIR-mm extinction and emission and  
    \item a transition from aliphatic-dominated to aromatic-dominated CH and CC IR modes. 
  \end{itemize} 
  \item  The spectra for $\gtrsim 3$\,nm radius particles are rather invariant. 
  \item With decreasing size the CH modes increase in strength, with respect to CC modes, due increased surface hydrogenation.   
  \item In the  $6-8\,\mu$m region significant changes occur in the spectra for $a < 1$\,nm as the aliphatic CC component  declines.    
  \item A ``resilient'' $6.9\,\mu$m aliphatic CH$_2$ band is seen in the spectra, which is paired with a  $7.1\,\mu$m olefinic CH$_2$ band. 
  \item The $3.28\,\mu$m aromatic CH band will always be accompanied by aliphatic bands and /or a plateau in the $ 3.35-3.55\,\mu$m region.   
\end{enumerate}

\section{Limitations of the optEC$_{(s)}$(a) model}
\label{sect_limitations}

The limitations of optEC$_{\rm(s)}$ model that were presented in papers I and II, are also present in the optEC$_{(s)}$(a) model. However, other limitations are specific to the particle size-dependence incorporated into the optEC$_{\rm(s)}$(a) model:
\begin{itemize}
  \item Caution should be exercised in the use of the spectral bands long-wards of $\approx 7.3\,\mu$m in the interpretation of astrophysical spectra because these bands are not yet well-determined.  
  \item The model does not include $sp^2$ clustering into chains, cage-like structures or C$_5$ pentagons, which could be particularly important for small particle evolution/fullerene formation. 
  \item The model does not take into account any possible size-dependent modifications of the IR CH and CC band profiles. 
  \item The possible contribution of  long wavelength modes to the mm-cm emission cannot yet be discounted.
\end{itemize}

\section{Conclusions and summary}
\label{sect_conclusions}

Here and in the two preceding papers a self-consistent unification model was developed for the band gap- and size-dependent evolution of amorphous hydrocarbon, a-C(:H), grains in the ISM. The physical and chemical transformations of this apparently simple bi-atomic solid phase could be at the heart of many of the observables attributed to carbonaceous matter in the ISM, circumstellar media and in the Solar System. 

This work presents a model, the optical property prediction tool for the Evolution of Carbonaceous (s)olids as a function of r(a)dius, optEC$_{\rm(s)}$(a), that can be used to explore the nature of hydrocarbon solids in a wide variety of environments. The  optEC$_{\rm(s)}$(a) model is used to derive  a coherent set of optical properties [the complex refractive indices, $m = n(a,E_{\rm g},\lambda)+ik(a,E_{\rm g},\lambda)$] for a-C(:H) materials as a function of wavelength (from EUV to cm wavelengths), band gap (from $-0.1$ to $2.7$\,eV) and radius (from 0.33 to 100\,nm).
For  larger particles ($a \geqslant10$\,nm) the properties are essentially the same as for bulk materials. For the smallest  particles, with as few as a hundred or so atoms, the properties are a strong function of size because of surface hydrogenation and the key role of the particle-size-limited, aromatic clusters (containing only a few rings for the smallest consider particles). 

As it currently stands  the use of the optEC$_{\rm(s)}$(a) data to study the observable characteristics of interstellar hydrocarbon particles appears to be, at least qualitatively, consistent with the:
\begin{itemize}
  \item   evolution of the size-dependent spectra of ISM a-C(:H) dust, 
  \item spectra of the   AIB carriers in the 3\,$\mu$m region,  
  \item FUV extinction and the UV bump,  
  \item major observed extinction and AIB (non-)correlations, and
  \item variations in the FIR-mm emissivity index ($\beta \simeq 1.3$ to 3.1). 
\end{itemize}
Further the optEC$_{\rm(s)}$(a) model data makes some key predictions about the nature of the interstellar carbonaceous dust, {\it i.e.}:
\begin{itemize}
  \item the $3.28\,\mu$m aromatic CH band will always be accompanied by aliphatic CH$_n$ bands and/or plateau in the $3.35-3.55\,\mu$m region,  
   \item the ``end of the road'' evolution for small a-C(:H) particles is probably aromatic/aliphatic cage-like structures, which could possibly provide a route to fullerene formation,   
    \item the UV-photolytic fragmentation of small a-C:H grains will lead to the formation of small hydrocarbon molecules (CCH, c-C$_3$H$_2$, C$_4$H, {\it etc}.) in PDR regions, and 
    \item ``pure'' graphite grains and ``perfect'' PAHs are probably not important components of dust in the ISM. 
\end{itemize}

The optEC$_{\rm(s)}$(a) optical constant data are provided in the accompanying pairs of ASCII files for $n$ and $k$ for particle radii of 100, 30, 10, 3, 1, 0.5 and 0.33\,nm and band gaps of -0.1,  0.0,  0.1,  0.25,  0.5,  0.75,  1.0,  1.25,  1.5,  1.75,  2.0,  2.25,  2.5 and  2.67\,eV. 
It is hoped that these data will provide an interim, test-able but modifiable set of data that can be used in the interpretation of astronomical data. 
The user of these data is, however, cautioned to carefully consider the highlighted limitations of these data in the interpretation of astronomical data until such time as suitable laboratory data is available to replace them. 
The current tabulated optEC$_{\rm(s)}$(a) optical constant data should therefore be seen as a ``stop gap'' solution until that time. 

The optEC$_{\rm(s)}$(a) model (with its inherent size-dependent, band-gap limitations) has been used to predict the long-wavelength (FIR-cm) properties of a-C(:H) particles with the same number of carbon atoms as interstellar PAHs. The results for the optEC$_{\rm(s)}$(a) model are not consistent  with the predictions for interstellar PAH models in that, in order to absorb strongly at FIR-cm wavelengths, interstellar PAHs would seemingly need to be rather large {\it i.e.}, $N_{\rm C} \gtrsim 10^4 \equiv a \gtrsim 9$\,nm.  However, in the interstellar PAH models, absorption at long wavelengths is considered to be due to the overlapping contributions of broad, low-energy ``flopping'' modes, which have not been incorporated in the optEC$_{\rm(s)}$(a) model.  

A full investigation of these optEC$_{\rm(s)}$(a) data, in the interpretation of the observed interstellar dust extinction and emission properties, is now under way using the DustEM model \citep{2011A&A...525A.103C}.


\begin{acknowledgements} 
I would like to thank my colleagues at the IAS and elsewhere for many, many years of stimulating discussions on interstellar dust modelling, experiment, theory and observations. 
In particular, I am indebted to Laurent Verstraete for inumerable discussions on interstellar nano-particle physics. 
Once again, I would also like to thank the referee of this series of papers, Walt Duley, for his valuable and insightful remarks and suggestions. 

This research was, in part, made possible through the financial support of the Agence National de la Recherche (ANR) through the program {\it Cold Dust} (ANR-07-BLAN-0364-01). \\

\noindent This work is dedicated to the memory of John Bibby ( 1957 --- 2011 ).\\ 
So long John, thanks for 36 years of smiles and shared laughs. 
\end{acknowledgements}


\bibliographystyle{aa} 
\bibliography{biblio_HAC} 



\appendix

\section{Size-dependent structural properties and surface hydrogenation effects}
\label{appendix_surface_H_effects}

\begin{figure}
 \resizebox{\hsize}{!}{\includegraphics{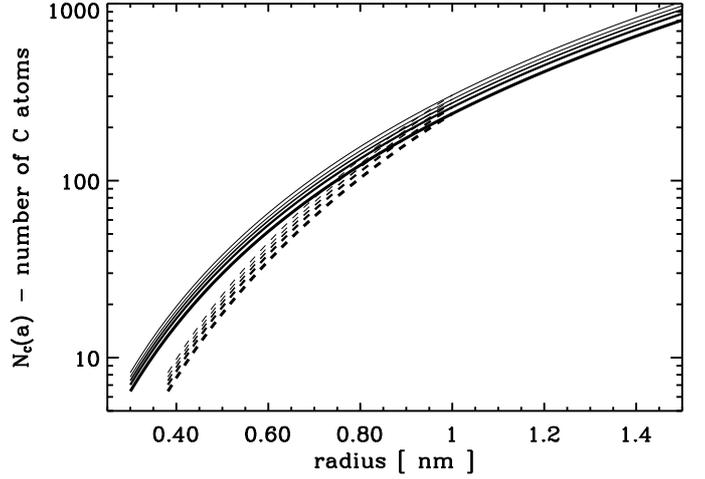}}
 \caption{The number of carbon atoms in a particle, $N_{\rm C}$, as a function of the particle radius  for $X_{\rm H} = 0.17$ (thin), 0.3, 0.4, 0.5 and 0.62 (thick). The dashed lines show $N_{\rm C}$ for effective radii $a_{\rm eff} = a^{0.8}$, which simulate lower effective densities for sub-nm particles,  {\it i.e.}, larger effective radii for a given number of carbon atoms. }
 \label{fig_NCatom_vs_a} 
\end{figure}

For an amorphous hydrocarbon material the mean atomic mass, ${\bar A}$, is given by
\[
{\bar A} = X_{\rm H} + 12(X_{sp^2}+X_{sp^3}) 
\]
\begin{equation}
\ \ \ \, = X_{\rm H} + 12(1-X_{\rm H}) = 12-11X_{\rm H}, 
\end{equation} 
where $X_{sp^2}$ and $X_{sp^3}$ are the atomic fractions of $sp^2$ and $sp^3$ carbon atoms, respectively. 
For a bulk material density $\rho$ and a grain radius $a$ the total number of atoms per particle, $N_{\rm atom}$, is then  
\begin{equation}
N_{\rm atom} = \frac{4 \pi a^3 \ \rho(X_{\rm H}) \ {\rm N}_{\rm A}}{3{\bar A}(X_{\rm H})},
\end{equation}
where N$_{\rm A}$ is Avogadro's number. 
The total number of atoms (C+H) in the bulk of the particle can then be estimated from the simplified expression
\begin{equation}
N_{\rm atom} \simeq 2500 \ \left( \frac{a}{\rm 1\,nm} \right)^3 \ \left( \frac{\rho(X_{\rm H})}{12-11X_{\rm H}} \right), 
\label{eq_Natom_estimator}
\end{equation}
which indicates that a 1\,nm radius a-C:H particle with $X_{\rm H} \simeq 0.5$ contains approximately 600 atoms, half of them being hydrogen atoms [$N_{\rm H} =  X_{\rm H} \, N_{\rm atom}$] and half carbon atoms [$N_{\rm C} = (1-X_{\rm H}) \, N_{\rm atom}$], {\it i.e.}, 
\begin{equation}
N_{\rm C} =  \frac{4}{3} \pi a^3 \ \rho(X_{\rm H}) \ {\rm N}_{\rm A} \ \frac{(1-X_{\rm H})}{(12-11X_{\rm H})} 
\label{eq_NC_vs_a}
\end{equation} 
or
\begin{equation}
N_{\rm C} \simeq 2500 \ \left( \frac{a}{\rm 1\,nm} \right)^3 \ \left( \frac{\rho(X_{\rm H}) \ (1-X_{\rm H})}{12-11X_{\rm H}} \right).  
\label{eq_NC_estimator}
\end{equation}
Fig.~\ref{fig_NCatom_vs_a} shows $N_{\rm C}$ as a function of radius for typical nano-particle sizes, and also indicates the effect of adopting a lower effective density for the smallest particles with $a \leqslant 1$\,nm (dashed lines). 
%

\begin{figure}
 \resizebox{\hsize}{!}{\includegraphics{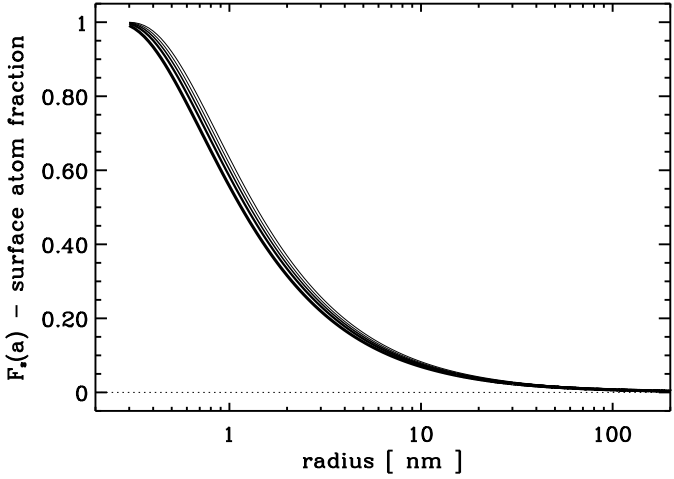}}
 \caption{The fraction of atoms in the surface, $F_s$, as a function of the particle radius  for $X_{\rm H} = 0.17$ (thin), 0.3, 0.4, 0.5 and 0.62 (thick).}
 \label{fig_Fs_vs_a} 
\end{figure}

The effective volume of a constituent atom within such a particle is
\begin{equation}
V_{\rm atom} = \frac{{\bar A}(X_{\rm H})}{\rho(X_{\rm H}) \ {\rm N}_{\rm A}} = \frac{4}{3} \pi r_{\rm atom}^3, 
\end{equation}
and the effective radius of an atom within the solid is then 
\begin{equation}
r_{\rm atom} = \left( \frac{3 \, {\bar A}(X_{\rm H})}{4 \pi \ \rho(X_{\rm H}) \ {\rm N}_{\rm A}} \right)^{\frac{1}{3}}. 
\end{equation}
The number of surface atoms per particle is given by the volume of the surface layer one atom deep divided by the atomic volume, {\it i.e.},
\begin{equation}
N_s = \frac{\frac{4}{3}\pi a^3-\frac{4}{3} \pi (a-2 r_{\rm atom})^3}{\frac{4}{3}\pi r_{\rm atom}^3} = \frac{a^3\{1-[1-(2r_{\rm atom} /a)]^3\}}{r_{\rm atom}^3}. 
\end{equation}
The number of atoms per particle $N_{\rm atom} = \frac{4}{3}\pi a^3/\frac{4}{3}\pi r_{\rm atom}^3 = (a/r_{\rm atom})^3$ and the fraction of the atoms that are ``in'' the surface, $F_s$, is then given by $N_s/N_{\rm atom}$, {\it {\it i.e.}},
\begin{equation}
F_s = \Bigg\{ 1-\left[ 1-\left(\frac{2 \, r_{\rm atom}}{a}\right) \right]^3 \Bigg\}. 
\end{equation}
Fig.~\ref{fig_Fs_vs_a} shows $F_s$ as a function of $a$ for various values of $X_{\rm H}$ and indicates that $F_s$ is not particularly sensitive to $X_{\rm H}$ but that the surface contribution is important for sizes less than 10\,nm and is dominant for particles as small as 0.5\,nm in radius. For particles with $a \gtrsim 30$\,nm surface effects can be safely ignored. 

The average number of hydrogen atoms associated with a carbon atom, based on the bulk composition, is
\begin{equation}
\frac{N_{\rm H}}{N_{\rm C}} = \left( \frac{X_{\rm H}}{1-X_{\rm H}} \right)
\end{equation}
and this must also be the minimum hydrogenation at the surface. The coordination of carbon atoms at the particle surface will be incomplete and result in  ``dangling'' C(---)$_n$ bonds ($n \leq 3$), which are assumed to hydrogenate to form additional C---H bonds at the particle surface. Assuming homogeneous mixing of the atoms, within the particle and at the surface, the number of carbon atoms in the particle is 
\begin{equation}
N_{\rm C} = (1-X_{\rm H})N_{\rm atom}
\end{equation}
and the number carbon atoms at the surface is 
\begin{equation}
N_{\rm C}^s = N_{\rm C} F_s = (1-X_{\rm H}) N_{\rm atom} F_s= (1-X_{\rm H})N_s.
\end{equation}
The fraction of carbon atoms at the surface is then $F_C^s = N_C^s/N_C = N_C F_s / N_C = N_s/N_{\rm atom} = F_s$ because the same bulk and surface composition are assumed. The possibility that the surface may reconstruct in some way without the addition of hydrogen is ignored here. Such a re-construction would alter the bulk properties, which are assumed to be fixed here, and is not allowed in the model because this would lead to structural inconsistencies. 

\begin{figure}
 \resizebox{\hsize}{!}{\includegraphics{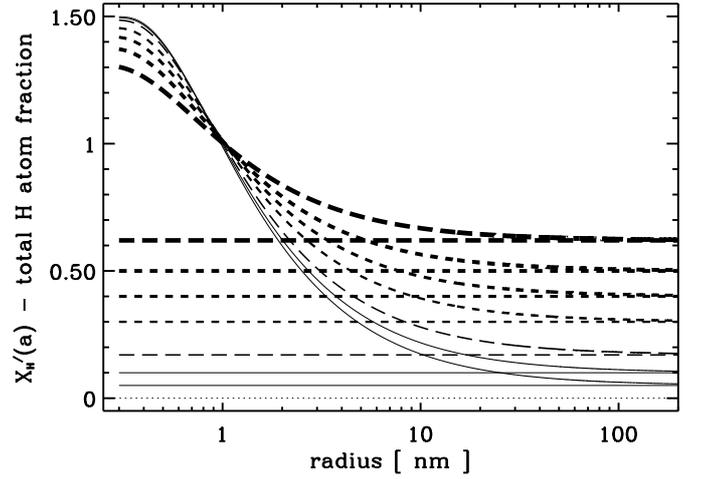}}
 \caption{The total hydrogen atom ``fraction'', $X_{\rm H}^\prime(a)$, as a function of the particle radius, $a$, in nm for various values of $X_{\rm H}$ (DG model: 0.05 and 0.1 - solid lines; eRCN model: 0.17 - thinnest dashed line, 0.3, 0.4, 0.5 and 0.62 - thickest dashed line). The horizontal lines show the cases where surface hydrogenation effects are ignored.}
 \label{fig_XHprime_vs_a} 
\end{figure}

A key question here is: How many extra hydrogen atoms are needed to bond with, or equivalently to ``passivate'', all of the ``dangling'' C(---)$_n$ bonds at the particle surface? In a statistical sense, for a particle large compared to the mean inter-atomic bond length, the particle surface presents a semi-infinite, planar cut through the network/bulk structure. Thus, a surface carbon atom can only ``use'' approximately half of its normal bulk coordination and so each hydrogen-passivated, surface carbon atom can have an extra $\frac{1}{2} m_{\rm C}$ hydrogen atoms attached, where $m_{\rm C}$ is the mean bulk carbon atom coordination number (see paper~I). The number of extra surface-passivating hydrogen atoms, attached to the surface carbon atoms,  is then 
\[
X_{\rm H}^s(a) = (1-X_{\rm H}) \, F_s \frac{m_{\rm C}}{2}  
\]
\begin{equation}
\ \ \ \ \ \ \ \ \ \ \ =  (1-X_{\rm H}) \, \Bigg\{ 1-\left[ 1-\left(\frac{2 \, r_{\rm atom}}{a}\right) \right]^3 \Bigg\} \, \frac{m_{\rm C}}{2}, 
\end{equation}
where the factor $(1-X_{\rm H}) = X_{\rm C}$ allows for the fact that only that fraction of the surface atoms are carbon. 
Substituting into Eq.~(\ref{eq_surf_H}), the particle size-dependent total (surface+bulk) hydrogen atom fraction is given by 
\begin{equation}
X_{\rm H}^\prime = X_{\rm H} + (1-X_{\rm H}) \, F_s \, \frac{m_{\rm C}}{2}. 
\label{XH_size_dep}
\end{equation}
In effect, and for small particles, this expression leads to total atomic fractions for the particles 
\begin{equation}
X_{\rm total} = X_{\rm H}^\prime + X_{\rm C} = X_{\rm H} + X_{\rm H}^s(a) + (1-X_{\rm H}) = 1 + X_{\rm H}^s(a)
\label{eq_total_atomic_fractions_as_a_func_of_a}
\end{equation}
that are greater than unity because the surface H atoms are counted as an ``extra'' atomic fraction, which seems a reasonable approach given that the surface H atoms do not affect the bulk structure. 
Fig.~\ref{fig_XHprime_vs_a} shows the surface and bulk hydrogen atom fractions as a function of the particle size for values of $X_{\rm H}$ typical of the eRCN model (dashed lines) and the DG model (solid lines). 

The surface passivation of the particles, by the addition of extra H atoms, occurs for all types of CH$_n$ groups ($n \leq 3$) and thus their fractional concentration,  $X_{\rm CH_{\it n}}$, must be augmented in the same way as for the hydrogen atom fraction $X_{\rm H}$. However, specific surface CH groups, CH$_{(n-1)}$, need to be considered in terms of how they transform to CH$_n$ groups by the addition of extra H atoms at the particle surface, {\it i.e.},
\begin{equation}
X_{\rm CH_n}^\prime = X_{\rm CH_n} + X_ {\rm CH_{(n-1)}} \, F_s ,  
\end{equation}
which can, as explained above, lead to total atomic fractions greater that unity because of the ``extra'' surface H atom convention adopted here. 

The bulk composition eRCN and DG models therefore turn out to be good representations for particles with radii $\gtrsim 100$\, nm. Only for smaller particles need surface passivation and particle size effects be of concern. Thus, small, stochastically-heated, amorphous hydrocarbon grains in the ISM are likely to show important size and surface effects, in addition to their properties that will be very strongly dependent on annealing (thermal processing) to more aromatic structures (in the absence of an opposing re-hydrogenation process).

\section{The size-dependent visible-UV electronic properties of a-C(:H)}
\label{appendix_visUV_el_props}

This Appendix considers, in detail, the necessary solid-state physics at the heart of the derivation of the size-dependent, EUV-cm wavelength, complex refractive indices of a-C(:H) materials. 

\subsection{The $\pi$--$\pi^\star$, C$_6$, $\sigma$--$\sigma^\star$ bands and the a-C(:H) band gap} 
\label{appendix_all_bands}

In order to aid the reader to negotiate the ``maze'' of the relevant solid-state physics, {\em in the following paragraph the salient points relating to the band gap properties of a-C:H and a-C as, principally, reviewed in the seminal paper by \cite{1986AdPhy..35..317R}} are briefly summarised here. The interested is encouraged to delve into the \cite{1986AdPhy..35..317R} review for a deeper understanding of these complex materials. 

Within the energy gap, between the valence and conduction bands, of a-C(:H) materials there exist states that control the electronic properties, conductivities, luminescence and doping in these amorphous, semi-conducting materials. The gap in an amorphous, semi-conductor is strictly a pseudo-gap or mobility gap, resulting from a localisation of the contributing states. The mobility edge in such a system is the energy separating these localised states from the extended states (states well within the valence and conduction bands), where increasing disorder leads to the localisation of the contributing states ({\it e.g.}, the segregation of aromatic clusters). In a-C the valence band and conduction band edges are due to $\pi$ states, which can support extended states. However, in a-C:H, with decreasing $sp^2$ content, the $\pi$ density of states decreases and the $\pi$--$\pi^\star$ bands will be localised, {\it i.e.}, isolated from one another. The optical absorption edges of a-C are consistent with broad valence and conduction band tails (pseudo gap $0.4 - 0.7$\,eV), which overlap at the gap centre. The tail states in a-C are therefore expected to be the $\pi$ states of larger than average aromatic clusters. Wider gap a-C:Hs likely belong to the class of materials where deep gap states can be distinguished, {\it i.e.}, states that are more localised than the tail states and that are associated with ``defect'' sites where the bonding is different from that of the bulk. These defect states are rather complex and likely associated with a breaking of the weaker $\pi$ bonds (rather than the $\sigma$ bonds), which is aided by the fact that $\pi$ defects can delocalise into conjugated, olefinic, $\pi$ electron systems (see also the postulate in Sect.~\ref{appendix_C6} below).  The defect density in a-C:H correlates inversely with the optical gap and the effect of increasing the hydrogen content in a-C:H materials is to lower the defect density by reducing (aromatic) cluster sizes.
This leads to an increase in the defect energy and thereby a decrease in the probability for the occurrence of these  defect states within the structure.  The creation energy for ``dangling bond'' sites is $\sim 1.8$\,eV and their character, analogous to that of the planar methyl radical (:CH$_3$), will be $\pi$-like in character, as for the olefinic $sp^2$ and aromatic clusters. {\em [End of the band gap physics summary taken from \cite{1986AdPhy..35..317R}.]}

The wide-band optical spectra of a-C:H materials show two clear, separated bands: a $\pi-\pi^\ast$ peak at $\sim 4$\, eV and a $\sigma-\sigma^\ast$ peak at $\sim 13$\, eV \citep{1986AdPhy..35..317R}. The energy positions of these $\pi-\pi^\ast$ and $\sigma-\sigma^\ast$ bands are practically independent of the hydrogen atom content \citep{2007DiamondaRM...16.1813K}.  Another band at $\sim 6.5$\, eV has been attributed to small, aromatic, ``benzene-like'' clusters in the structure \citep[{\it e.g.},][]{1986AdPhy..35..317R}. Such small aromatic clusters are an intrinsic part of the covalent network and are not fully-hydrogenated like benzene. Here the $\sim 6.5$\, eV band is, nevertheless, designated C$_6$. However, it is much broader than the width predicted for a single, six-fold, aromatic ring and so, as hypothesised below (see Sect.~\ref{appendix_C6}), clusters other than simple six-fold, aromatic rings probably contribute to the broadening of the C$_6$ band. With thermal annealing or photo-darkening the $\pi-\pi^\ast$ peak strengthens with respect to both the $\sigma-\sigma^\ast$ and C$_6$ bands; trends that result from an increase in aromaticity with  annealing. In this work the following characteristic band energies have been used: 4.0\,eV for the $\pi-\pi^\ast$, 6.5\,eV for the C$_6$ and 13.0\,eV for the $\sigma-\sigma^\ast$ bands.

\subsubsection{The $\pi$--$\pi^\star$ band}
\label{appendix_pi_pi}

\begin{figure} 
 \resizebox{\hsize}{!}{\includegraphics{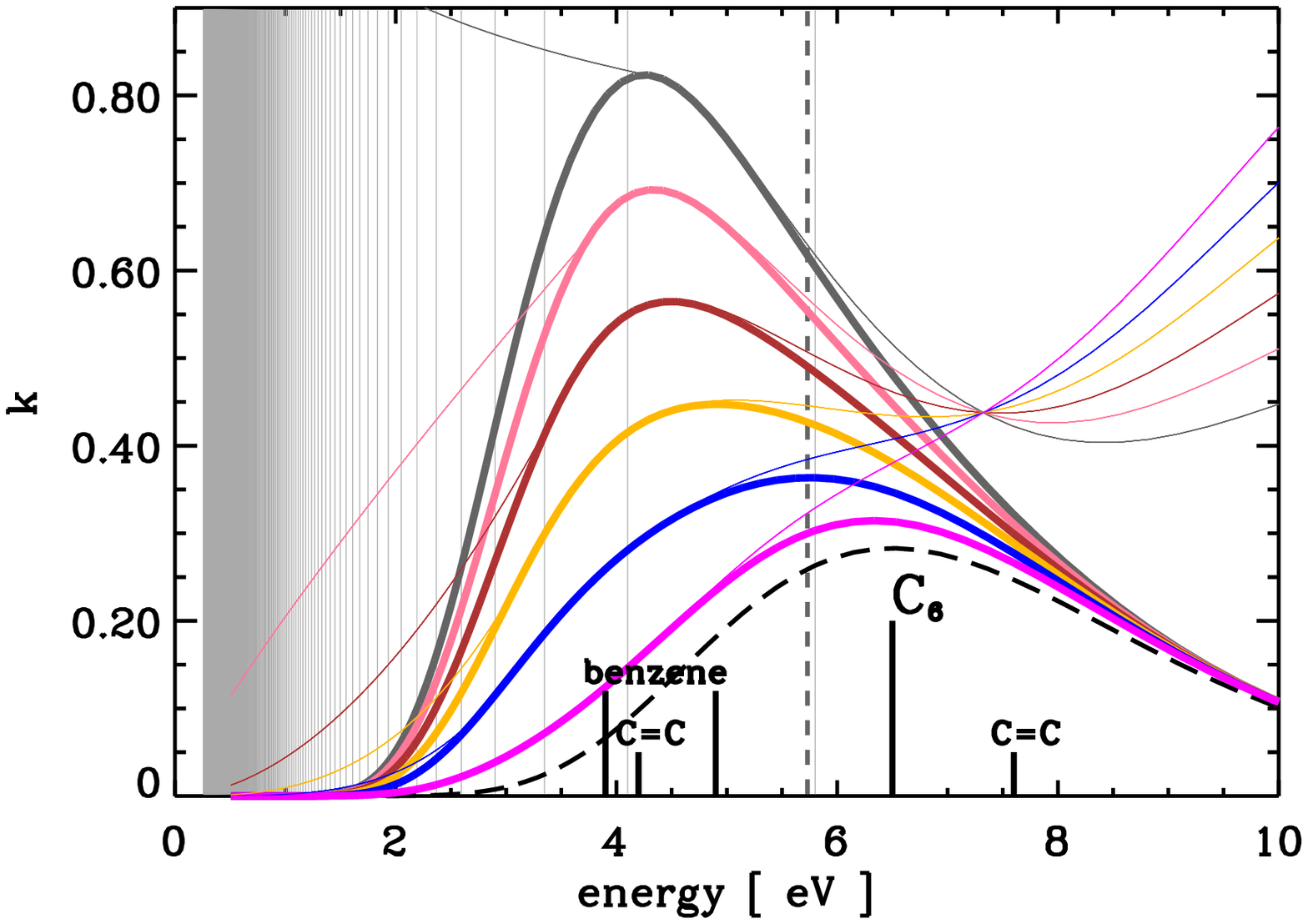}}
 \resizebox{\hsize}{!}{\includegraphics{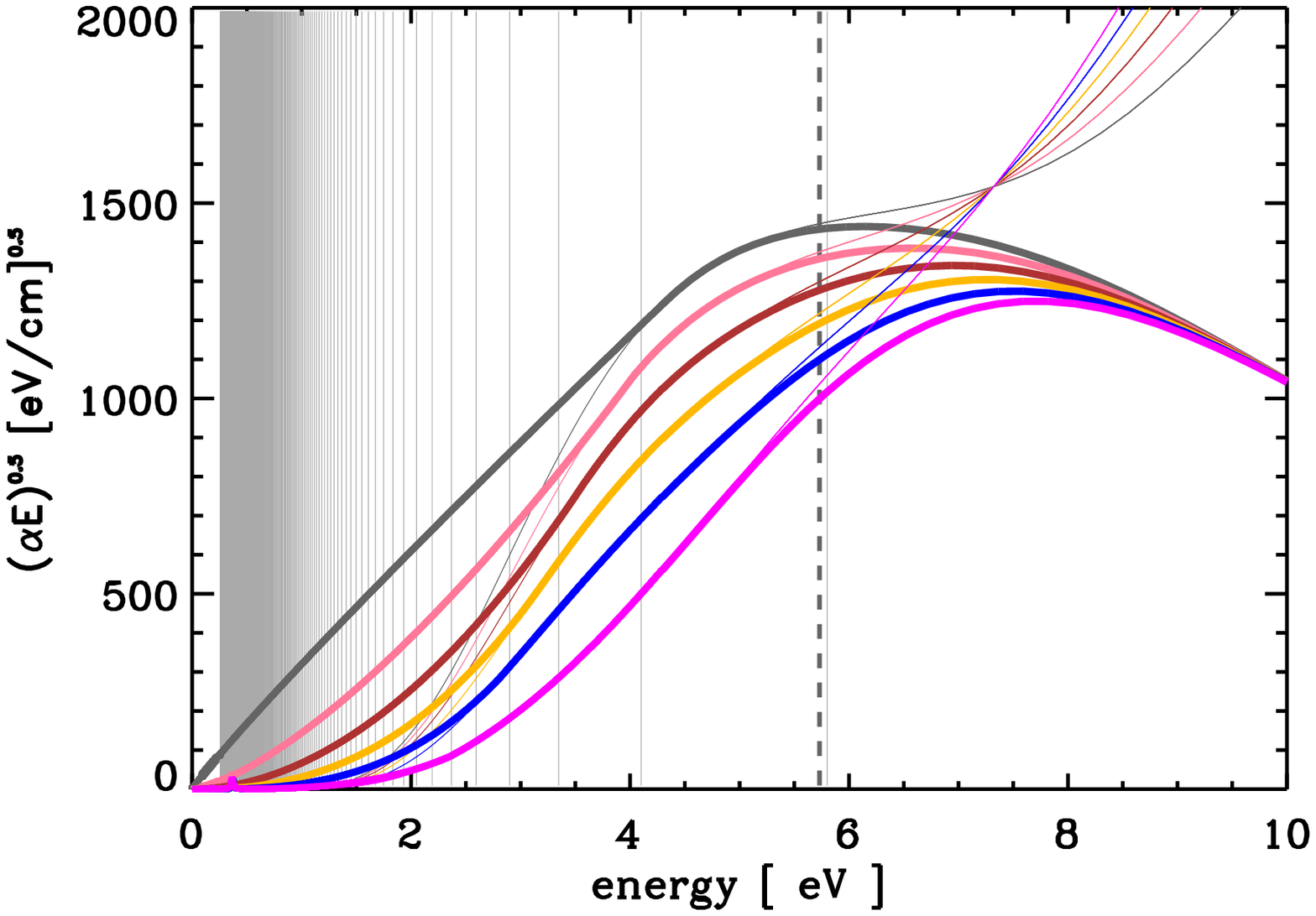}}
 \caption{Upper figure: the aromatic component contribution to the imaginary part of the refractive index, $k$, for the optEC$_{(\rm s)}$ model materials with $E_{\rm g} = 0, 0.5, 1.0, 1.5, 2.0$ and 2.5\,eV (dark grey, pink, brown, yellow, blue  and purple lines, from top to bottom, respectively). The dark grey long-dashed curve indicates the adopted C$_6$ band profile peaking at 6.5\,eV, as indicated. Lower figure: the same data presented as $(\alpha E)^{0.5}$ but with the low energy extrapolation now included. See text for details.}
 \label{fig_e2_aromatics}
\end{figure}

Fig.~\ref{fig_e2_aromatics} shows the aromatic component, {\rm i.e.}, the sum of the $\pi$--$\pi^\star$ and C$_6$ bands, contribution to the imaginary part of the index of refraction (upper figure; $\pi$--$\pi^\star$ + C$_6$) and in a Tauc plot (lower figure; $\pi$--$\pi^\star$ + C$_6$ with the low-energy addition given by Eqs.~(21) and (22) in paper~II). 
Also indicated by the vertical grey lines are the band gap positions for aromatic clusters as a function of $N_R$ ($= 1, 2, 3, 4, 5, \ldots$ from right to left) as predicted by Eq.~(\ref{Eg4XH}). The position of the UV extinction bump at 217.5\,nm (the vertical, dashed, grey line) is also indicated, which lies very close to the predicted ``band gap'' for $N_R = 1$, {\it i.e.}, a C$_6$ or benzene-like aromatic cluster. 

The $\pi$--$\pi^\star$ band has been modelled as arising from a power-law distribution of aromatic ring systems. The maximum possible size for the aromatic clusters, $N_R$(max), is determined using the values expected for a bulk material \cite[{\it e.g.}, Eq.~(2) in paper~II and/or Fig.~23 from][]{1986AdPhy..35..317R} or their size-limited values, $N_R(a)$, which can be obtained by solving for $N_R$ in Eqs.~(3) and (4) given in paper~II. For $E_{\rm g} < 0.5$\,eV and for the smallest considered particles, {\it i.e.}, for $a = 0.33$\,nm with $\approx 40$ C atoms, $N_R$(max) $ = 6$. Steric considerations imply that the aromatic clusters should principally consist of ``isolated'' two- and three-ring systems containing about three quarters of the carbon atoms.  The remaining carbon atoms must be in olefinic and aliphatic bridging structures such as --CH=CH-- and --CH$_2$--, and these must be mostly in the form of short aliphatic chains with paired carbon atoms, {\it i.e.}, --CH$_2$--CH$_2$--. Indeed, the spectra of sub-nm, H-poor a-C(:H) particles (with $E_{\rm g} = 0.1-0.5$\,eV) in Figs.~\ref{fig_spectra_0.5nm} to \ref{fig_spectra_0.33nm}  (see Sect.~\ref{sect_3mic_spectra}) clearly reveal only aromatic CH and aliphatic CH$_n$ bands, which seem to be consistent with this hypothesis.  

\begin{figure} 
 \resizebox{\hsize}{!}{\includegraphics{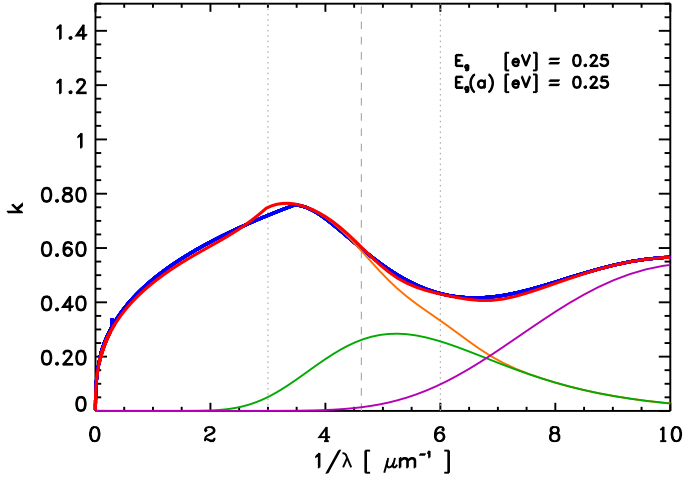}}
 \resizebox{\hsize}{!}{\includegraphics{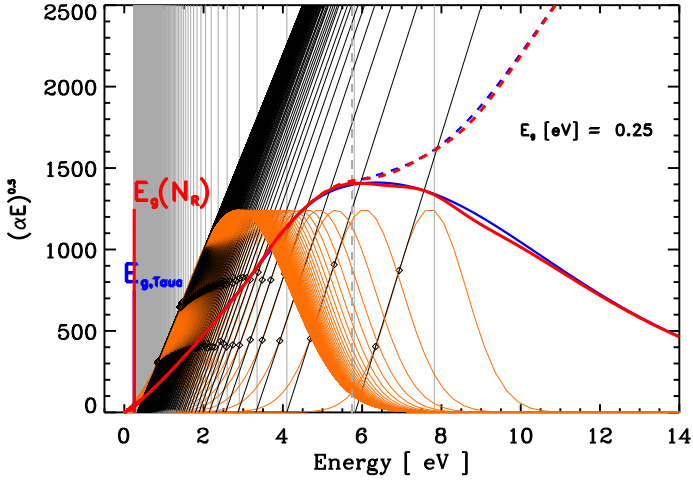}}
 \caption{Upper figure: the $\pi-\pi^\star$ and low energy adjustment ($L+\pi-\pi^\star$, orange), C6 (green) and $\sigma-\sigma^\ast$ (purple) component contributions to the imaginary part of the refractive index, $k$ (blue line), for the optEC$_{\rm(s)}$(a) model material with $E_{\rm g}{\rm (bulk)} = 0.25$\,eV and particle radius of 3\,nm 
 (The red line shows the summed $L+\pi-\pi^\star$, C6 and $\sigma-\sigma^\ast$ contributions). 
 Lower figure: the same data as in the upper figure  presented as $(\alpha E)^{0.5}$ (the blue and red lines, where the solid lines show only the summed aromatic, $L+\pi-\pi^\star$ and C6 components) but where the aromatic cluster band profiles are shown for $N_R = 2, 3, 4,$ \ldots (thin orange line profiles right to left), the extrapolations to band gap (black lines with the band gap-determining extrapolation data points indicated by the open diamond symbols), the band gap values (thin grey vertical lines), the limiting $E_{\rm g}(N_R)$ value and the bulk material band gap or Tauc gap, $E_{\rm g,Tauc}$. See text for details.}
 \label{fig_aromatics_00}
\end{figure}
\begin{figure} 
 \resizebox{\hsize}{!}{\includegraphics{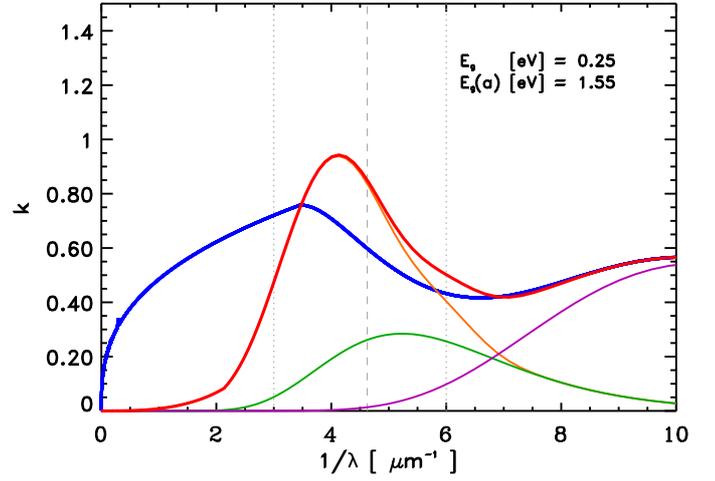}}
 \resizebox{\hsize}{!}{\includegraphics{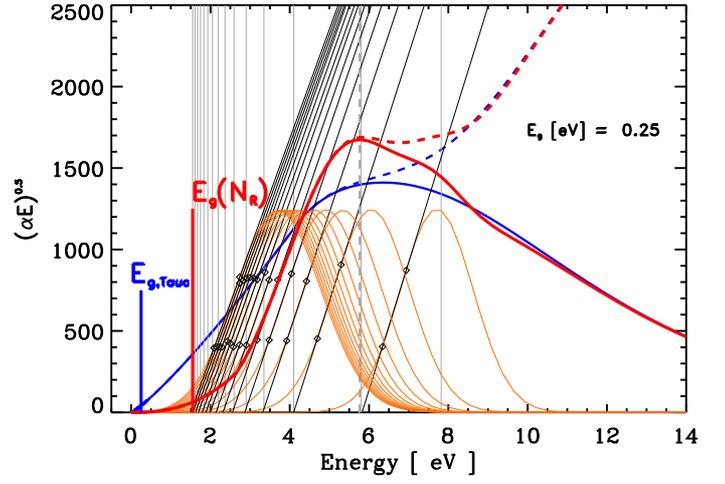}}
 \caption{Upper figure: the $L+\pi-\pi^\star$, C6 and $\sigma-\sigma^\ast$ component contributions to the imaginary part of the refractive index, $k$, for  $E_{\rm g}{\rm (bulk)} = 0.25$\,eV and radius 0.5\,nm.    Lower figure: the same data presented as $(\alpha E)^{0.5}$. See text and previous figure for details.The black lines show the aromatic cluster profile extrapolations (determined by the open diamond data points) to the band gap values (thin grey vertical lines).}
 \label{fig_aromatics_01}
\end{figure}

Paper~II gives the derivation of the imaginary part of the complex index of refraction, $k$, using: 
\begin{equation}
k(E,E_{\rm g} ) = \sum_i \sigma_{{\rm C},i}(E,E_{\rm g}) \ N_{{\rm C},i}(E_{\rm g}) \ \bigg\{ \frac{ h\, c }{ 4 \pi \, E } \bigg\}, 
\label{eq_k_decomposition}
\end{equation}
where $\sigma_{{\rm C},i}(E,E_{\rm g})$ is the specific carbon atom cross-section and $N_{{\rm C},i}(E_{\rm g})$ is the number of carbon atoms per unit volume.  In fitting the size-dependent $\pi-\pi^\star$ contribution to $k$ the material band gap needs to be taken into account and it is therefore convenient to fit the $k$ data in the Tauc form, {\it i.e.}, 
\begin{equation}
\Sigma_{\rm Tauc} = \bigg\{ \left( \frac{ 4 \pi \, k_{\pi-\pi^\star} }{ \lambda} \right) E \bigg\}^{0.5}= ( \alpha_{\pi-\pi^\star} \, E)^{0.5}, 
\label{eq_k_pipistar_Tauc}
\end{equation}
and then derive $k$ from,
\begin{equation}
k_{\pi-\pi^\star} = \frac{ \lambda  \, \Sigma_{\rm Tauc}^2 }{ 4 \pi \, E} 
\label{eq_k_pipistar}
\end{equation}
The $\pi-\pi^\star$ band contribution to the cross-section, $\sigma_{{\rm C},\pi-\pi^\star}$, is given by 
\begin{equation}
\sigma_{{\rm C},\pi-\pi^\star} = \sum_{N_{\rm R}=1}^{N_{\rm R}(a)} n(N_{\rm R}) \, V_{\rm C}  \, \sigma_{\rm C} \, S_{\pi-\pi^\star}, 
\label{eq_k_contrib_xsect}
\end{equation}
where $n(N_{\rm R})$ is the relative abundance of aromatic ring systems with $N_{\rm R}$ rings, $V_{\rm C} = 10^{23} \, f_{\rm arom.}(E_{\rm g})$ is the number of carbon atoms per unit volume, $f_{\rm arom.}(E_{\rm g})$ is the fraction of $sp^2$ carbon atoms in aromatic clusters, $\sigma_{\rm C}$ is the assumed aromatic carbon atom cross-section ($\simeq 10^{-18}$\,cm$^2$) and $S_{\pi-\pi^\star}$ is a scaling factor (see Table~\ref{table_params_colour_code}).
The aromatic cluster size distribution has been modelled using the following power law description, 
\begin{equation}
n(N_{\rm R}) = \frac{ N_{\rm R}^{-p} \ n_{\rm C}(N_{\rm R}) }{ \sum n_{\rm C,arom.} },
\label{eq_aromatics_size_dist}
\end{equation}
where the power law exponent, $p$, varies from 2.5 to 3.55 for the highest to lowest band gap materials, respectively, and  $n_{\rm C}(N_{\rm R})$ is the number of carbon atoms in an aromatic cluster with $N_{\rm R}$ rings (see Eq.~3 in paper~II). Note that the cluster abundance is normalised by $\sum n_{\rm C,arom.}$, the total number of carbon atoms in aromatic clusters per unit volume. For the smallest aromatic cluster, $N_{\rm R} = 1$, a benzene-like ring, a smaller relative abundance, before normalisation, of 0.08 (rather than unity) is assumed because the contribution of this structure is already included, in part, within the C6 band. 

For the time being, and until laboratory data give better constraints, ``reasonable'' Gaussian profiles have been assumed for the aromatic cluster band profiles because the exact form of the bands is unknown. Lorentz and Drude band profiles give wings that are too strong and do not allow a fit to the low-energy behaviour. 

The derived Gaussian band peak energy for each given aromatic cluster is given by  
\begin{equation}
E_0 =  \{ \ 5.8(N_{\rm R})^{-0.5} + 1.8) \ \} \ {\rm eV}, 
\label{eq_gaussian_E0}
\end{equation}
and for the width, $\sigma_{\rm G}$, a band width for a cluster of given $N_{\rm R}$ is assumed so as to yield the correct band gap for that cluster,  {\it i.e.}, $E_{\rm g}(N_{\rm R}) = 5.8 N_{\rm R}^{-0.5}$ (as per Eq.~\ref{eq_NR_Eg}). 
With this imposition, a fit was obtained using 
\begin{equation}
\sigma_{\rm G} = 0.650 \, N_{\rm R}^{0.08}. 
\label{eq_gaussian_sigma}
\end{equation}
The band gap for the given $N_{\rm R}$ cluster is obtained by extrapolation of the band profile at energies $(E_0-0.6)$\,eV and $(E_0-1.2)$\,eV (indicated by the diamonds in Figs.~\ref{fig_aromatics_00} and \ref{fig_aromatics_01}), using the above-described width expression, so as to be consistent with the required value.
The resulting aromatic cluster band gaps (vertical grey lines), band shapes (orange gaussian profiles) and the band gap extrapolations (sloping black lines) are shown in the lower panels of Figs.~\ref{fig_aromatics_00} and \ref{fig_aromatics_01}. 

\begin{table*}
\caption{The optEC$_{\rm (s)}$(a) material band gap, $E_{\rm g}$, fraction of $sp^2$ C atoms in aromatic clusters, $f_{\rm arom.}$, maximum number of rings per aromatic cluster, $N_{\rm R}$(max), aromatic cluster power-law, $p$, scaling factor, $S_{\pi-\pi^\star}$, hydrogen atom fraction, $X_{\rm H}$, and colour coding scheme.}
\begin{center}
\begin{tabular}{lcccccll}
                                                              &                           &                     &         &                                 &                          &               &             \\[-0.35cm]
\hline
\hline
                                                              &                           &                      &         &                                &                         &                &             \\[-0.35cm]
 $E_{\rm g}$  [ eV ]                               &  $f_{\rm arom.}$ & $N_{\rm R}$(max) &  $p$ &  $S_{\pi-\pi^\star}\times 10^6$ &    $X_{\rm H}$  &   colour  &               \\[0.05cm]
\hline
                                                             &                            &                      &          &                                &                        &                &              \\[-0.35cm]
   \hspace*{-0.3cm} $-0.1$  [$E_{g-}$]  &       1.00             &     $10^6$      &  3.55 &  1.70  &       0.00   &   black              &                             \\
    0.0                                                    &       1.00              &    $10^5$      &  3.53 &   1.60  &        0.00            &   dark grey        &                   \\
    0.1                                                    &       0.96              &    $10^4$      &  3.53 &   1.57  &        0.02             &   mid grey        &                    \\
    0.25                                                  &       0.80              &    $10^3$      &  3.50 &   1.90  &        0.05             &   lightgrey        &   $|$             \\
    0.5                                                    &       0.60              &      120          &  3.40 &   2.20  &        0.11             &   pink               &   $|$  a-C     \\
    0.75                                                  &       0.55              &         50         &  3.20 &   2.00  &        0.17             &   red                &   $|$              \\
    1.0                                                    &       0.45              &         35         &  3.03 &   2.10  &        0.23             &   brown            &                      \\
    1.25                                                  &       0.36              &         12         &  3.25 &   2.20  &        0.29             &   orange          &   $|$              \\
    1.5                                                    &       0.26              &           8         &  3.10 &   2.30  &        0.35             &   yellow            &   $|$              \\
    1.75                                                  &       0.13              &           7         &  2.93 &   3.20  &        0.41             &   green             &   $|$             \\
    2.0                                                    &       0.07              &           6         &  2.80 &   4.00  &        0.47             &   blue               &    $|$  a-C:H  \\
    2.25                                                  &       0.04              &           5         &  2.74 &   4.70  &        0.52             &   cobalt            &    $|$             \\
    2.5                                                    &       0.02              &           4         &  2.60 &   3.60  &        0.58             &   violet             &    $|$             \\
    2.67 [$E_{g+}$]                                &       0.01              &           3         &  2.50 &   1.50  &        0.62             &   purple           &     $|$            \\\hline
\hline
                                                             &                            &                      &  &   &                            &                        &                 \\[-0.25cm]
\end{tabular}
\end{center}
\label{table_params_colour_code}
\end{table*}

Table~\ref{table_params_colour_code} shows the parameters used in the aromatic cluster size distribution fitting  and the $E_{\rm g}$-$X_{\rm H}$ colour coding scheme. The fraction of $sp^2$ C atoms in aromatic clusters, $f_{\rm arom.}$, is taken from paper~I, the maximum number of rings per aromatic cluster, $N_{\rm R}$(max), is estimated from \cite[][Fig.~23]{1986AdPhy..35..317R}.

\begin{figure} 
 \resizebox{\hsize}{!}{\includegraphics{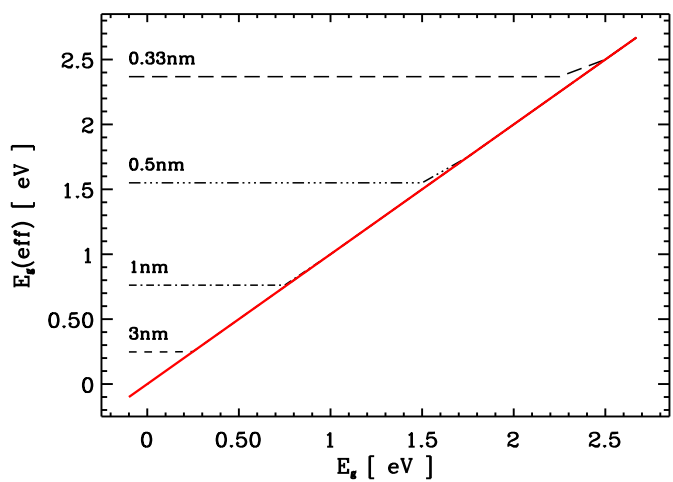}}
 \caption{The size-dependent or effective band gap, $E_{\rm g}$(eff) $\equiv E_{\rm g}\{a,N_R({\rm max})\}$, {\it vs.} the expected bulk material band gap, $E_{\rm g}$(bulk), for 0.33, 0.5, 1 amd 3\,nm radius particles. The sloping red line shows the behaviour for infinite particle size.}
 \label{fig_Eg_eff_vs_Eg}
\end{figure}

For particle radii larger than 3\,nm (for $E_{\rm g} \geqslant 0.25$\,eV), where there is no need to take into account particle size limitations on $N_R$(max), this limiting size depends only on the band gap of the material under consideration. Empirically, the $\pi-\pi^\star$-$N_R$(max) size-limitations on the particle optical properties are only of concern for 
\begin{equation}
| \, (E_{\rm g}+0.2){\rm [eV]} \times a{\rm [nm]} \, | < 1, 
\label{eq_Eg_a_emp_limit}
\end{equation}
{\it i.e.}, principally for low band gap, small particles (see Figs.~\ref{Eg_size_effects} and \ref{fig_Eg_eff_vs_Eg}). 
The aromatic cluster power-law, $p$, and the scaling factor, $S_{\pi-\pi^\star}$, can be rather well fit by the following analytical expressions,
\begin{equation}
p = 3.53 - 0.19 \times E_{\rm g}^2
\label{eq_NR_power_law}
\end{equation}
\begin{equation}
S_{\pi-\pi^\star} = \{ \ [0.8 + 0.17 \times E_{\rm g}^3] - [0.0025 \times {\rm e}^{E_{\rm g}^2}] \ \}, 
\label{eq_normalisation_factor}
\end{equation}
where $S_{\pi-\pi^\star}(E_{\rm g} = 0) = 0.8$. However, in the fitting and derivation of the $E_{\rm g}$- and size-dependent values of $k$ the values given in Table~\ref{table_params_colour_code} were used, these give a more exact representation of the bulk material data. 

Figs.~\ref{fig_aromatics_00} and \ref{fig_aromatics_01} show the results of modelling the $\pi$--$\pi^\star$ band as a power-law distribution of aromatic ring systems for the $E_{\rm g} = 0.25$\,eV and for two particle radii, $a =3$\,nm and $ a = 0.5$\,nm, where small particle effects {\em are not} and {\em are} important, respectively.  Clearly, the aromatic cluster size-distribution approach (red line) allows a very satisfactory fit to the derived $k$ data (blue line) when cluster size-limiting effects are unimportant (Fig.~\ref{fig_aromatics_00}). The principal effect to note here, as the particle size is reduced ({\it c.f.}, Figs.~\ref{fig_aromatics_00} and \ref{fig_aromatics_01}), is a decrease in $k$ at low energies ($E < 4$\,eV, $1/\lambda < 3\,\mu$m$^{-1}$) and an increase in $k$ in the UV ($E \sim 4-9$\,eV, $1/\lambda \sim 3-9\,\mu$m) resulting in a bump in the $\sim4\,\mu$m$^{-1}$ ($\sim6$\,eV) region.  At higher energies ($E > 9$\,eV), where the $\sigma-\sigma^\star$ band dominates, the behaviour of $k$ is size independent in this model.  For small particle sizes, large aromatic ring systems are not allowed and so that the optical activity associated with these small band gap entities is not present. The aromatic clusters being shifted towards lower $N_R$ systems, with higher band gaps, leads to a shift in the optical activity to higher energies.  

In order to complete the fitting of the long-wavelength behaviour of the bulk material $\pi-\pi^\star$ band, with the aromatic cluster size-distribution approach, a linear portion to $k$ is added at energies below $E^\prime_1$, as per paper~II (Sect.~4.1.2), with a size-dependent behaviour of $E^\prime_1$  given by 
\begin{equation}
E^\prime_1 = 4.0 - \left( \ \frac{E_{\rm g}}{1.2} \ \right) \ \ \ {\rm eV}, 
\label{eq_size_dep_E1prime}
\end{equation}
which is a slightly modified from of Eq.~(21) in paper~II (the constant 4.0 replaces the value of 4.5 used in paper~II). 
Then, for the size-dependent behaviour the band gap predicted by the largest-possible aromatic cluster, $E_{\rm g}\{a,N_R({\rm max})\}$, is used rather than the expected band gap for the bulk material, $E_{\rm g}$(bulk), in determining the appropriate size-dependent value of $E^\prime_1$. The relationship between the effective and bulk material band gaps is shown in Fig.~\ref{fig_Eg_eff_vs_Eg}. 

For the power-law index at long wavelengths, $\gamma$, that given by Eq.~(22) in paper~II is used, except for the the two lowest values of $E_{\rm g}$, {\it i.e.}, $-0.1$ and $0.0$\,eV, where setting $\gamma = -0.35$ and $-0.14$, respectively, gives a better to fit the bulk material data. 

The particle size-dependent adjustment of the summed $\pi$--$\pi^\star$ and C$_6$ aromatic structure contributions to the determination of $k$ imposes an effective scaling down because of the reduced effects at long-wavelengths. Therefore the $\pi$--$\pi^\star$ and C$_6$ contributions are scaled-up to ensure the conservation of $n_{\rm eff}$, {\it i.e.}, their total integrated contribution to $k$ is taken to be constant. The required $\pi$--$\pi^\star\,+$ C$_6$ enhancement factor, {\it i.e.}, $n_{\rm eff}(a)/n_{\rm eff}$(bulk), is found to be at most $\simeq 1.9$, 2.5 and 3.1 for 1, 0.5 and 0.33\,nm radius particles, with $E_{\rm g} = -0.1$\,eV, respectively.

\subsubsection{The C$_6$ band}
\label{appendix_C6}

The C$_6$ band contribution to the wide-gap optical properties, as derived in paper~II, is shown by the dark grey dashed line in the upper part of Fig.~\ref{fig_e2_aromatics}.  Given that the adopted C$_6$ band is apparently larger than would be predicted by the $E_{\rm g}$ relation (Eq.~\ref{eq_NR_Eg}), it is likely, for low-symmetry a-C(:H) materials, that there could be an important contribution from the normally weak $\pi-\sigma^{\star}$ transitions involving the promotion of electrons from bonding $\pi$ orbitals to anti-bonding $\sigma^{\star}$ orbitals. Here it is hypothesised that the C$_6$ band also contains an important contribution from olefinic $sp^2$-rich clusters as well. In this respect, note that the lowest-lying triplet (singlet) transitions lie at 4.2\,eV (7.6\,eV) for ethylene, $>$C$=$C$<$, and at 3.9\,eV (4.9\,eV) for benzene \citep{1986AdPhy..35..317R}. These triplet and singlet band positions are indicated in the upper plot of Fig.~\ref{fig_e2_aromatics}. In particular, it is postulated that conjugated olefinic chains and cyclic precursor-aromatic configurations, which upon aromatisation can transform into the aromatic cluster states with $N_R = 0$ to 3 ({\it i.e.}, benzene-, naphthalene- and anthracene-like structures) contribute to the C$_6$ band.  This effect will become weaker as $N_R$ increases and hence this is why the C$_6$ band is likely to be particularly broadened by the presence of linked, conjugated olefinic cycles.  In support of this \cite{2004PhilTransRSocLondA..362.2477F} point out that the clustering of the $sp^2$ phase into chains or even cage-like structures plays a key role in the determination of the properties of amorphous (hydro)carbon materials. 

\subsubsection{The $\sigma$--$\sigma^\star$ band}
\label{appendix_sig_sig}

The $\sigma$--$\sigma^\star$ band might also exhibit some size-dependence effects, which might perhaps manifest as a narrowing of the band as the particle size decreases. This could result from a reduction in the range of $sp^3$ C atom structural configurations that would then lead to a profile more characteristic of polyethylene, (CH$_2$)$_n$, or even diamond. This work does not consider this effect but it would be interesting to see if any such a change is seen in laboratory data. However, in order to explain this, the experimental conditions would require the isolation of a-C(:H) nano-particles ($a < 3$\,nm) with a range of H atom concentrations.

\section{The size-dependent optical properties of a-C(:H) materials}
\label{appendix_n_and_k}

\begin{figure} 
 \resizebox{\hsize}{!}{\includegraphics{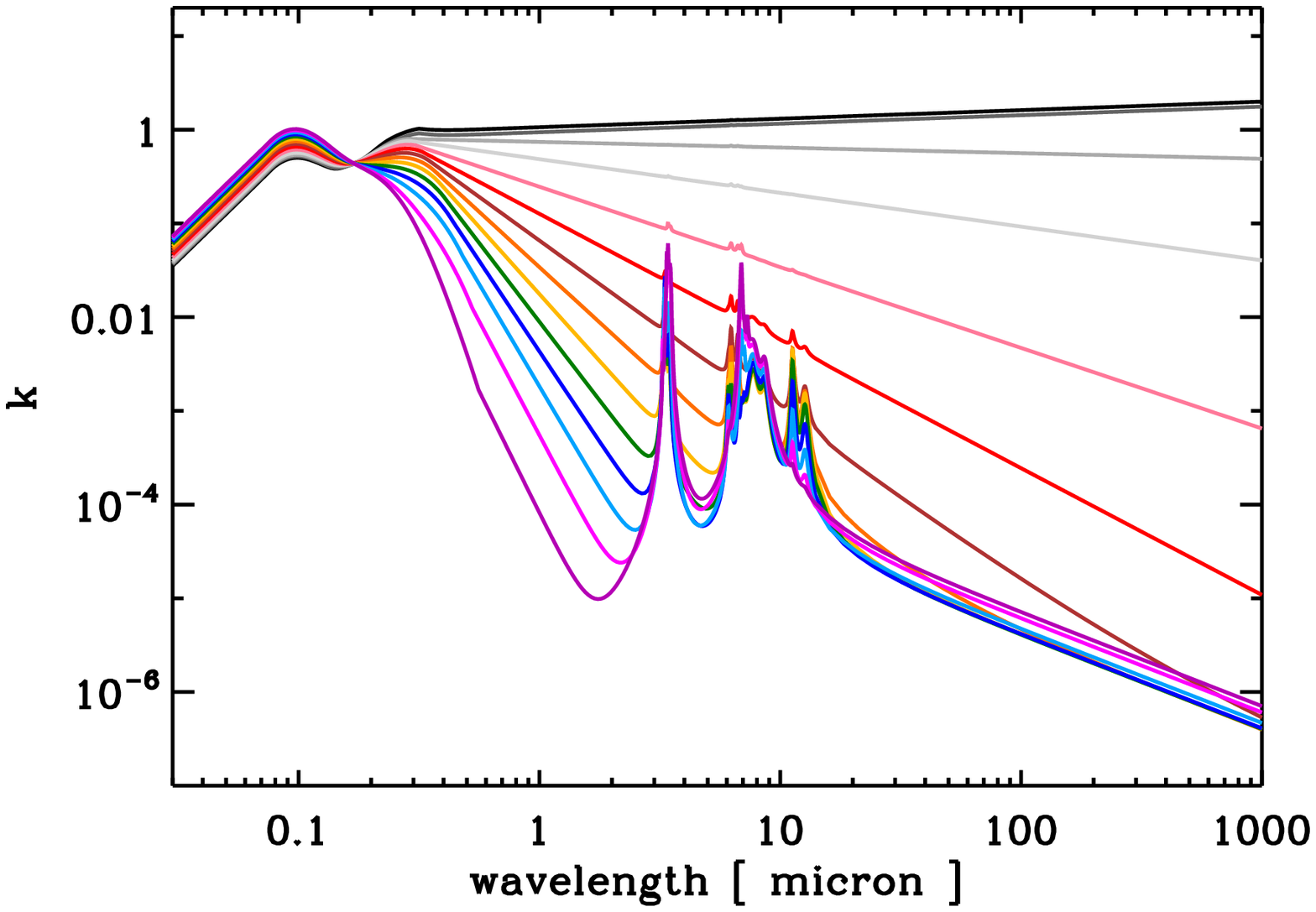}}
 \resizebox{\hsize}{!}{\includegraphics{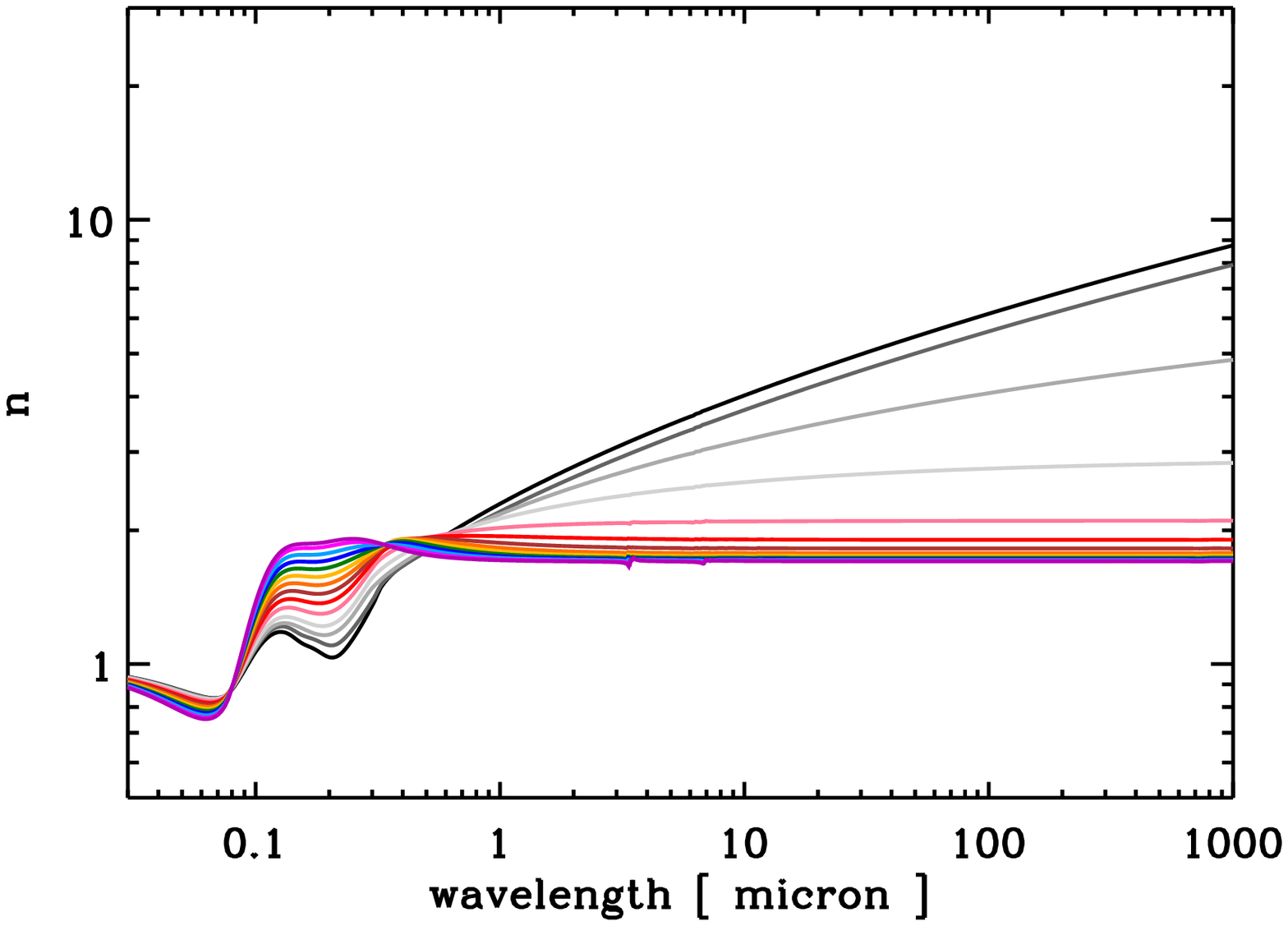}}
 \caption{The real part, $n$ (upper), and imaginary part, $k$ (lower), of the complex index of refraction, $m=n(a,E_{\rm g},\lambda)+ik(a,E_{\rm g},\lambda)$,  for  30\,nm radius a-C(:H) particles as a function of $E_{\rm g}$, as predicted by the optEC$_{\rm(s)}$(a) model presented in this paper (see Table~B.1 
 for the line colour-coding).}
 \label{fig_nk_30nm}
\end{figure}

\begin{figure} 
 \resizebox{\hsize}{!}{\includegraphics{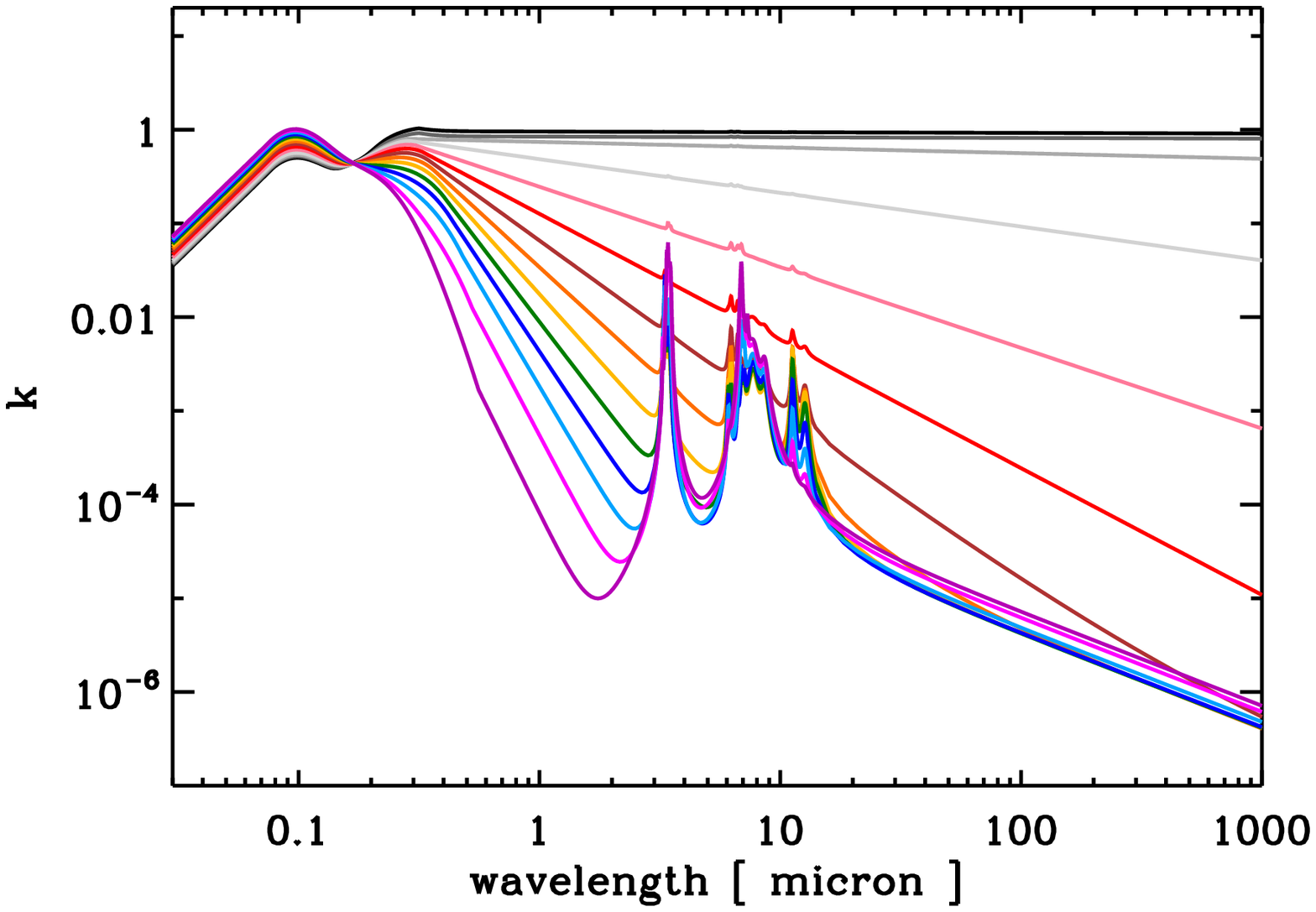}}
 \resizebox{\hsize}{!}{\includegraphics{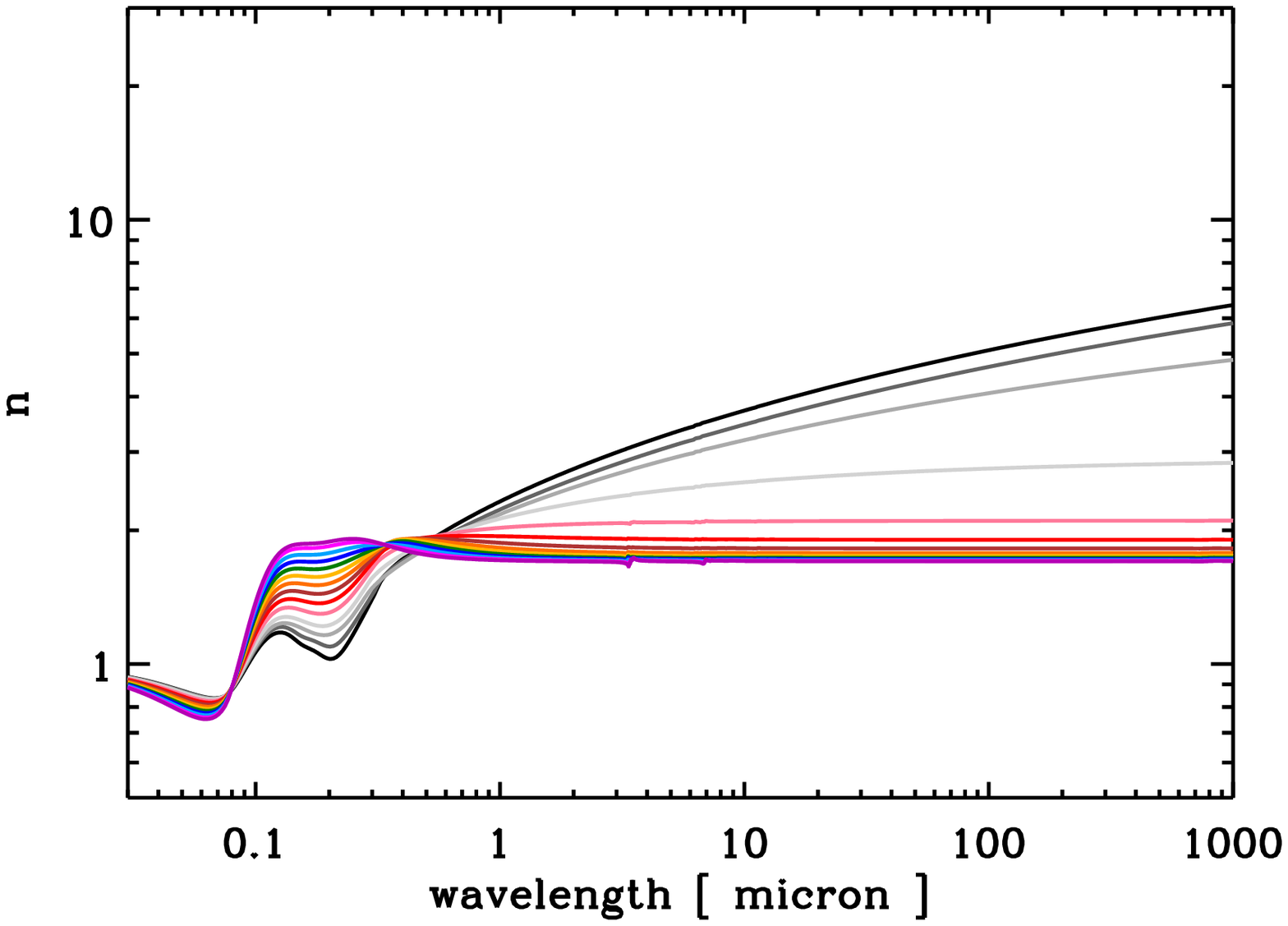}}
 \caption{As per Fig.~\ref{fig_nk_30nm} but for particles of radius 10\,nm.}
 \label{fig_nk_10nm}
\end{figure}

\begin{figure} 
 \resizebox{\hsize}{!}{\includegraphics{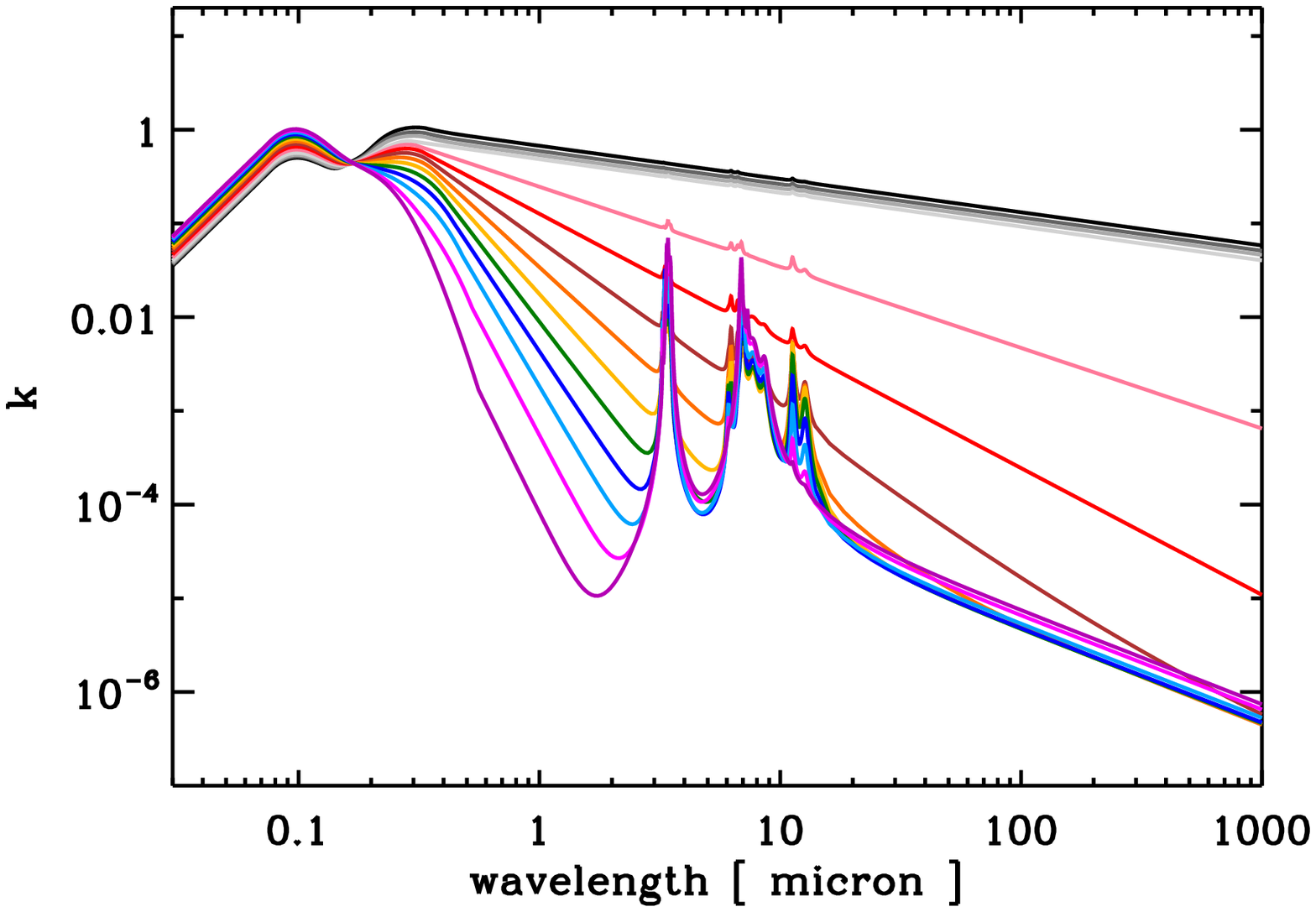}}
 \resizebox{\hsize}{!}{\includegraphics{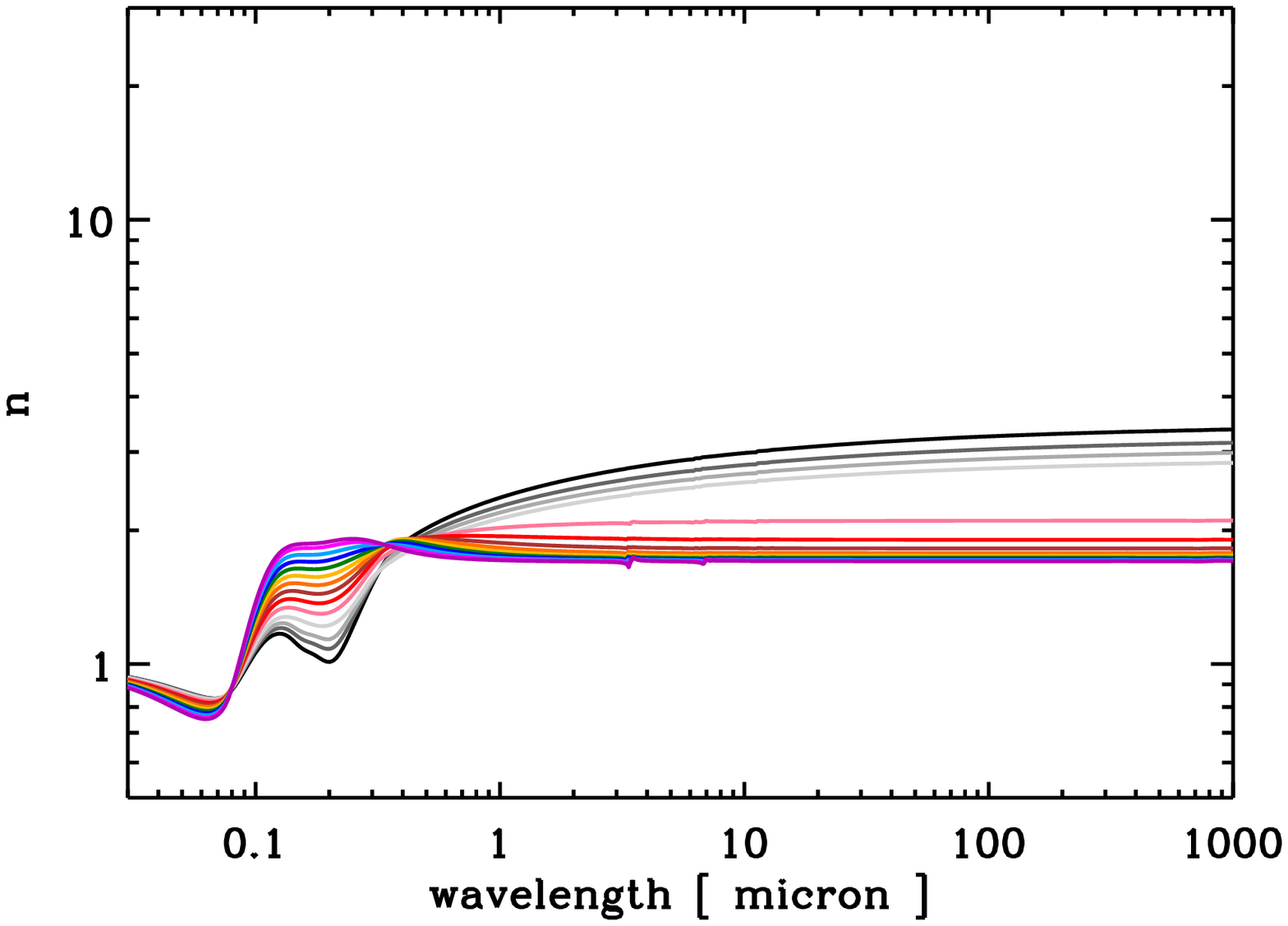}}
 \caption{As per Fig.~\ref{fig_nk_30nm} but for particles of radius 3\,nm.}
 \label{fig_nk_3nm}
\end{figure}

\begin{figure} 
 \resizebox{\hsize}{!}{\includegraphics{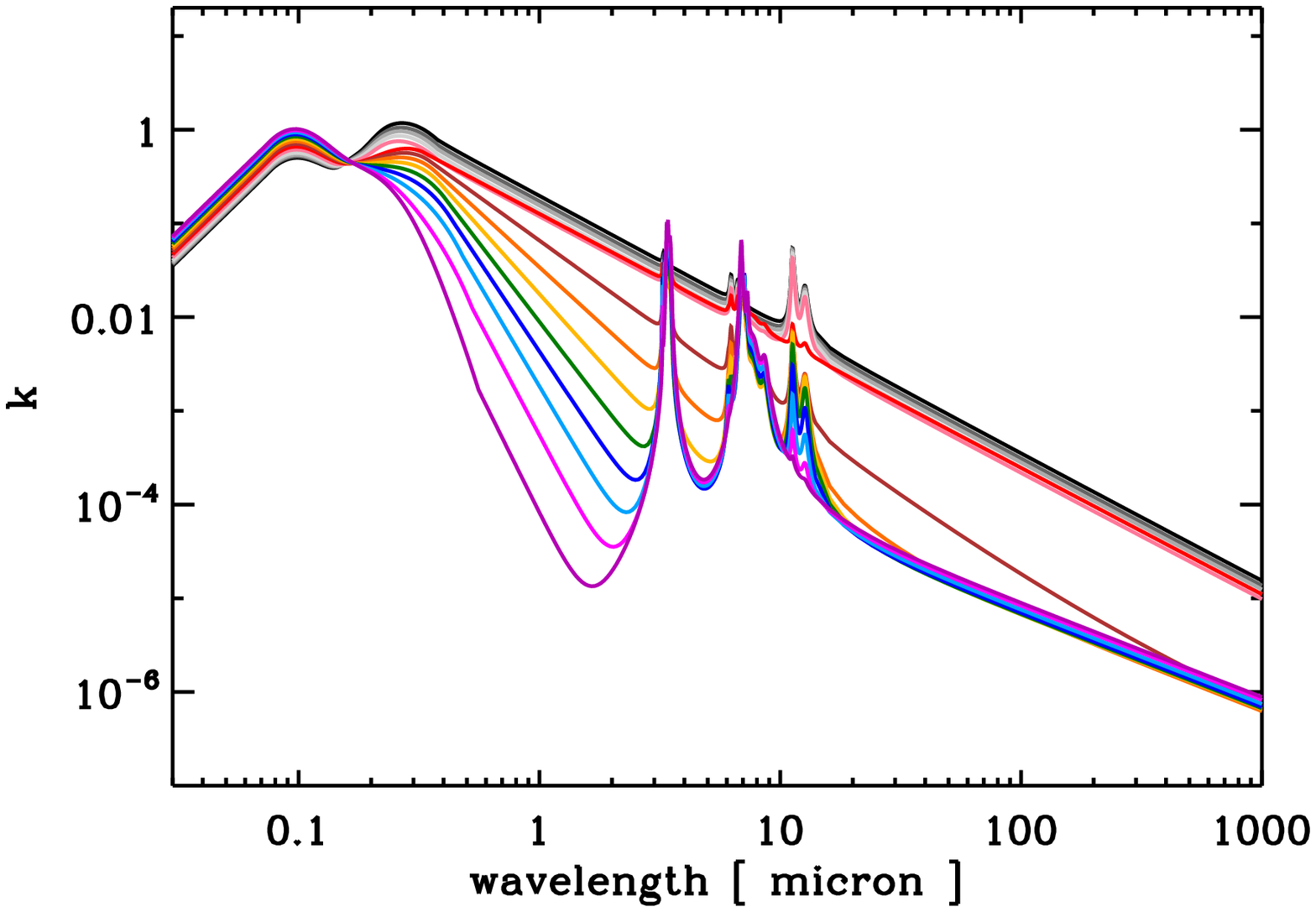}}
 \resizebox{\hsize}{!}{\includegraphics{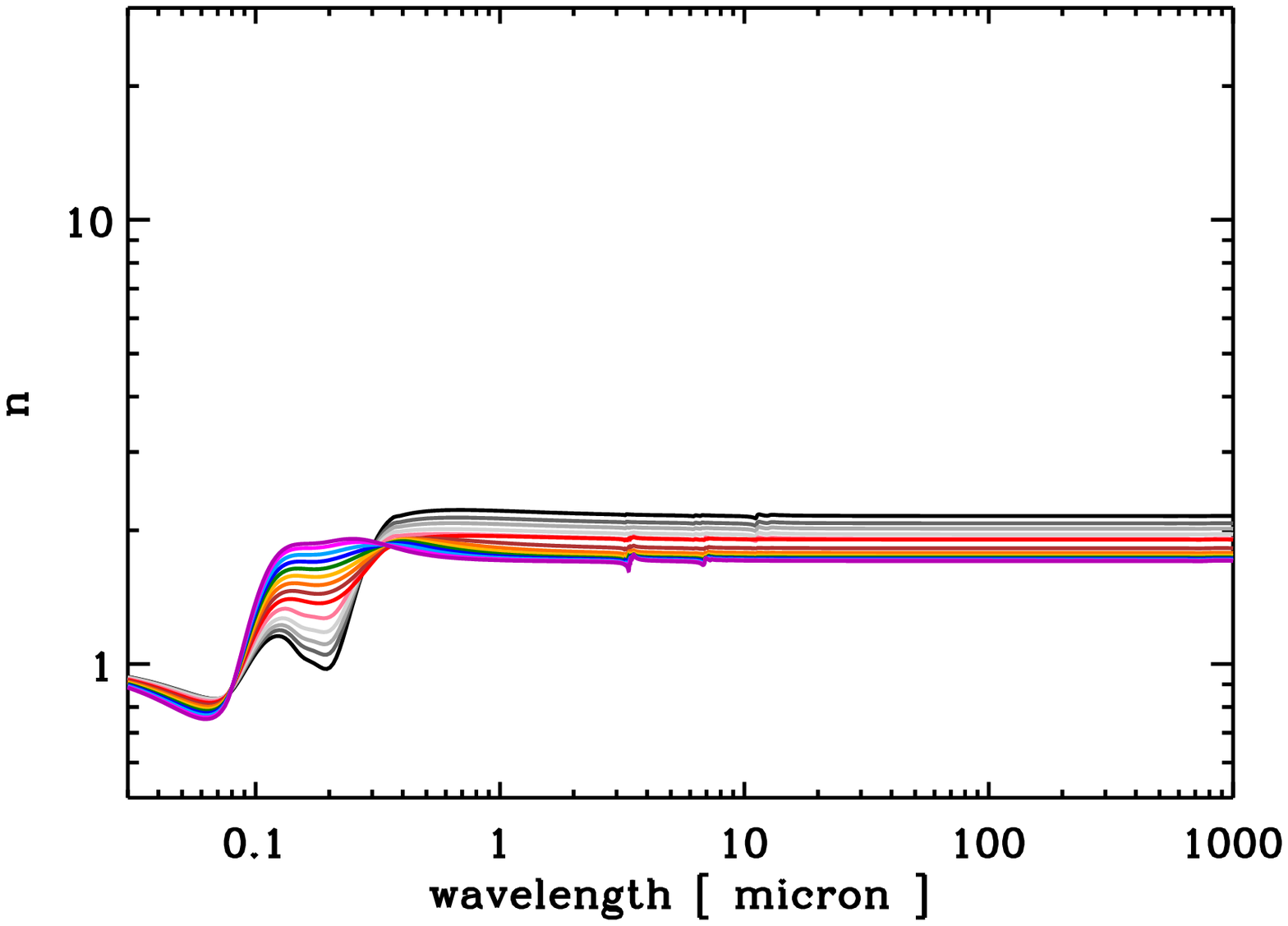}}
 \caption{As per Fig.~\ref{fig_nk_30nm} but for particles of radius 1\,nm.}
 \label{fig_nk_1nm}
\end{figure}

\begin{figure} 
 \resizebox{\hsize}{!}{\includegraphics{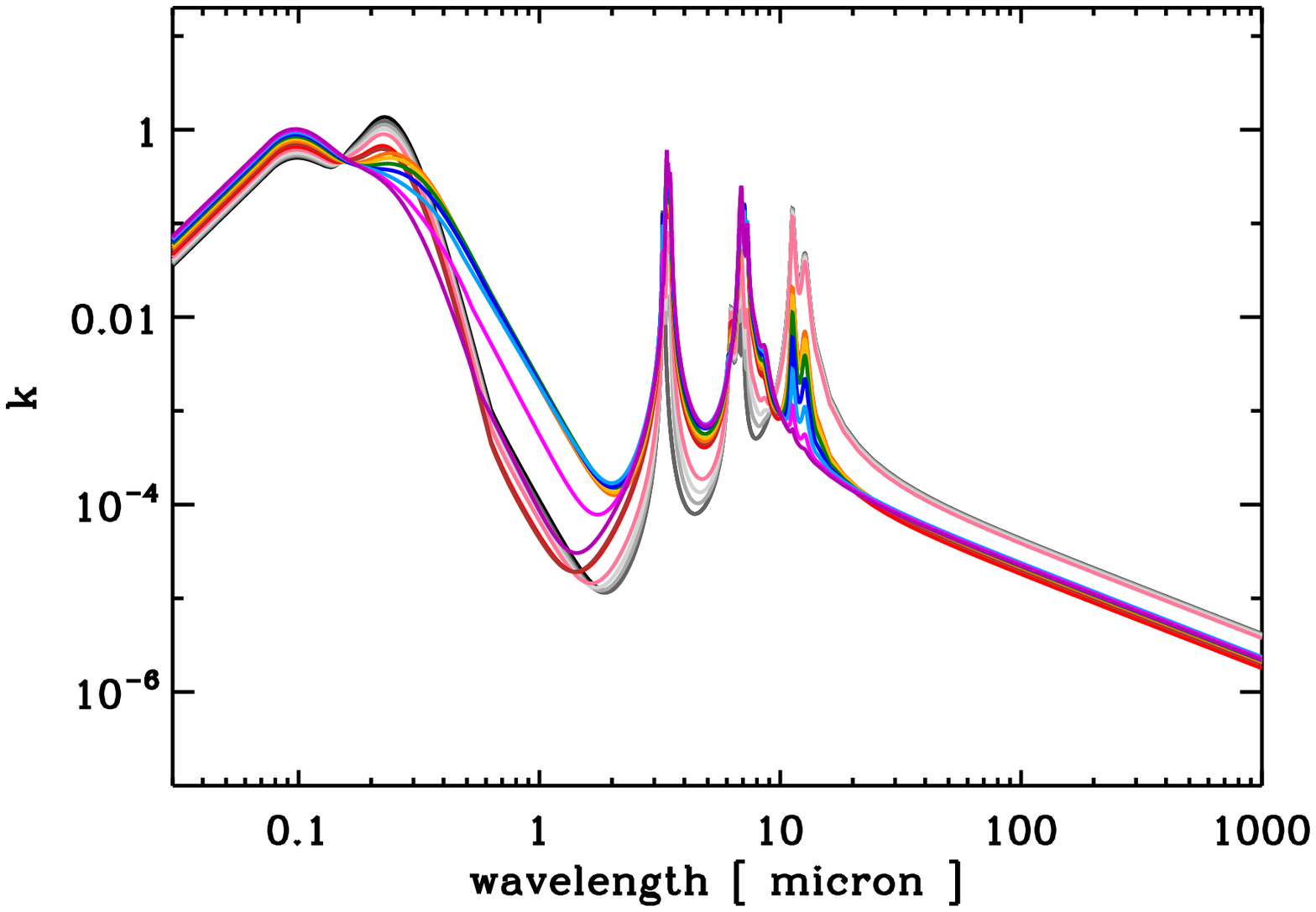}}
 \resizebox{\hsize}{!}{\includegraphics{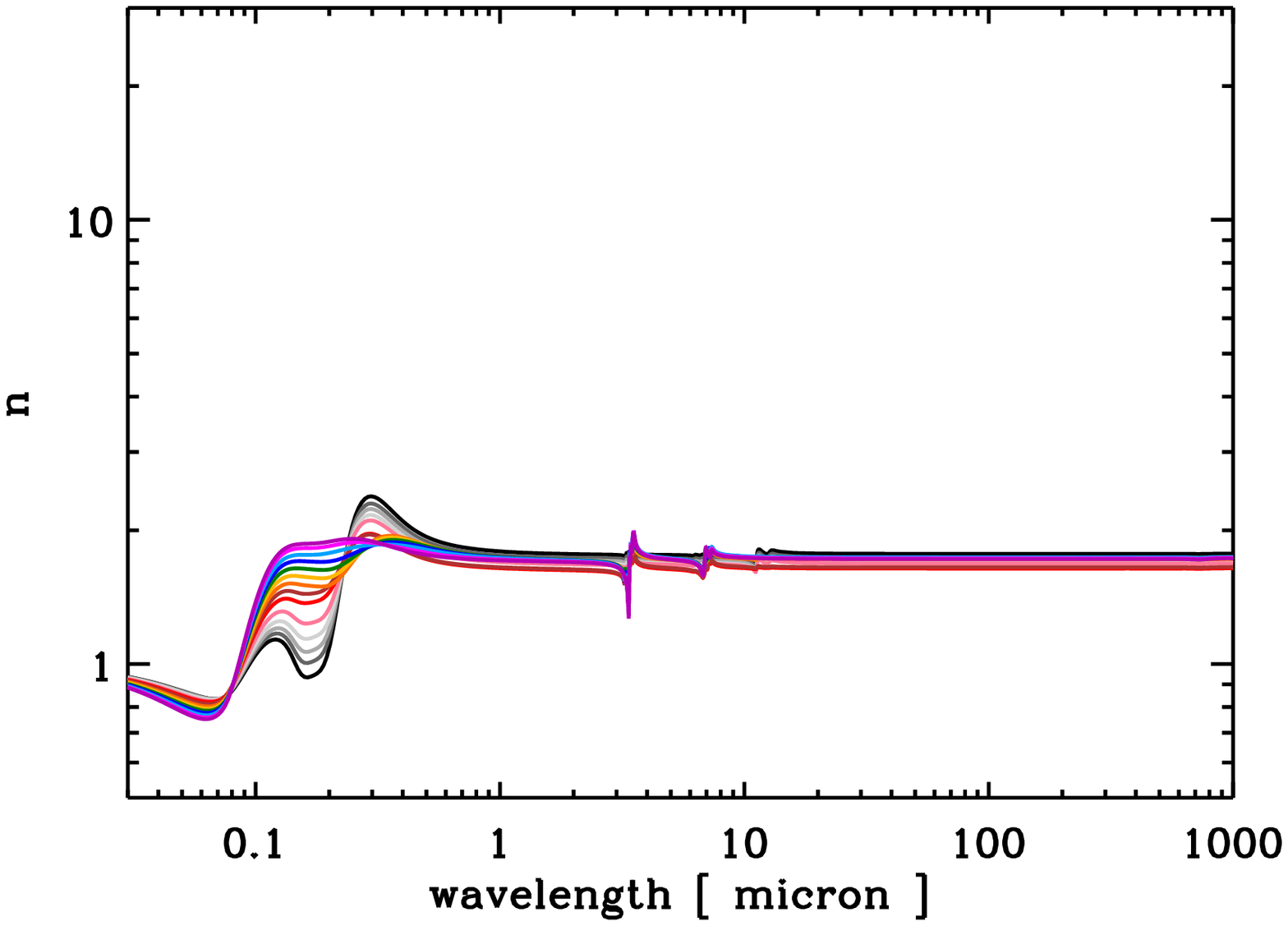}}
 \caption{As per Fig.~\ref{fig_nk_30nm} but for particles of radius 0.33\,nm.}
 \label{fig_nk_0.33nm}
\end{figure}

Figs.~\ref{fig_nk_30nm} to \ref{fig_nk_0.33nm} show the the imaginary parts ($k$: upper panels) and the real parts ($n$: lower panels) of the complex indices of refraction for a-C(:H) materials as a function of wavelength, from EUV to mm wavelengths.  Each coloured line represents one of the 14 different band gap materials ($E_{\rm g} = -0.1$ to 2.67\,eV: see Table~\ref{table_params_colour_code} for the full colour code).These figures show the data for particles with radii of 30, 10, 3, 1 and 0.33\,nm, respectively, and complement Figs.\ref{fig_nk_100nm_eg} and \ref{fig_nk_0.5nm_eg} given in Sect.~\ref{sect_optECsa_data}. Note that in all of the figures the same $x$- and $y$-axis ranges are used in order to allow a direct comparison of the various data. 

It is found that for particles with radii $\gtrsim 10$\,nm ({\it i.e.}, Figs.~\ref{fig_nk_100nm_eg} and \ref{fig_nk_30nm} to \ref{fig_nk_10nm}) that $k$ and $n$ are rather invariant but that they vary significantly for smaller particles, where the most obvious effect is a dramatic decrease in the ``continua'' at wavelengths longward of $\approx 0.5\, \mu$m. However, the underlying spectral bands remain but vary in form as indicated in the figures in Appendix~\ref{appendix_size_dep_spectra}. This is a direct result of the decreasing maximum aromatic cluster size as particle size decreases. 

Note that, as the particles become smaller, there is a ``piling-up'' or ``convergence'' effect, in that the optical properties for all band gap materials start to look the same. This is, in major part, a result of the restrictions placed on the largest possible aromatic clusters by the particle size, which leads to an opening-up of the gap with decreasing particle radius. It also indicates a transition towards more molecular properties with prominent bands and weak continua. 

%
%
%
%

\section{Size-dependent optical-UV properties}
\label{appendix_vis_UV}

\begin{figure} 
 \resizebox{\hsize}{!}{\includegraphics{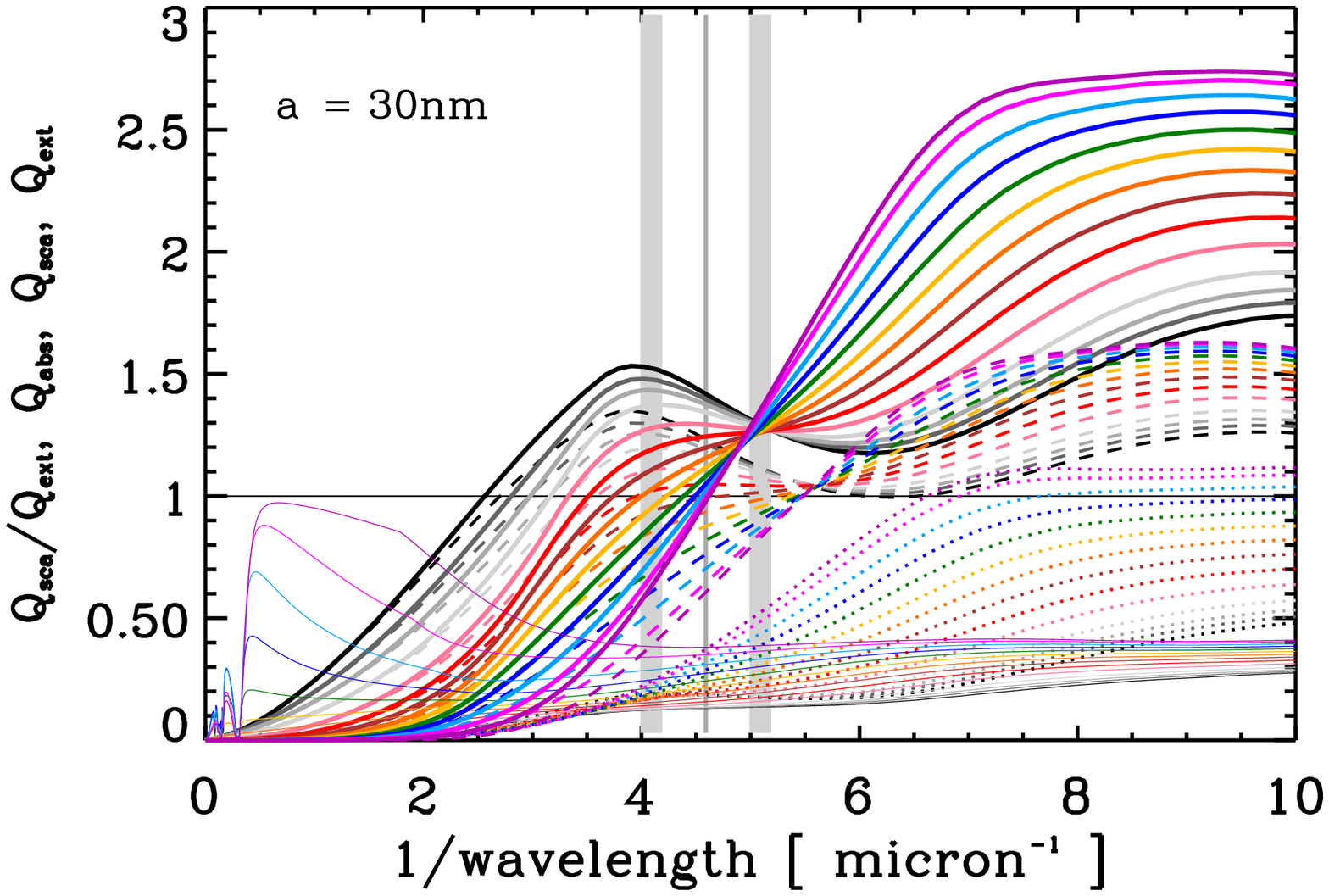}}
 \caption{Q$_{\rm ext}$ (thick solid lines),  Q$_{\rm sca}$ (dotted lines), Q$_{\rm abs}$ (dashed lines) and Q$_{\rm sca}$/Q$_{\rm ext}$ (thin solid lines) {\it vs.} $\lambda^{-1}$ for 30\,nm radius particles as a function of band gap.The vertical grey line shows the central position of the UV bump at 217\,nm and the vertical, lighter grey bands on either side indicate the full width of the observed UV bump \citep{2004ASPC..309...33F,2007ApJ...663..320F}. The thin horizontal black line shows the limit where Q$_{\rm sca}$/Q$_{\rm ext} = 1$. }
\label{fig_Qs_vs_size_2}
\end{figure}

\begin{figure} 
 \resizebox{\hsize}{!}{\includegraphics{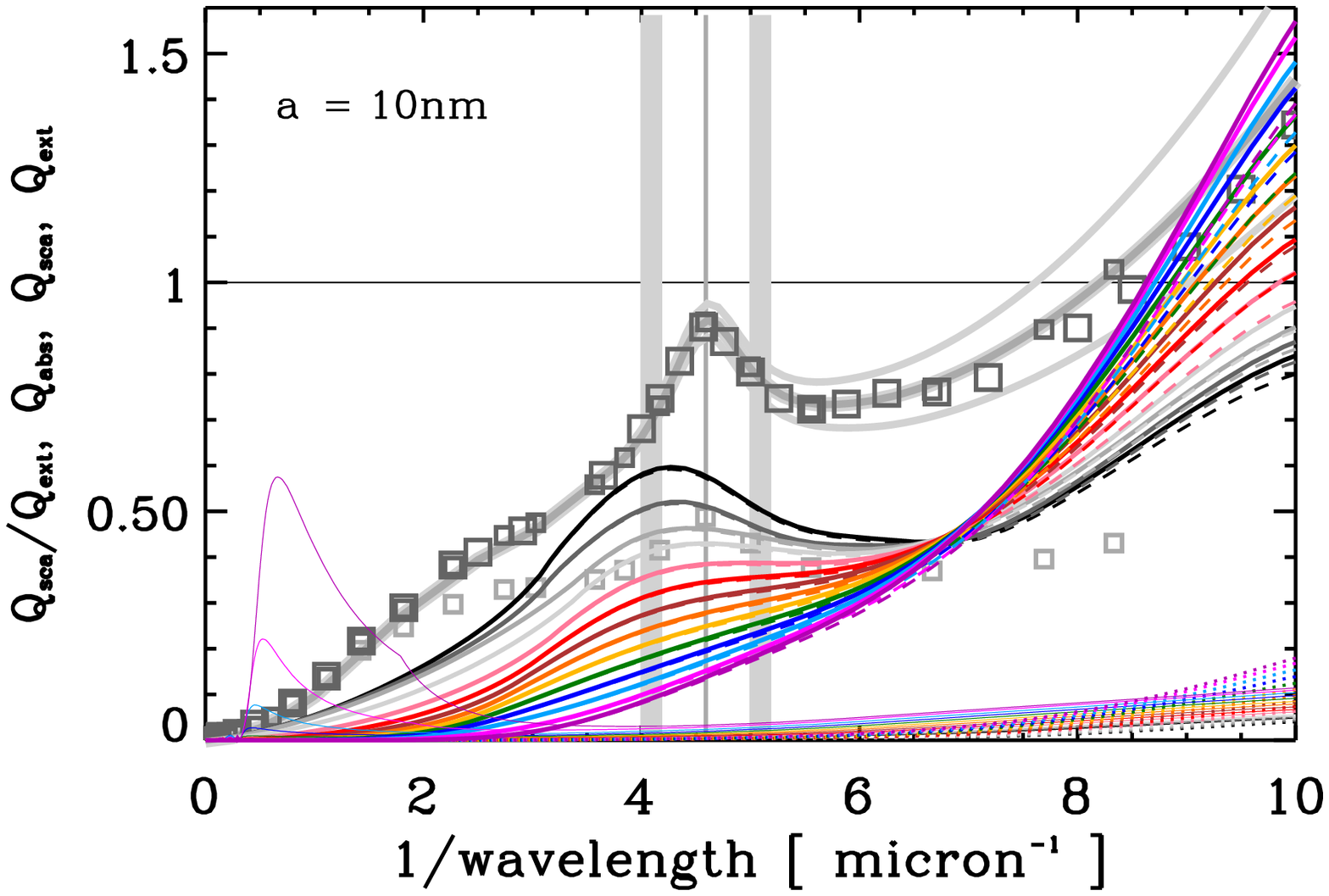}}
 \caption{Same as for Fig.~D.1 
 but for 10\,nm radius particles. The grey squares show the diffuse ISM extinction curve for $R_{\rm V} = 3.1$ (dark) and 5.1 (light) from \cite[][large squares]{1979ARA&A..17...73S} and \citep[][small squares]{1990ARA&A..28...37M}. The grey curves indicate the average galactic extinction, and its variation (upper and lower grey curves), as derived by \cite{2007ApJ...663..320F}.}
\label{fig_Qs_vs_size_3}
\end{figure}

\begin{figure} 
 \resizebox{\hsize}{!}{\includegraphics{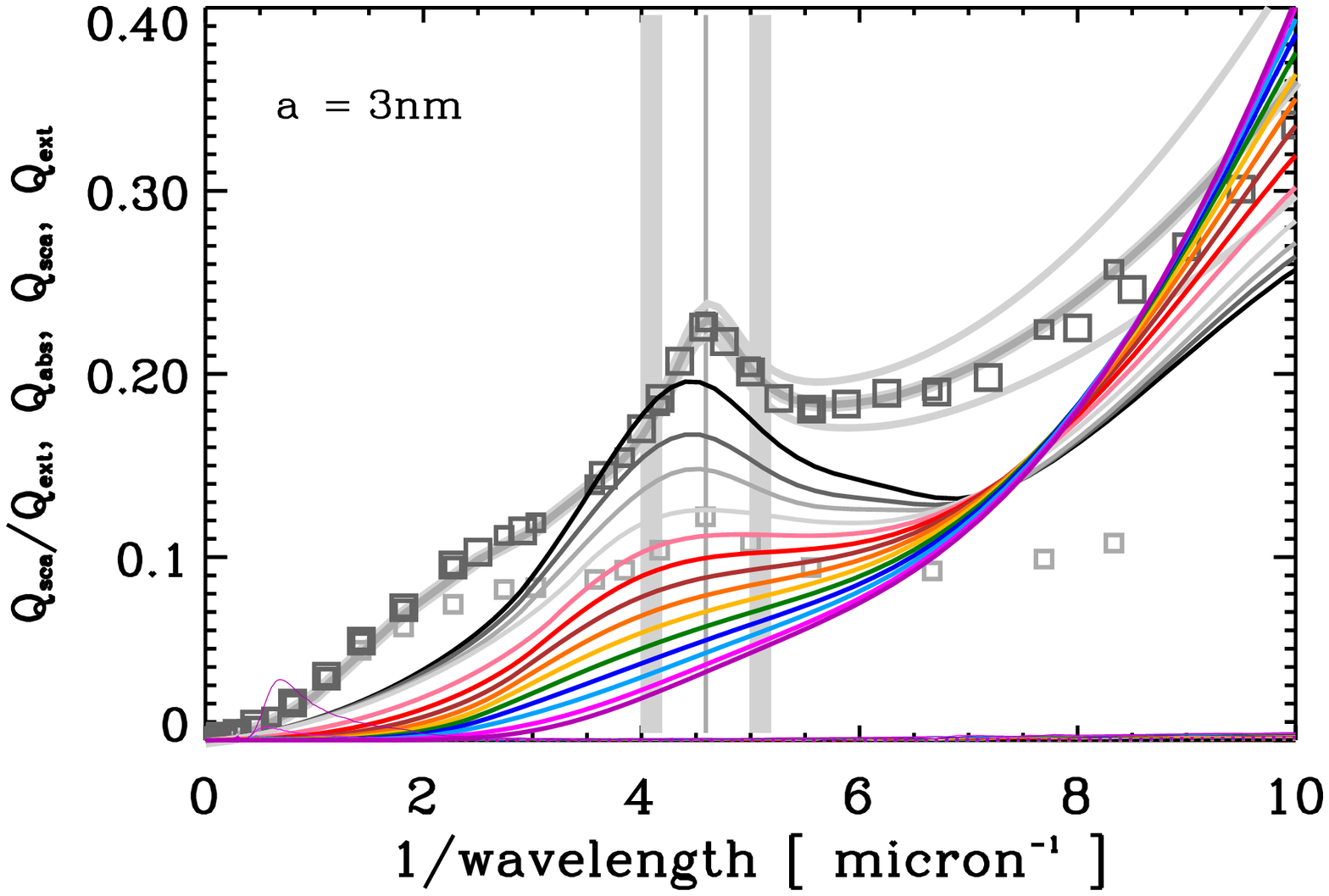}}
 \caption{Same as for Fig.~\ref{fig_Qs_vs_size_2} but for 3\,nm radius particles.}
\label{fig_Qs_vs_size_4}
\end{figure}

\begin{figure} 
 \resizebox{\hsize}{!}{\includegraphics{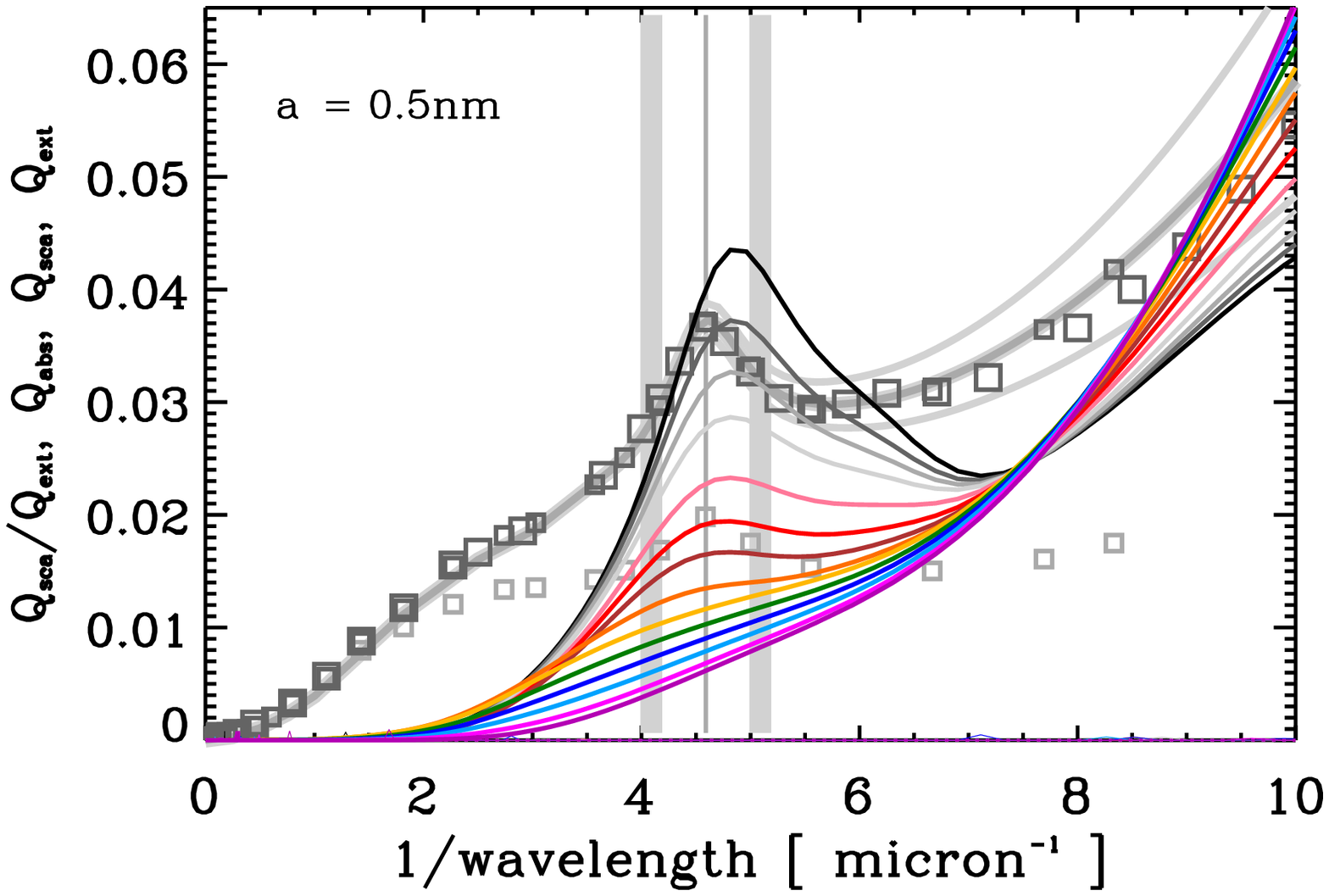}}
 \caption{Same as for Fig.~\ref{fig_Qs_vs_size_2} but for 0.5\,nm radius particles.}
\label{fig_Qs_vs_size_6}
\end{figure}

\begin{figure}
 \resizebox{\hsize}{!}{\includegraphics{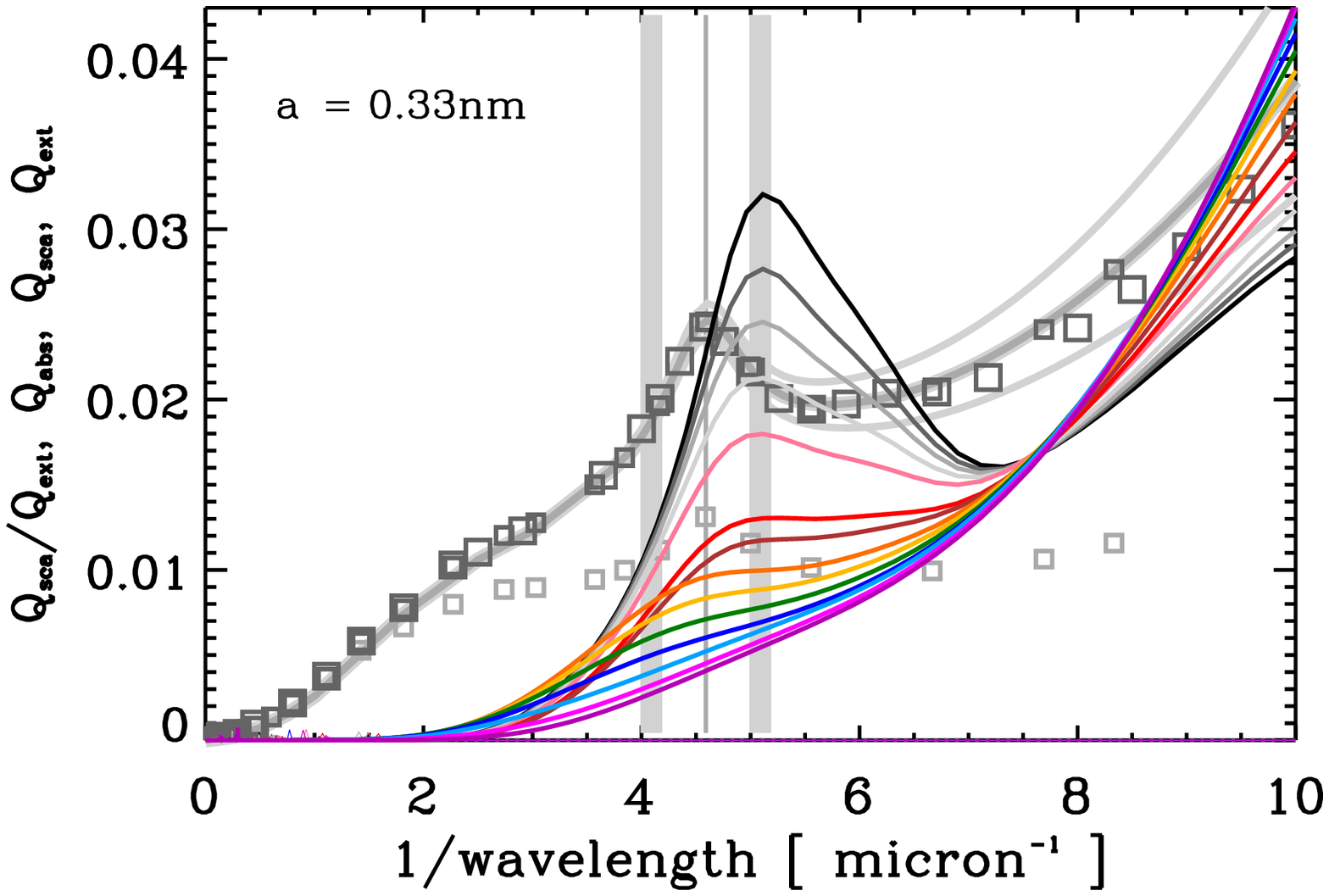}}
 \caption{Same as for Fig.~D.1 
 but for 0.33\,nm radius particles.}
\label{fig_Qs_vs_size_7}
\end{figure}

Figs.~\ref{fig_Qs_vs_size_2} to \ref{fig_Qs_vs_size_7} present Q$_{\rm ext}$ (thick solid),  Q$_{\rm sca}$ (dotted), Q$_{\rm abs}$ (dashed) and Q$_{\rm sca}$/Q$_{\rm ext}$ (thin solid), as a function of inverse wavelength, for particle radii of 30, 10, 3, 0.5 and 0.33\,nm, and for all considered band gaps. See Tables~\ref{table_colour_code} and \ref{table_params_colour_code} for the line colour-coding; from large gap, purple (2.67\,eV), to low band gap, black.  These figures complement Figs.~\ref{fig_Qs_vs_size_1_eg} and \ref{fig_Qs_vs_size_5_eg} for 100 and 1\,nm radii particles.  The thin horizontal black lines in each figure show the maximum albedo, {\it i.e.}, the limit where Q$_{\rm sca}$/Q$_{\rm ext} = 1$ and the extinction would be due to pure scattering. The vertical, dark grey band shows the observed variation in the central position of the 217\,nm UV bump and the vertical, light grey bands on either side indicate observed variation in its observed full width at half maximum \citep[{\it e.g.},][]{2004ASPC..309...33F,2007ApJ...663..320F}.

As can clearly be seen in Figs.~\ref{fig_Qs_vs_size_1_eg} and \ref{fig_Qs_vs_size_2}, the extinction of hydrogen-rich a-C:H particles ($X_{\rm H} \gtrsim 0.5$ or equivalently $E_{\rm g} \gtrsim 2$\,eV), with radii greater than 30\,nm, will be increasingly dominated by scattering over a wider wavelength region as the particle size increases. In particular, it can be seen that 100\,nm radius particles have albedos $\geq 0.5$ for $X_{\rm H} \gtrsim0.2$ ($\equiv E_{\rm g} \geq 1$\,eV) over a wavelength range of at least $\approx 0.3-1\,\mu$m ($\approx 1-3\,\mu$m$^{-1}$). For larger hydrogen contents even higher albedos are to be found and over a much wider wavelength range. 

%
%
%
%
%

\section{The size-dependent spectra of optEC$_{\rm(s)}$(a)-modelled particles}
\label{appendix_size_dep_spectra}

\begin{figure} 
 \vspace*{-2cm}
 \resizebox{\hsize}{!}{\includegraphics{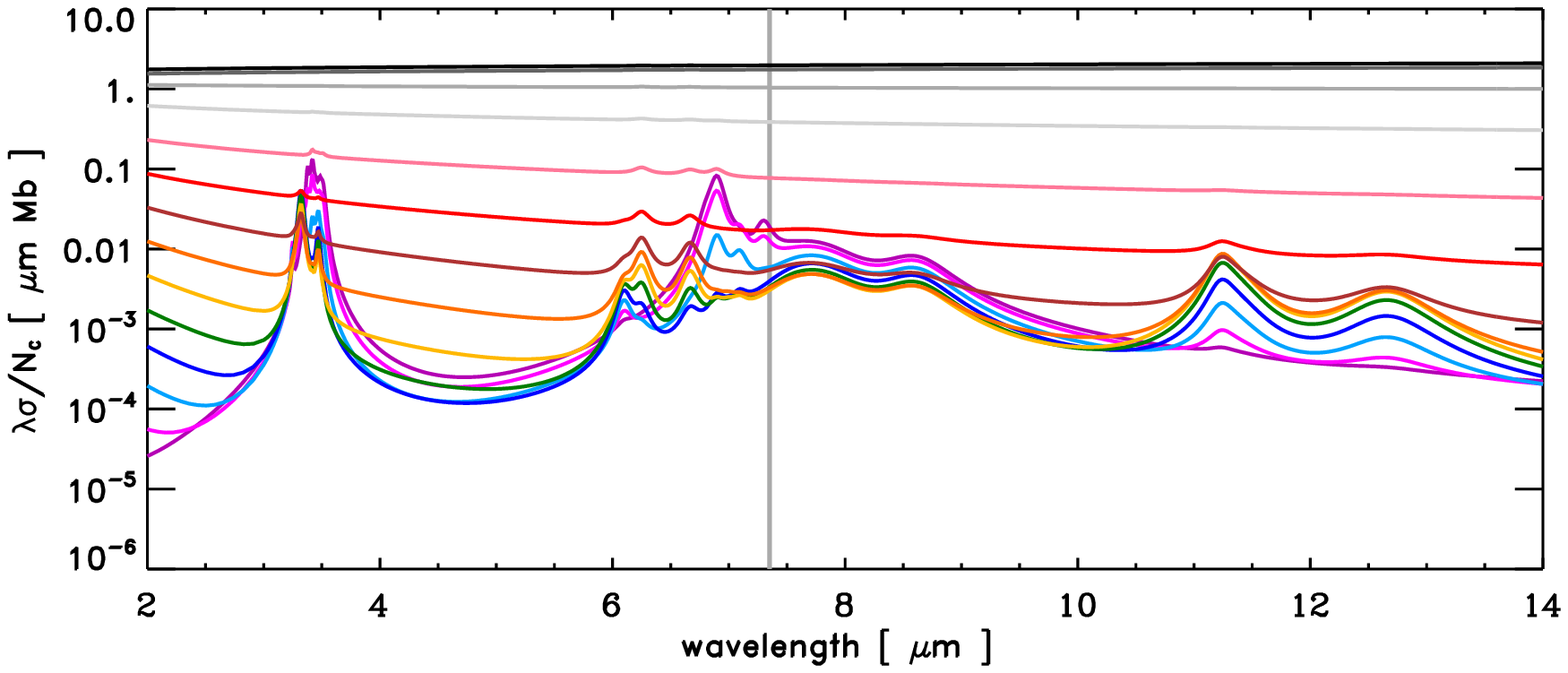}}
 \caption{The predicted spectra of optEC$_{\rm(s)}$(a)-modelled particles of radius 30\,nm, presented as the wavelength times absorption coefficient per carbon atom, $\lambda \alpha / N_{\rm C}$, in the $2-14\,\mu$m region (1\,Mb$= 10^{-18}$\,cm$^{2}$). N.B. The bands with central positions long-ward of the vertical grey line ($\lambda(\nu_0) > 7.3\,\mu$m)  are not yet well-determined by laboratory measurements.}
 \label{fig_alpha_spectrum_30nm}
\end{figure}

\begin{figure} 
\vspace*{-2cm}
 \resizebox{\hsize}{!}{\includegraphics{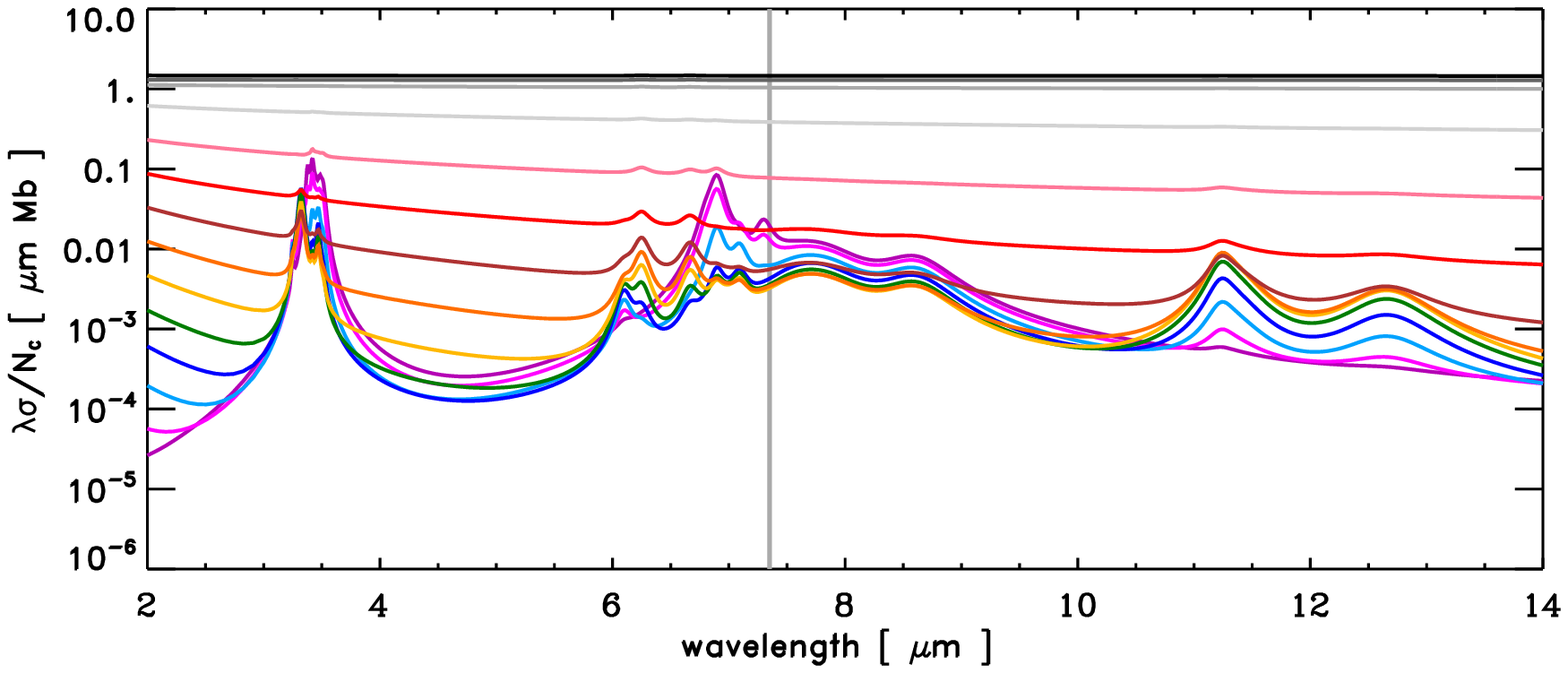}}
 \caption{As per Fig.~\ref{fig_alpha_spectrum_30nm} but for particles of radius 10\,nm.}
 \label{fig_alpha_spectrum_10nm}
\end{figure}

\begin{figure} 
\vspace*{-2cm}
 \resizebox{\hsize}{!}{\includegraphics{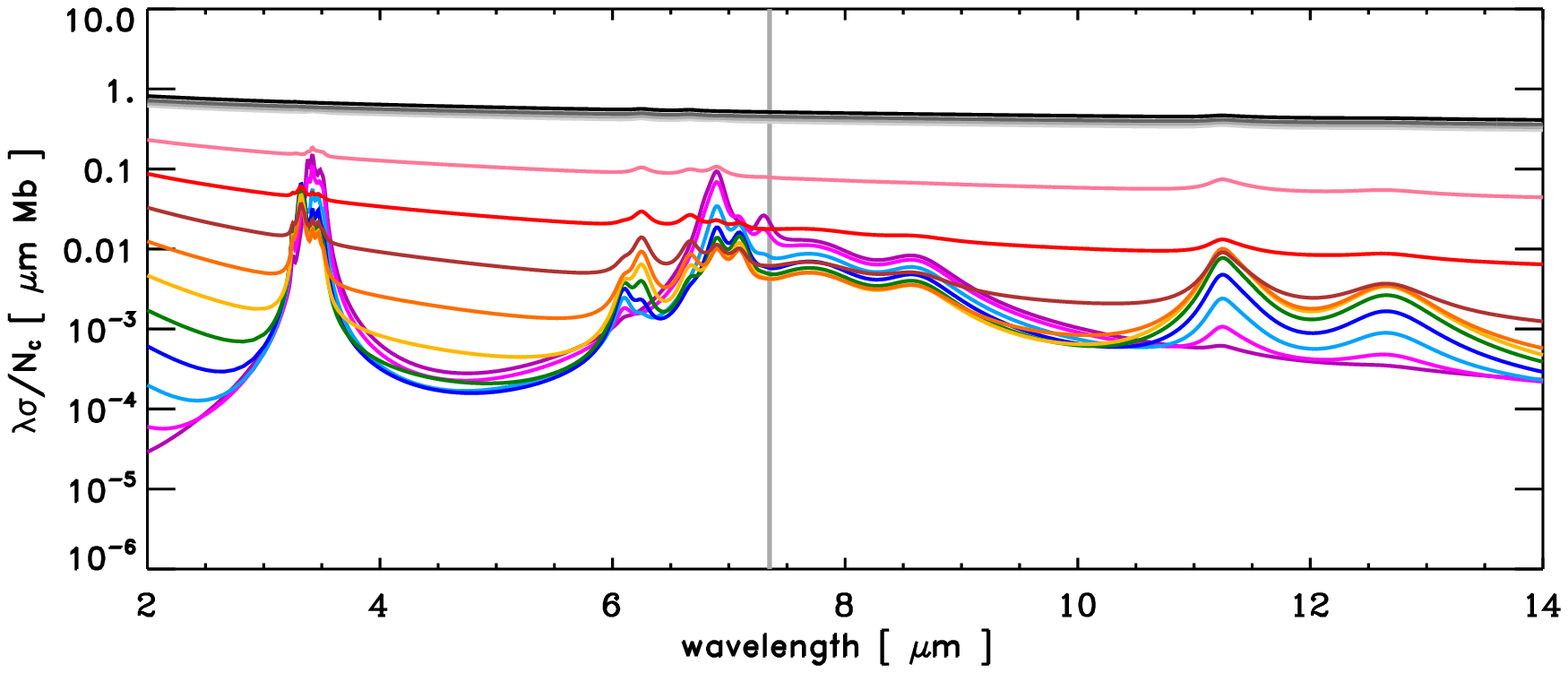}}
 \caption{As per Fig.~\ref{fig_alpha_spectrum_30nm} but for particles of radius 3\,nm.}
 \label{fig_alpha_spectrum_3nm}
\end{figure}

\begin{figure} 
\vspace*{-2cm}
 \resizebox{\hsize}{!}{\includegraphics{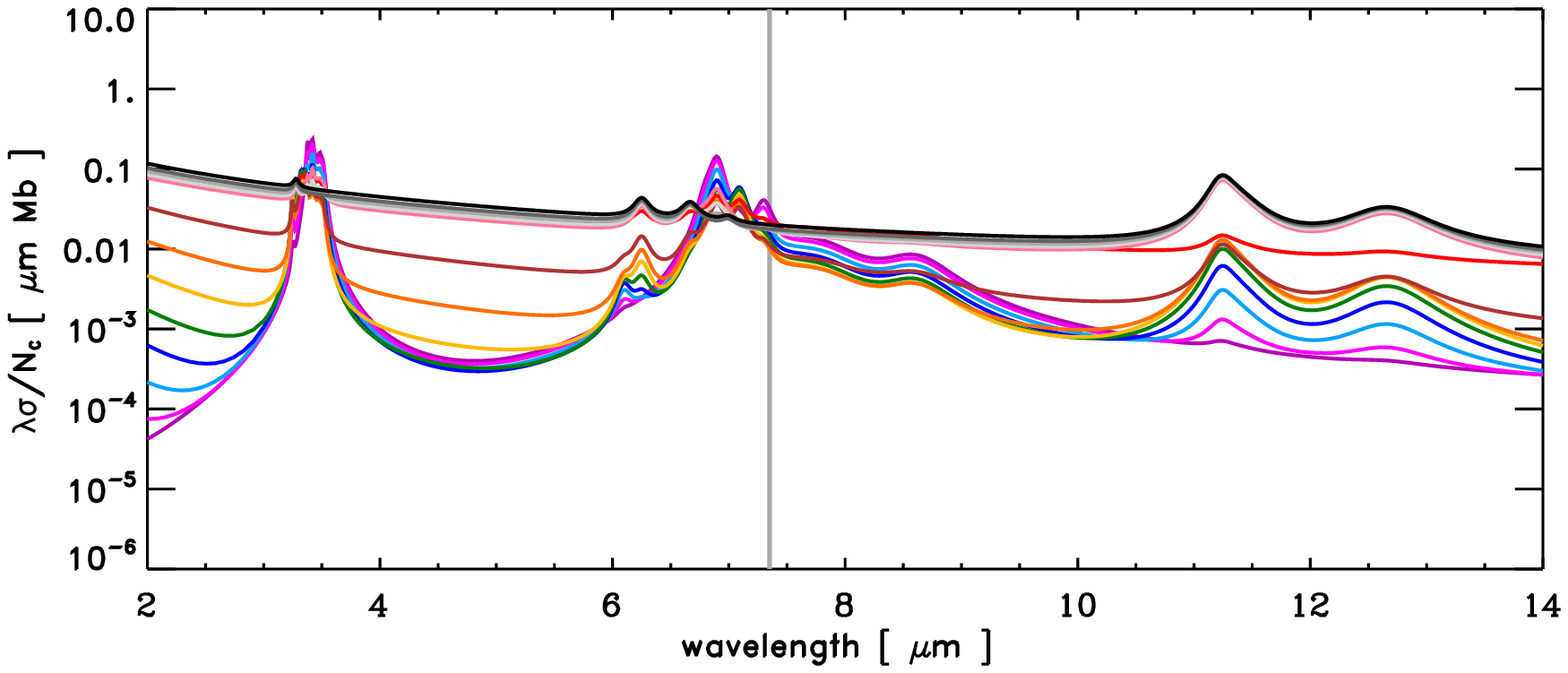}}
 \caption{As per Fig.~\ref{fig_alpha_spectrum_30nm} but for particles of radius 1\,nm.}
 \label{fig_alpha_spectrum_1nm}
\end{figure}

\begin{figure} 
\vspace*{-2cm}
 \resizebox{\hsize}{!}{\includegraphics{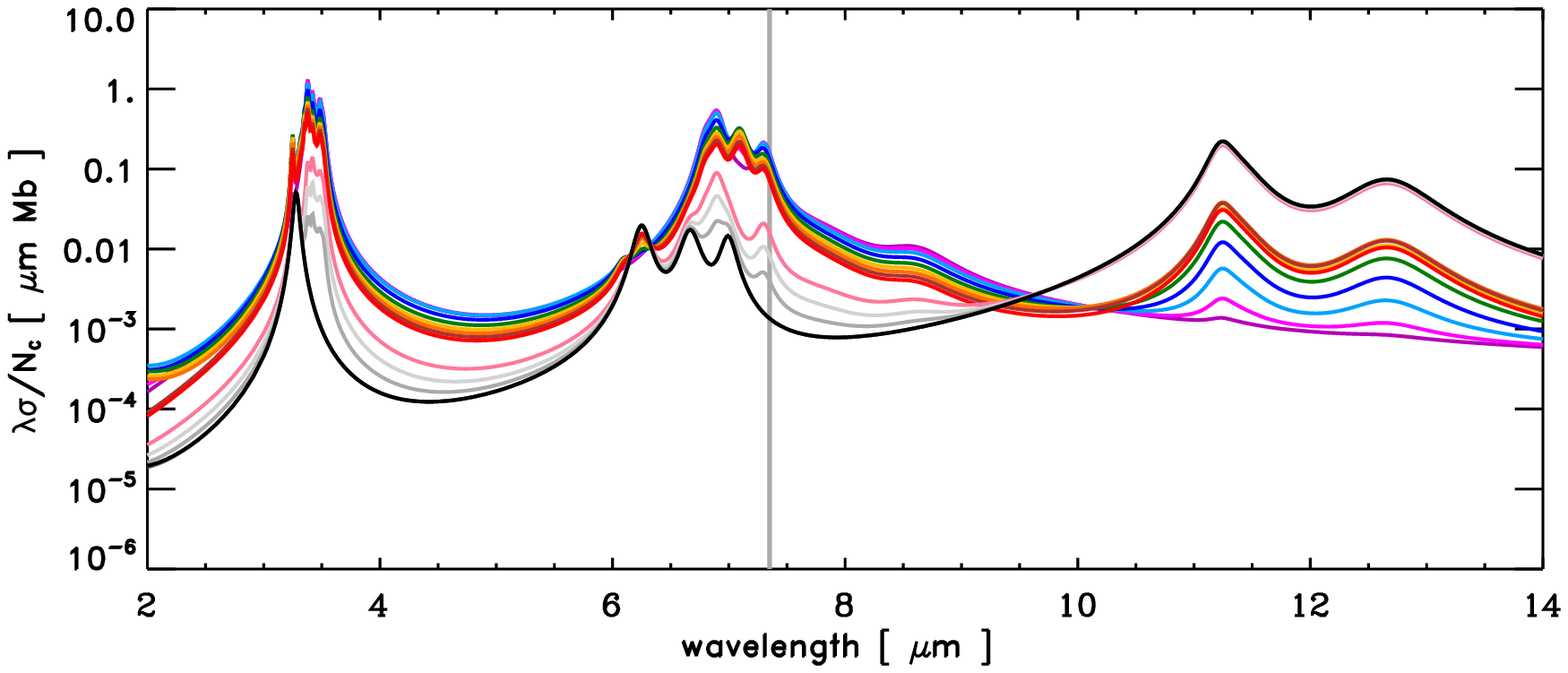}}
 \caption{As per Fig.~\ref{fig_alpha_spectrum_30nm} but for particles of radius 0.33\,nm.}
 \label{fig_alpha_spectrum_0.33nm}
\end{figure}

Figs.~\ref{fig_alpha_spectrum_30nm} to \ref{fig_alpha_spectrum_0.33nm} present the predicted spectra, based on the eRCN and DG models, for optEC$_{\rm(s)}$(a)-modelled particles of radii $30, 10, 3, 1$, and $0.33$\,nm, presented as wavelength times absorption coefficient per carbon atom, $\lambda \alpha / N_{\rm C}$, in the $2-14\,\mu$m region, as a function of wavelength and $E_{\rm g}$. The complementary spectra for 100 and 0.5\,nm radius particles are given in Figs.~\ref{fig_alpha_spectrum_100nm_eg} and \ref{fig_alpha_spectrum_0.5nm_eg}.\footnote{Note that the spectra plotted here are for different values of $X_{\rm H}$ than were given in paper~I.}  The adopted colour-coding for this plot is from large band gap (2.67\,eV, purple) to low band gap (-0.1\,eV, black) with intermediate values in steps of 0.25\,eV from 2.5 (violet) to 0\,eV (grey) with the addition of the $E_{\rm g} = 0.1$ and $-0.1$\,eV cases (light grey and black, respectively), {\it i.e.}, as per Table~\ref{table_params_colour_code}. 
 
The spectra presented here show a systematic evolution with both the band gap, $E_{\rm g}$, as already highlighted in papers~I and II, and perhaps more importantly with particle size. With decreasing band gap, $E_{\rm g}$, and as described in paper~I, the intrinsic CC and CH$_n$ modes responsible for the modelled absorption features show a clear transformation from $sp^3$-dominated CH$_n$ bands to $sp^2$-dominated CH$_n$ bands as aromatisation proceeds. Note that the  spectra for low band gap materials ($E_{\rm g} \leqslant1.5$\,eV) derived from the DG model now include the expected aromatic CH bands, which are due to the passivation/hydrogenation of the aromatic surface.

The spectra for particle radii down to $\simeq 3$\,nm really do not change substantially.
However, descending in radius from 3 to 0.33\,nm the following evolutionary sequence(s) is found: 
\begin{itemize}
  \item a general increase in CH modes with respect to CC modes as surface hydrogenation becomes more important, 
  \item a significant change in the spectra in the $7-9\,\mu$m region for particle radii smaller than 1\,nm as the aliphatic CC component contribution declines and  
  \item a surprisingly ``resilient'' aliphatic CH$_2$ band at $6.9\,\mu$m, accompanied by an olefinic CH$_2$ band at $7.1\,\mu$m, is seen in the spectra of materials with $E_{\rm g} \geqslant 0.1$\,eV.   
\end{itemize}

Note that bands with central positions long-ward of the vertical grey line ($\lambda > 7.3\,\mu$m) are not yet well-determined by laboratory measurements and should therefore be treated with some caution. 
It is also likely that the bands in the $6.5-7.5\,\mu$m region could be too strong in this model and may require re-calibration using better laboratory data when they becomes available.

%
%
%
%

\section{a-C:H heat capacities}
\label{appendix_heat_cap}

To zero order it can be assumed that the heat capacity of {\em bulk} a-C:H/a-C materials is the linear sum of the abundance-weighted heat capacities for the given atom in the given bonding configuration, $C^\prime_{\rm V}(x)_i$, of the constituent components or chemical groups as per \cite{1997ApJ...475..565D}, {\it i.e.}, 
\begin{equation}
C_{\rm V} = \sum_i \bigg\{ C^\prime_{\rm V}(\rm CH)_i + C^\prime_{\rm V}(\rm CC)_i \bigg\}.
\label{eq_Cv_1}
\end{equation}
Assuming a single heat capacity for the CH component and decomposing the CC component into its $sp^2$ and $sp^3$ constituents and normalising by the number of C atoms per unit volume, $N_{\rm C}$, gives
\begin{equation}
\frac{C_{\rm V}}{N_{\rm C}} = \frac{X_{\rm H} C_{\rm V}({\rm CH}) + X_{sp2} C_{\rm V}({\rm CC})_{sp2} + X_{sp3} C_{\rm V}({\rm CC})_{sp3}}{(1-X_{\rm H})}, 
\label{eq_Cv_2}
\end{equation}
which with the substitution $R = X_{sp3}/X_{sp2}$, and recalling that $[{\rm H}]/[{\rm C}] = X_{\rm H}/(1-X_{\rm H})$, the above equation can be re-written as 
\[
\frac{C_{\rm V}}{N_{\rm C}} = \frac{[{\rm H}]}{[{\rm C}]} C_{\rm V}({\rm CH}) \ + 
\] 
\begin{equation}
\ \ \ \ \ \ \ \ \ \ \frac{X_{sp3}}{(1-X_{\rm H})} \left[ \frac{C_{\rm V}({\rm CC})_{sp2}}{R} + C_{\rm V}({\rm CC})_{sp3} \right].
\label{eq_Cv_3}
\end{equation}
The key question is then: what heat capacities should be used for the different components?  For $C_{\rm V}({\rm CH})$ the values from \cite{1997ApJ...475..565D}, which were used to determine the heat capacities of PAHs in the ISM,  would seem to provide a reasonable approach. 
For the heat capacity of the $sp^2$ CC component, $C_{\rm V}({\rm CC})_{sp2}$, a value appropriate for graphite, basically as per \cite{1997ApJ...475..565D}, could be used for the aromatic component or a material typical of an olefinic structure, however, the former approach is rather well-established and is probably to be preferred.  
For $sp^3$ component heat capacity, $C_{\rm V}({\rm CC})_{sp3}$,  values for diamond or polyethylene could be used; the latter is probably to be preferred. 

For finite-sized hydrocarbon particles the above  expressions need to be modified to include the contribution from the surface-bond terminating, or surface passivating, H atoms. In this case the [H]/[C] term in Eq.~(\ref{eq_Cv_2}) has to be replaced by $X_{\rm H}^\prime/(1-X_{\rm H})$ from Eq.~(\ref{XH_size_dep}) to give 
\[
\frac{C_{\rm V}}{N_{\rm C}} = \frac{1}{(1-X_{\rm H})} \ \times 
\] 
\begin{equation} 
\ \ \ \ \ \ \ \ \ \ \Bigg\{ X_{\rm H}^\prime C_{\rm V}({\rm CH}) \ + \ X_{sp3} \left[ \frac{C_{\rm V}({\rm CC})_{sp2}}{R} + C_{\rm V}({\rm CC})_{sp3} \right] \Bigg\}.
\label{eq_Cv_4}
\end{equation}



\end{document}